\pdfoutput=1
\documentclass[3p]{elsarticle}

\usepackage[english]{babel}
\usepackage[bb=boondox,bbscaled=1.05,scr=dutchcal]{mathalfa}	%
\usepackage{dsfont}
\usepackage[T1]{fontenc}

\usepackage[dvipsnames]{xcolor}
\usepackage[hypertexnames=false,colorlinks=true,linkcolor=blue,%
citecolor=purple,filecolor=magenta,urlcolor=cyan]{hyperref}
\DeclareRobustCommand{\SkipTocEntry}[5]{} 

\usepackage{amssymb,amsmath,amsthm,anyfontsize,enumerate,layout,mathrsfs,mdwlist,stmaryrd,thmtools,doi}
\usepackage[textsize=footnotesize]{todonotes}
\presetkeys%
    {todonotes}%
    {color=Apricot}{}%
\usepackage{mathtools}

\usepackage[english,noabbrev]{cleveref}
\crefname{lemma}{lemma}{lemmata}
\Crefname{lemma}{Lemma}{Lemmata}

\theoremstyle{plain}                          
\newtheorem{theorem}{Theorem}[section]
\newtheorem{proposition}[theorem]{Proposition}    
\newtheorem{lemma}[theorem]{Lemma}
\newtheorem{corollary}[theorem]{Corollary}
\newtheorem{conjecture}[theorem]{Conjecture}
\theoremstyle{definition}
\newtheorem{definition}[theorem]{Definition}
\newtheorem{prop-defin}[theorem]{Proposition-definition} 
\newtheorem{example}[theorem]{Example}
\theoremstyle{remark}
\newtheorem{remark}[theorem]{Remark}

\renewcommand{\theta}{\vartheta}
\renewcommand{\phi}{\varphi}
\renewcommand{\epsilon}{\varepsilon}

\newcommand{\mb}[1]{\mathbb{#1}} 
\newcommand{\mf}[1]{\mathfrak{#1}}
\newcommand{\mc}[1]{\mathcal{#1}}

\newcommand{\R}{\mb{R}} 
\newcommand{\N}{\mb{N}} 
\newcommand{\C}{\mb{C}} 
\newcommand{\Z}{\mb{Z}} 

\renewcommand{\P}{\mb{P}}

\newcommand{\del}{\partial}

\DeclareMathOperator{\Tr}{Tr}

\DeclareMathOperator{\Erd}{Erd}

\DeclareMathOperator*{\Res}{Res}

\DeclareMathOperator{\Div}{div}

\DeclareRobustCommand{\Stircyc}{\genfrac[]{0pt}{}}

\begingroup\expandafter\expandafter\expandafter\endgroup
\expandafter\ifx\csname pdfsuppresswarningpagegroup\endcsname\relax
\else
\fi
{}

\begin{document}

\title{Topological recursion on transalgebraic spectral curves and Atlantes Hurwitz numbers}

\author[1]{Vincent Bouchard}
\ead{vincent.bouchard@ualberta.ca}

\author[1,2]{Reinier Kramer\texorpdfstring{\corref{cor1}}{cor}}
\ead{reinier.kramer@unimib.it}

\author[3]{Quinten Weller}
\ead{qgw1@nyu.edu}

\cortext[cor1]{Corresponding author}

\affiliation[1]{%
	\organization={University of Alberta}, 
	\addressline={632 Central Academic Building},
	\city={Edmonton},
	\postcode={AB T6G 2G1}, 
	\country={Canada}}

\affiliation[2]{
	\organization={Università di Milano-Bicocca}, 
	\addressline={Dipartimento di Matematica e Applicazioni, Via Roberto Cozzi, 55},
	\city={Milan},
	\postcode={20126}, 
	\country={Italy}}

\affiliation[3]{%
	\organization={New York University},
	\addressline={726 Broadway},
	\city={New York},
	\postcode={NY 10003}, 
	\country={USA}}

\begin{abstract}
	Given a spectral curve with exponential singularities (which we call a ``transalgebraic spectral curve''), we extend the definition of topological recursion to include contributions from the exponential singularities in a way that is compatible with limits of sequences of spectral curves. This allows us to prove the topological recursion/quantum curve correspondence for a large class of transalgebraic spectral curves. As an application, we find that Atlantes Hurwitz numbers, which were previously thought to fall outside the scope of topological recursion, satisfy (our extended version of) topological recursion, and we construct the corresponding quantum curve directly from topological recursion.
\end{abstract}


\maketitle

\tableofcontents


\section{Introduction}

\subsection{Motivation}

Topological recursion~\cite{EO07} is a method to recursively define a collection of multidifferentials $ \omega_{g,n} $ on a given object, called a spectral curve. It was originally developed to solve loop equations coming from matrix models \cite{CEO06}, but has applications to many other areas of mathematics as well: among others intersection theory on moduli spaces of curves~\cite{EO07,Eyn11,DOSS14}, volumes of moduli spaces~\cite{EO07a,ABCDGLW23}, Hurwitz theory~\cite{BM08,EMS11,BHLM14,DKPS23}, Gromov--Witten theory~\cite{BKMP08,EO15,DOSS14,NS14,FLZ17,FLZ20,GKLS22}, maps~\cite{BCDG19}, free probability~\cite{BCGLS21}, supersymmetric gauge theories \cite{BBCC24}, and integrable systems of various kinds~\cite{BEM17,EGMO21,BDKS20}.\par

The spectral curve of topological recursion is usually taken to be a Riemann surface $ \Sigma$, together with two meromorphic functions $x$ and $ y$ on it, and a symmetric bidifferential $ B$. The recursion only depends on the local behaviour of the spectral curve near the ramification points of $x$, which were originally required to be simple and regular points of $y$. These conditions on the ramification points have been lifted to a far higher generality~\cite{BHLMR13,BE13,CN19,DN18a,BBCCN18,BKS24}, allowing for higher ramification orders as well as certain poles of $y$.\par
Although topological recursion itself is local in nature, it behaves better if the spectral curve is global, i.e. if $x$ and $y$ are meromorphic functions on a compact Riemann surface. In this case, the functions $ x$ and $ y$ satisfy a polynomial relation $ P(x,y) = 0$, and, according to the topological recursion/quantum curve correspondence conjecture~\cite{BE09,GS11} (see also \cite{N15}), one should be able to quantise such an equation. Explicitly, there should exist an operator $ \hat{P}(\hat{x}, \hat{y},\hslash )$, where $ \hat{x} = x \cdot $ and $ \hat{y} = \hslash \frac{d}{dx}$, such that $ P = \hat{P} (x,y,0)$, and such that it annihilates the wave function
\begin{equation}\label{e:wf1}
	\psi(x(z)) = \exp\left[ \sum_{n=1}^{\infty} \sum_{g=0}^{\infty} \frac{\hslash^{2g+n-2}}{n!} \int^z\cdots\int^z \left( \omega_{g,n}  - \delta_{g,0} \delta_{n,2} \frac{dx(z_1) dx(z_2)}{(x(z_1) - x(z_2))^2} \right) \right] \,,
\end{equation}
that is, $ \hat{P} \psi = 0$, where $ z $ is a coordinate on a fundamental domain of the spectral curve. This is a subtle issue, in part due to non-unique quantisation and integration paths. Nevertheless, the correspondence was proven to hold for a large class of genus zero algebraic spectral curves with arbitrary ramification in~\cite{BE17}, and for all algebraic spectral curves with simple ramification in~\cite{EG23,MO22,EGMO21}.\par
Compact spectral curves exhibit more nice features: topological recursion is related to intersection theory of the moduli spaces of curves in a general setup~\cite{Eyn11,DOSS14}, and in case the spectral curve is compact, the intersection theory is well-behaved and largely independent of the choice of bidifferential~\cite{DNOPS15}.\par

But what if the spectral curve is of a form where $ x$ and $y$ do not satisfy an algebraic relation? This occurs in a large class of examples, mostly related to hypergeometric tau-functions and Hurwitz theory, see e.g. \cite{BEMS10,EMS11,DKPS23,BDKS22,BDKS20}, where the spectral curve is usually of the form $ x(z) = z e^{-\psi (y (z))}$, for some series $ \psi(y)$ and $ y(z)$. In important examples, both $ \psi$ and $ y$ are polynomial, so the function $ x$ has an essential singularity. Does the topological recursion/quantum curve correspondence conjecture hold for these cases?

In certain cases, for instance in~\cite{MSS13}, and also the more general setup of \cite{ALS16}, it is shown that the differentials $\omega_{g,n}$ produced by topological recursion are generating series for certain types of Hurwitz numbers. Using this interpretation, it is then proven that a quantum curve exists, albeit a fairly complicated one. However, the quantum curves were constructed from the enumerative interpretation in terms of Hurwitz numbers, rather than from the spectral curve itself, which is somewhat unsatisfactory. Can we construct the quantum curves directly from the spectral curve using topological recursion, in the spirit of \cite{BE17,EG23,MO22,EGMO21}?

\subsubsection{An observation}

In fact, this project started with the following observation. Consider topological recursion on the spectral curve $\mathcal{S}$ given by the equation
\begin{equation}\label{e:sc1}
y - e^{x^r y^r} = 0,
\end{equation}
for $r \in \mathbb{Z}_{\geq 1}$.\footnote{Our spectral curve looks a bit different from \cite{MSS13,DKPS23}, but this is because we are taking the one-form $\omega_{0,1}$ to be $y dx$ instead of $y \frac{dx}{x}$. Ultimately, it is the same spectral curve.} As shown in \cite{DKPS23}, the differentials $\omega_{g,n}$ produced by topological recursion are generating series for $r$-completed cycles (also called $r$-spin) Hurwitz numbers. Moreover, using the semi-infinite wedge space interpretation of Hurwitz numbers, it was proven in \cite{MSS13} that these Hurwitz numbers satisfy a quantum curve in the sense above. The resulting quantum curve is a quantisation of the spectral curve, but a rather complicated one:
\begin{equation}\label{e:qchur}
	\hat{P} = \hat{y} - \hat{x}^{1/2}{\rm e}^{\frac{1}{r+1} \sum_{i=0}^{r} \hat{x}^{-1}(\hat{x}\hat{y})^i\hat{x}(\hat{x}\hat{y})^{r-i}}\hat{x}^{-1/2},
\end{equation}
with $\hat{x} = x \cdot$ and $\hat{y} = \hslash \frac{d}{dx}$.

To obtain this quantum curve  directly from topological recursion, in the spirit of \cite{BE17}, one approach is to consider a sequence of genus zero compact spectral curves $\mathcal{S}_N$, specified by rational functions $x_N$ and $y_N$, such that $x_N$ and $y_N$ go to the above $x$ and $y$ in the limit $N \to \infty$. (Schematically, $\lim_{N \to \infty} \mathcal{S}_N = \mathcal{S}$.) For all positive integers $N$, topological recursion produces differentials $\omega_{g,n}^N$. From these differentials, one can construct a wave function $\psi_N$ as in \eqref{e:wf1}. If the genus zero spectral curves for finite $N$ fall within the class studied in \cite{BE17}, then we know right away that there exist quantum curves $\hat{P}_N$ such that $\hat{P}_N \psi_N = 0$ for all positive integers $N$, and we can construct $\hat{P}_N$ explicitly. Finally, we can take the limit $N \to \infty$ to get a quantum curve $\hat{P}_\infty$. 

Assuming that the $N\to \infty$ limit of the differentials $\omega_{g,n}^N$ recovers the differentials $\omega_{g,n}$ of the $N\to \infty$ spectral curve, which we could rewrite schematically as the condition
\begin{equation}\label{e:trcommute}
\lim_{N \to \infty} \left( \omega_{g,n}^N[ \mathcal{S}_N]  \right) = \omega_{g,n} \left[ \lim_{N \to \infty} \mathcal{S}_N \right],
\end{equation}
then the limiting quantum curve $\hat{P}_\infty$ should annihilate the wave function $\psi$ constructed from the differentials $\omega_{g,n}$ by \eqref{e:wf1}. However, the quantum curve $\hat{P}_\infty$ that we obtain using a specific sequence in \cref{s:QC} in this way reads
\begin{equation}\label{e:qcat}
\hat{P}_\infty = \hat{y} - e^{(\hat{x} \hat{y})^r},
\end{equation}
which is not the same as the quantum curve \eqref{e:qchur} that was obtained for $r$-completed cycles Hurwitz numbers! Both are quantisations of the same spectral curve, but they are certainly different, and annihilate different wave functions. What is going on?

Meanwhile, the quantum curve $\hat{P}_\infty$ from \eqref{e:qcat} already appeared in the work of \cite{ALS16}, where it was proved to annihilate the wave function constructed from differentials that are generating series for another type of Hurwitz numbers, known as Atlantes Hurwitz numbers. In fact, it was already noticed in that paper that the quantum curve $\hat{P}_\infty$ from \eqref{e:qcat} and the quantum curve $\hat{P}$ from \eqref{e:qchur} are both quantisations of the same spectral curve. Since it was known that topological recursion on this spectral curve produces generating series for $r$-completed cycle Hurwitz numbers, this observation was taken as an indication that Atlantes Hurwitz numbers fall outside the scope of topological recursion. To quote \cite{ALS16}: ``We have an example where the dequantization of the quantum curve doesn't give a spectral curve suitable for the corresponding topological recursion.'' They also state: ``We can conclude that the dequantization of $\hat{y} - e^{\hat{x}^r \hat{y}^r}$ cannot be the spectral curve for the atlantes Hurwitz numbers, suitable for the construction of the topological recursion.'' 

But... is this really the end of the story?

\subsubsection{A resolution}

In this paper we resolve this conundrum and propose an explanation for this observation. The key is that for topological recursion to commute with limits of sequences of spectral curves as in \eqref{e:trcommute}, exponential singularities of the limiting spectral curve must be taken into account. More precisely, given a spectral curve with exponential singularities, one can construct differentials $\omega_{g,n}$ by using topological recursion ignoring the exponential singularities, as has been done so far in the literature. But, as we propose in this paper, one can also construct another set of differentials, call them $\omega_{g,n}^\infty$, using an extension of topological recursion that includes contributions from exponential singularities (informally considered as ``ramification points of infinite order''). In general, for a given spectral curve with exponential singularities, the differentials $\omega_{g,n}^\infty$ and $\omega_{g,n}$ will be distinct. It turns out that, as we prove in this paper, topological recursion commutes with limits of sequences of spectral curves as in \eqref{e:trcommute} only if the differentials on the right-hand-side are the differentials $\omega_{g,n}^\infty$ that include contributions from exponential singularities.

This explains the observation above. For the spectral curve \eqref{e:sc1}, the differentials $\omega_{g,n}$ that ignore the exponential singularities are generating series for $r$-completed cycles Hurwitz numbers, as shown in \cite{DKPS23}. However, as we show in the current paper, the differentials $\omega_{g,n}^\infty$ that include contributions from the exponential singularity are generating series for Atlantes Hurwitz numbers. This shows that Atlantes Hurwitz numbers do fall within the scope of topological recursion, once the formalism is properly extended to include contributions from exponential singularities.

Furthermore, since we show that topological recursion (properly extended to include contributions from exponential singularities) commutes with limits of sequences of spectral curves, we obtain directly that the wave function $\psi_\infty$ constructed from the differentials $\omega_{g,n}^\infty$ is annihilated by the quantum curve $\hat{P}_\infty$ from \eqref{e:qcat}. This provides a construction of the quantum curve for Atlantes Hurwitz numbers directly from topological recursion, and explains why it differs from the quantum curve for $r$-completed cycles Hurwitz numbers that was obtained in \cite{MSS13}.

\subsection{Main results}

We propose an extension of topological recursion that includes contributions from exponential singularities of the spectral curves, i.e. points $ p \in \Sigma $ where $ x(z) \sim M_0(z) e^{M_1(z)} $ as $ z \to p$ and $M_0$, $M_1$ are meromorphic functions with $ p$ a pole of $ M_1$. We will consider these singularities as ramification points of infinite order. Such functions are called ``transalgebraic'', and hence we will call these spectral curves ``transalgebraic spectral curves''.\par

As the topological recursion formula involves sums over local deck transformations, infinite order ramification points require infinite sums, and this leads to multiple issues; chief amongst these is the definition of the residue at what may not be an isolated singularity. Instead of dealing with these issues directly, we construct topological recursion on transalgebraic spectral curves as a limit of topological recursion on sequences of finite degree, meromorphic, spectral curves, as in \eqref{e:trcommute}. For this definition to make sense, we need to make sure that the $N \to \infty$ limit of the differentials exists and satisfies desired properties, which we do.

Furthermore, we show that while the definition of topological recursion on transalgebraic spectral curves is fairly complicated (having to do with limits of sequences of spectral curves, although we do provide a formula applicable in some specific cases), in the end it is not too bad: for any given transalgebraic spectral curve, the formal definition has to be used only finitely many times, after which the exponential singularity does not contribute any further terms.\par

With this construction in hand, we study the topological recursion/quantum curve correspondence, with the aim of constructing quantum curves directly from topological recursion for transalgebraic spectral curves. For a subclass of transalgebraic curves, which we call regular, we adapt the argument of \cite{BE17} to construct the quantum curves associated to these transalgebraic spectral curves.\par

As an application, we show that Atlantes Hurwitz numbers, which were introduced in \cite{ALS16} as an example of Hurwitz numbers not satisfying topological recursion, do fit in our transalgebraic framework. We show that the differentials constructed from topological recursion (suitably extended to include contributions from exponential singularities) on the spectral curve \eqref{e:sc1} are generating series for Atlantes Hurwitz numbers (while the differentials constructed from the usual topological recursion that ignores exponential singularities on the ``same'' spectral curve\footnote{In fact, as we will explain, it is better to think of these two spectral curves as distinct spectral curves, in the following sense. While the functions $x$ and $y$ satisfy the same plane curve equation, and the symmetric bidifferential $B$ is given by the same formula for both, the two spectral curves have two different Riemann surfaces. For the spectral curve for $r$-completed cycles Hurwitz numbers, we take the Riemann surface to be $\Sigma=\mathbb{C}$, with the exponential singularity of $x$ and $y$ at infinity removed, while for the spectral curve for Atlantes Hurwitz numbers we take the Riemann surface to be $\Sigma = \mathbb{P}^1$, which includes the exponential singularity (see Examples \ref{ex:rspin} and \ref{ex:atlantes}).} are generating series for $r$-completed cycles Hurwitz numbers). Finally, we prove that the corresponding wave function is annihilated by the quantum curve \eqref{e:qcat} directly from topological recursion. 

\subsection{New developments}

Recently, in \cite{ABDKS24}, Alexandrov, Bychkov, Dunin-Barkowski, Kazarian, and Shadrin provided a new topological recursion based method that yields correlators $\omega_{g,n}$ that are generating functions of Atlantes Hurwitz numbers; \cite{ABDKS24} then shows that these $\omega_{g, n}$ are exactly the ones produced by our transalgebraic topological recursion with $x(z) = ze^{-z^r}$ and $y(z) = z$ (the curve of \cref{ex:atlantes}). We outline this approach below.

There is a duality in the topological recursion that comes from switching the roles of $x$ and $y$. In \cite{ABDKS22}, the same authors proved a formula which explicitly relates the correlators of the $x$-$y$ swapped spectral curve with those of the original curve, provided that $x$ and $y$ are meromorphic. In \cite{ABDKS23}, they realised this relation can be extended to cases when only $dx$ and $dy$ were meromorphic, provided one slightly adjusts the definition of the topological recursion at the simple poles of $dy$ (this adjustment was called `LogTR').

In \cite{ABDKS24}, they studied the spectral curve with
\begin{equation}
	x(z) = \log z - z^r \,, \qquad y(z) = z \,.
\end{equation}
Then, they swapped the roles of $x$ and $y$ to obtain the dual spectral curve. Next, the $n=1$ dual correlators were shifted by some prescribed terms and the $x$-$y$ swap formula was applied to bring them back to the `original' curve, but with new correlators. These correlators were not shown to satisfy any topological recursion relation, but were just defined formally through the $x$-$y$ swap formula as described above. However, the above process did yield explicit formulas for the correlators \cite[equation (91)]{ABDKS24}, due to the simplicity of the dual spectral curve ($y$ is unramified).


\subsection{Outline of the paper}

In Section \ref{s:TR} we define spectral curves and review the topological recursion framework for meromorphic spectral curves with arbitrary ramification. In Section \ref{s:transalgebraic} we define transalgebraic spectral curves and explain how they can be realized as limits of sequences of meromorphic spectral curves. We then proceed in Section \ref{s:TRtransalgebraic} with the definition of topological recursion on transalgebraic spectral curves and prove various properties of this extension of topological recursion, including the fact that essential singularities contribute to topological recursion only in a finite number of steps. 

In Section \ref{s:QC} we prove the topological recursion/quantum curve correspondence for a large class of transalgebraic spectral curves, which we call regular. This class includes the spectral curve for $r$-completed cycles and Atlantes Hurwitz numbers. We focus on this particular curve in Section \ref{TRAtlantes}, where we show that topological recursion on this transalgebraic spectral curve (including the exponential singularity) produces generating functions for Atlantes Hurwitz numbers. We conclude with open questions in Section \ref{s:open}.

Appendix \ref{QCForTransAlgTR} provides the extension of the results of \cite{BE17} needed to study the topological recursion/quantum curve correspondence for transalgebraic spectral curves, while Appendix \ref{ProofOfAtlantesBDKS} contains the proof of \cref{NoPolesForn>1} about Atlantes Hurwitz numbers.

\subsection{Notation} \label{notation}

We set $ \mc{S}(z) = \frac{\sinh (z/2)}{z/2}$.\par
For a natural number $ n$, we define $ \llbracket n \rrbracket \coloneqq \{ 1, \dotsc, n\}$. Given a set $S$ indexed by another set $I$, i.e. $ S = \{ s_i \, | \, i \in I \}$, and given a subset $ J \subseteq I$, we denote $ s_J \coloneqq \{ s_i \, | \, i \in J \}$. In particular, $ s_I = S$. For a set $Z$, we use the notation $\mu \vdash Z$ to indicate that $\mu$ is a set partition of $Z$; the length of this partition is then denoted by $l(\mu)$.\par
For a curve $ \Sigma $ with local coordinate $ z$, we denote the induced coordinates on $ \Sigma^n$ by $ ( z_1, \dotsc, z_n )$. With the previously given notation, this can be denoted $ z_{\llbracket n \rrbracket } $ -- this is a slight abuse of notation, as we confuse sets and ordered lists, but we will only use it for arguments of symmetric multidifferentials.\par
Given $C=\{z\} \cup z_{\llbracket n \rrbracket}\subset\Sigma$ and $C'=C\setminus\{z\}$ sets of cardinality $n+1$ and $n$, respectively, take a symmetric $n$-differential $\eta$ and define
\begin{equation}
	\Res_{C=z}\eta(z_{\llbracket n \rrbracket}) \coloneq \Res_{C'=z}\eta(z_{\llbracket n \rrbracket}) \coloneq \Res_{z_1=z} \cdots \Res_{z_n=t} \eta(z_{\llbracket n \rrbracket})\,.
\end{equation}
Since $\eta$ is symmetric, this notation makes sense as the order in which we take the residues does not matter. In a similar spirit, we also define 
\begin{equation}
	\Res_{z \in C'} \coloneq \sum_{z_0\in C'} \Res_{z=z_0}\,.
\end{equation}
For any set of points $ C \subset \Sigma $ we denote by $ z^C $ one arbitrarily chosen point in this set. Lastly, as we will have to take many residues at once, we define, for fixed points $a_1,\dots,a_n \in \Sigma$ 
\begin{equation}
	\Res_{\substack{z_l=a_l\\l=1,\dots,n}} \coloneq \Res_{z_1=a_1} \cdots \Res_{z_n=a_n}\,,
\end{equation}
along with the obvious generalisations to our residues over sets notations.

\section{Topological recursion on meromorphic spectral curves}

\label{s:TR}

\subsection{Spectral curves}

One of the main goals of this paper is to extend the definition of topological recursion to spectral curves with exponential singularities, which we will call transalgebraic spectral curves. Let us start by recalling the usual formulation of topological recursion, in the Bouchard-Eynard formalism \cite{BE13}, which extends the original Eynard-Orantin formalism \cite{EO07} to higher order ramification. We start with the definition of a spectral curve. 

\begin{definition} \label{d:AlgSC+Adm}
	A \emph{spectral curve} is a quadruple $ \mc{S} = (\Sigma, x, y, B)$, where:
	\begin{enumerate}
	\item $\Sigma$ is a Riemann surface;
	\item $x$  and $y$ are functions on $\Sigma$ that are holomorphic except potentially at a finite number of points and that separate points, i.e. $ (x,y)$ is injective;
	\item $B$ is a symmetric bi-differential on $\Sigma \times \Sigma$ with a double pole on the diagonal with biresidue $1$ and no other poles.
	\end{enumerate}
	We say that the spectral curve is \emph{meromorphic} if $x$ and $y$ are meromorphic on $\Sigma$ and the ramification locus $R$ of $ x $, which is the set of all zeros of $ dx $ and all poles of $ x $ of order $\geq 2$, is finite.
\end{definition}

The usual Bouchard-Eynard formalism will correspond to the case of meromorphic spectral curves, whereas our extension will be to a certain class of non-meromorphic spectral curves. Let us now review some of the basic features of meromorphic spectral curves.

We write $ r_a $ for the ramification order of $ x $ at $a \in R$. For a point $ z \in \Sigma$, we write $ \mf{f}(z) = x^{-1}(x(z))$ for the fibre, and $ \mf{f}'(z) = \mf{f}(z) \setminus \{ z\}$. Also, if $ a \in R$ and $ z$ is close to $ a$, we write $ \mf{f}_a(z)$ for the local Galois conjugates of $ z$, and again $ \mf{f}'_a(z) = \mf{f}_a(z) \setminus \{ z\}$. We note that while $\mathfrak{f}_a(z)$ is always finite of cardinality $r_a$, $\mathfrak{f}(z)$ may be countably infinite, as $x : \Sigma \to \mathbb{P}^1$ is not necessarily a finite branched covering if $\Sigma$ is non-compact.
	
For $ a \in R$, define a local coordinate $\zeta_a$ on a neighbourhood of $ a$ by $x = x(a) + \zeta_a^{r_a} $ if $ x (a) \neq \infty $ and $x =  \zeta_a^{-r_a} $ if $ x(a) = \infty$. Then the one-form $\omega_{0,1} \coloneqq y dx$ has an expansion:
\begin{equation}
	\omega_{0,1}   \coloneqq y dx = \sum_{k = -l}^\infty t^a_k \zeta_a^{k-1} d\zeta_a
\end{equation}
for some $l$ and $ t^a_k$. Let $ s_a \coloneqq \min \{ k \, | \, t^a_k \neq 0 \text{ and } r_a \nmid k \} $. The following admissibility condition on spectral curves is required to make sense of topological recursion \cite{BBCCN18}:
	
\begin{definition}\label{d:AlgSCAdm}
A meromorphic spectral curve is \emph{admissible} if for every point $a\in R$, $s_a$ and $r_a$ are coprime, and either  $s_a \leq -1$, or $1 \leq s_a \leq r_a + 1$ with $ r_a = \pm 1 \pmod{s_a}$.
\end{definition}

\begin{remark}\label{r:free}
	The ramification points $a \in R$ with $ s_a \leq -1 $ never contribute to the Eynard-Bouchard topological recursion \cite{BBCCN18}. This means that any ramification point $a$ that is a pole of $x$ may be dropped from the topological recursion unless $dy$ has a zero at $a$. As a result it is common practice in the topological recursion literature to sloppily refer to the set of zeros of $dx$ as all the ramification points of $x$ \cite{EO07,BE13}. We will see that, for our purposes, it is critical to include the poles of $x$ of order greater than two as ramification points.
\end{remark}

A particularly interesting class of spectral curves is when the Riemann surface $\Sigma$ is compact.

\begin{definition}\label{d:AlgComp}
A spectral curve is \emph{compact} if $\Sigma$ is a compact connected Riemann surface.
\end{definition}

Compact meromorphic spectral curves have nice geometric properties. If $\Sigma$ is compact, $x$ and $y$ are two meromorphic functions on a compact Riemann surface, and hence they identically satisfy an algebraic equation
\begin{equation}
P(x,y) = 0,
\end{equation}
where $P$ is a polynomial. (Note that if $\Sigma$ is non-compact, $x$ and $y$ may still satisfy a relation as above, but $P$ could no longer be a polynomial -- see \cref{ex:rspin}.)

We also note that for compact meromorphic spectral curves, since $x$ is a meromorphic function on a compact Riemann surface $\Sigma$, $x: \Sigma \to \mathbb{P}^1$ is a finite degree branched covering. This means that $\mathfrak{f}(z)$ is finite and of cardinality given by the degree of $x$ if $ z \notin R$.

\begin{example}\label{ex:rs}
Consider the spectral curve $\mathcal{S}$ with $\Sigma = \P^1$, $x =  z^r$, $y = z^{s-r}$, and $B = \frac{d z_1 dz_2}{(z_1-z_2)^2}$, with $r,s$ integers such that $r \geq 2$, $1 \leq s \leq r + 1$, and $ r = \pm 1 \pmod{s}$. This is a compact meromorphic spectral curve, and $x$ and $y$ satisfy the algebraic equation $x^{r-s} y^r- 1 = 0$ if $s < r$, or $y^r - x = 0$ if $s = r+1$. One can also check that the spectral curve is admissible. In particular, this is the fundamental $(r,s)$ spectral curve studied in \cite{BBCCN18}.
\end{example}

\begin{example}\label{ex:rspin}
Consider the spectral curve $\mathcal{S}$ with $\Sigma = \mathbb{C}$, $x = z e^{-z ^r}$, $y = e^{z^r}$, and $B = \frac{d z_1 dz_2}{(z_1-z_2)^2}$, with $r \in \mathbb{Z}_{\geq 1}$. As the function $x$ is meromorphic on $\mathbb{C}$ (it is in fact holomorphic), this spectral curve is meromorphic, but it is not compact. One can check that it is admissible. The functions $x$ and $y$ satisfy the relation
\begin{equation}
y-e^{x^r y^r} = 0,
\end{equation}
which is not algebraic. This spectral curve will play an important role in the following. As proven in \cite{DKPS23}, the differentials $\omega_{g,n}$ produced by topological recursion on this spectral curve are generating functions for $r$-completed cycles Hurwitz numbers. We will come back to this enumerative geometric interpretation in \cref{TRAtlantes}.\footnote{To avoid confusion, we remark that for this spectral curve $y$ is often defined in the literature via $\omega_{0,1} = y d \log x$ instead of $\omega_{0,1} = y dx$, which gives $y=z$ instead of $y = e^{z^r}$. The two definitions are of course equivalent, as it simply amounts to redefining $y \mapsto x y$. In this paper, for all spectral curves, the one-form will be defined as $\omega_{0,1} = y  dx$.}
\end{example}

\begin{remark}\label{re:compact}
We emphasize here that the choice of Riemann surface $\Sigma$ in the definition of a spectral curve is very important. For instance, if we replace the Riemann surface $\P^1$ by $\mathbb{C}$ in \cref{ex:rs}, it should be considered as a different spectral curve, since the pole of $x$ at infinity is not included in the Riemann surface. In this case, as $x$ is meromorphic on $\P^1$ and holomorphic on $\mathbb{C}$, the usual topological recursion (which applies to meromorphic spectral curves) can be used to calculate correlators $\omega_{g,n}$ for both spectral curves, and it happens that the correlators coincide, since the pole at infinity does not contribute to the topological recursion. But this will not always be the case.

For instance, one may want to consider the spectral curve of \cref{ex:rspin}, but with the Riemann surface $\Sigma = \P^1$, which includes the exponential singularity of $x$ at infinity. We claim that the usual topological recursion does not apply in this case, as it only applies to meromorphic spectral curves; instead, one should use the extended version that we propose, in which the exponential singularity generally contributes. As we will see, with this extended definition topological recursion produces different correlators (for $r \geq 2$) depending on whether the essential singularity at infinity is included in the Riemann surface or not, i.e. whether $\Sigma$ is taken to be $\P^1$ or $\mathbb{C}$ in the spectral curve of \cref{ex:rspin}.
\end{remark}

\subsection{Topological recursion}

The standard definition of topological recursion applies to admissible meromorphic spectral curves. It does not however require the spectral curve to be compact; for instance, it can be applied to both spectral curves in \cref{ex:rs,ex:rspin}. Out of the data of the spectral curve, a collection of symmetric differentials $ \{ \omega_{g,n} \}_{g\geq 0, n \geq 1}$ are recursively computed. As mentioned in the introduction, topological recursion is interesting because for many spectral curves, the correlators $\omega_{g,n}$ that it produces are generating functions for interesting enumerative invariants, such as Hurwitz numbers, Gromov-Witten invariants, etc.\par

For the definition of (Bouchard--Eynard) topological recursion, recall the notation given in \cref{notation}.

\begin{definition} \label{BE-TR}
	Given an admissible meromorphic spectral curve $\mathcal{S} = (\Sigma, x, y, B)$, \emph{topological recursion} gives a procedure to define multi-differentials $ \{ \omega_{g,n} \}_{g\geq 0, n \geq 1}$, recursively on $ 2g-2+n$, as follows: the base cases are $\omega_{0,1} = y dx$ and $\omega_{0,2}  = B$, and the recursive step is
	\begin{equation} \label{TR-RecursiveStep}
		\omega_{g,n+1}(z_0, z_{\llbracket n\rrbracket}) \coloneqq \sum_{a\in R} \Res_{z = a}\sum_{\emptyset \neq Z' \subseteq \mf{f}'_a(z)} K_{|Z|+1}(z_0,z, Z') \mc{W}_{g,n,|Z|+1}(z, Z \mid z_{\llbracket n \rrbracket})\,,
	\end{equation}
	where
	\begin{equation}
		K_{|Z|+1} (z_0, z, Z )  \coloneqq \frac{\int_{*}^z \omega_{0,2}(z_0, \cdot)}{\prod_{z' \in Z} (y (z') -y(z))dx(z)}
	\end{equation}
	is the \emph{recursion kernel} and
	\begin{equation}
		\mc{W}_{g,n,|Z|} (Z \mid z_{\llbracket n \rrbracket}) \coloneqq \sum_{\substack{\mu \vdash Z\\ \bigsqcup_{k=1}^{\ell (\mu)} N_k = \llbracket n \rrbracket\\ \sum g_k = g + \ell (\mu) -n}}' \prod_{k=1}^{\ell (\mu)} \omega_{g_k,|\mu_k| + |N_k|}(\mu_k, z_{N_k})\,,
	\end{equation}
	where the prime on the summation means we omit any term containing $ \omega_{0,1}$. The differentials $\omega_{g,n}$ are often called \emph{correlators} due to their origin in matrix models.
\end{definition}

	For future reference, we also define
	\begin{equation}
		\mc{E}_{g,n,|Z|} (Z \mid z_{\llbracket n \rrbracket}) \coloneqq \sum_{\substack{\mu \vdash Z \\ \bigsqcup_{k=1}^{\ell (\mu)} N_k = \llbracket n \rrbracket\\ \sum g_k = g + \ell (\mu) -n}} \prod_{k=1}^{\ell (\mu)} \omega_{g_k,|\mu_k| + |N_k|}(\mu_k, z_{N_k})\,,
	\end{equation}
	where terms containing $\omega_{0,1}$ are now included.
	
	Topological recursion was originally found as a method to obtain the unique solution to the loop equations of matrix models in mathematical physics that satisfies a particular normalisation condition. This can be defined more mathematically, in terms of the so-called abstract loop equations, with the normalisation condition often called the ``projection property''.

\begin{proposition}[{\cite[Appendix~C]{BBCCN18}}] \label{AbsLoopEqns}
	A collection of multidifferentials $ \{ \omega_{g,n} \}_{g \geq 0, n \geq 1}$ satisfies topological recursion if and only if it satisfies the \emph{higher abstract loop equations}:
	\begin{equation}
		\sum_{\substack{Z \subset \mf{f}_a(z) \\ |Z | = i } } \mc{E}_{g,n, i} ( Z \, | \, z_{\llbracket n \rrbracket} ) = \mc{O} \Big( z^{-r_a \mf{d}_a^i} \big(\frac{dz}{z}\big)^i \Big) \quad \text{as } z \to a\,,
	\end{equation}
	where
	\begin{equation}
		\mf{d}_a^i \coloneqq - 1 - \Big\lfloor \frac{s_a (i-1)}{ r_a } \Big\rfloor\,,
	\end{equation}
	and the \emph{projection property}: if $ 2g - 2 + n \geq 0$,
	\begin{equation} \label{ProjProp}
		\omega_{g, n+1} (z_0, z_{\llbracket n \rrbracket} ) = \sum_{a \in R} \Res_{z = a} \Big( \int_a^z \omega_{0,2} ( z_0, \mathord{\cdot}) \Big) \omega_{g, n+1} (z, z_{\llbracket n \rrbracket} ) 
	\end{equation}
\end{proposition}

Any collection $ (\omega_{g, n} )_{g,n}$ satisfying the projection property may also be called \emph{polarised}, and the choice of $\omega_{0,2} = B$ is sometimes referred to as a choice of polarisation.

Another property of the multi-differentials, which seems not to have appeared in the literature in this generality, is a bound on their pole order. We give that property in the next lemma, which generalises \cite[Proposition 9]{BE13} and \cite[Section 7]{DN18a}.

\begin{lemma}\label{PoleOrder}
  Let the multi-differentials $ \{ \omega_{g,n} \}_{ g \geq 0, n \geq 1} $ be obtained by topological recursion on the admissible meromorphic spectral curve $ \mc{S} = ( \Sigma, x, y, B)$. Then, the pole order of $ \omega_{g,n}$ in each variable at a point $ a \in \Sigma $ is bounded by $(s_a-1) (2g - 2 + n ) + 2g$.
\end{lemma}

\begin{proof}
  The proof is exactly the same as that of \cite[Proposition 9]{BE13}, which is the $ s_a = r_a + 1 $ case, noting that the pole order of the $ k$-th recursion kernel in general is $ (k-1)(s_a-1)$, by the definition of $ s_a$.\par
  Note that the original proof does not handle the possibility of $ \omega_{0,2}$s in $ \mc{W}^g_{k,n}$ with both arguments coupled to the kernel in the recursive step. Each such occurrence adds a double pole, so if we call the number of these occurrences $ b$, then the pole order of each term in $ \mc{W}^g_{k,n} $ is at most
  \begin{equation*}
    (s_a-1)(2(g + \ell (\mu ) - k) - 2k + n + k) + 2(g + \ell(\mu) + b - k)\,.
  \end{equation*}
  Then we can use that $ \ell (\mu) + b \leq k$, by the definition of $b$, so this can be bounded by $ (s_a-1)(2g - k + n) + 2g$, which is the same as found without the diagonal poles, so this omission does not change the pole order.
\end{proof}

\section{Transalgebraic spectral curves}

\label{s:transalgebraic}

Our goal is to extend the definition of topological recursion to spectral curves where $x$ is not meromorphic on the Riemann surface $\Sigma$. More precisely, we wish to study spectral curves where $x$ has exponential singularities at some isolated points on $\Sigma$. The key example to keep in mind is \cref{ex:rspin}: we want to consider this spectral curve, but with the Riemann surface $\Sigma = \P^1$, which includes the exponential singularity of $x$ at infinity.

\subsection{Transalgebraic functions}

Let us define more precisely the class of functions that we are interested in. We first define exponential singularities.

\begin{definition}[\cite{BP15b,PM19}]
	Let $\Sigma $ be a Riemann surface and $ z \in \Sigma$. A function $f $ is said to have an \emph{exponential singularity} at $ z$ if it is holomorphic and non-zero on some punctured open neighbourhood $ U \setminus \{ z\}$ of $z$, but cannot be extended to a meromorphic function on $U$.\par
	The \emph{exponential order} of $f$ at $z$ is defined to be
	\begin{equation}
		\Erd_f(z)=\inf \big\{ d \in \mathbb{R}_{\geq 0} \, \big| \, \limsup_{w\to z} | w - z |^d \log |f(w)| < \infty \big\} \, .
	\end{equation}
\end{definition}

A transalgebraic function on $\Sigma$ is a function that is holomorphic on $\Sigma$ except potentially at a finite number of points, where it can have either poles or exponential singularities. More precisely:

\begin{definition}[\cite{PM19}]
	Let $\Sigma $ be a Riemann surface. Let $ \mc{T}_n(\Sigma) $ be the space of \emph{transalgebraic functions} on $\Sigma $ with at most $n\in\mathbb{Z}_{\geq 0}$ zeros, poles, and exponential singularities, which consists of all non-zero holomorphic functions on $\Sigma \setminus S$ for some $ S \subset \Sigma$ such that $ |S| \leq n$ and such that for any $ z \in S$, $ \Erd_f(z) < \infty$.
	
	We define the \emph{class of transalgebraic functions} on $\Sigma $ as
	\begin{equation}
		\mc{T}(\Sigma)=\bigcup_{n\in\mathbb{Z}_{\geq 0}}\mc{T}_n(\Sigma)
	\end{equation}
\end{definition}

It is a natural question to ask why we insist that there are finitely many zeros and poles in the previous definition, but it is rather straightforward to see that this condition is required for ramification points to be isolated. Of course, by the great Picard theorem, all points in $\mathbb{P}^1$ are obtained infinitely often as one approaches an essential singularity save for possibly two points and, fixing an affine coordinate on $\mathbb{P}^1$, we may put those two points (if they exist) at zero and infinity via a change of coordinates. The following proposition tells us that if we didn't have exactly two such points, occasionally called \emph{Picard points} in the literature, then the exponential singularity would be a cluster point of ramification points; it seems natural to exclude such things from the perspective of the topological recursion.
\begin{proposition}
	Let $ \pi : B_\epsilon(0)\setminus\{0\} \to \mathbb{P}^1 $ be a branched covering from the punctured disk of radius $\epsilon>0$ to $\mathbb{P}^1$ with an exponential singularity at zero. Assume the only $\infty $ may be a Picard point of $\pi$, and if it is, $\epsilon$ is small enough that $\pi$ never takes the value infinity. Then $\pi$ has a ramification point in $B_\epsilon(0)\setminus\{0\}$.
\end{proposition}
\begin{proof}
	Proceed by contradiction and assume $ \pi : B_\epsilon(0)\setminus\{0\} \to C $ is an honest covering map where $ C = \pi(B_\epsilon(0)\setminus\{0\}) $ is either $\mathbb{C}$ or $\mathbb{C}_\infty \cong \mathbb{P}^1$. As $C$ is simply connected the monodromy group of $\pi$ is trivial. Ergo, there exists a right inverse map (non-unique) $ \pi^{-1} : C \to B_\epsilon(0)\setminus\{0\} $ so that $\pi\circ \pi^{-1} = {\rm id}_{B_\epsilon(0)\setminus\{0\}} $. If we define the image of $ \pi^{-1} $ to be $B$ then the map $ \pi' : C \to B$ with $ \pi' \equiv \pi^{-1} $ is biholomorphic and, in particular, a homeomorphism so $B$ is simply connected; therefore, $ {\pi'}^{-1} : B \to C$ is the universal cover of $C$ so there exists a covering map $\phi : B \to B_\epsilon(0)\setminus\{0\}$ such that $ \pi\circ\phi = {\pi'}^{-1} $. As $ {\pi'}^{-1} $ has the inverse $\pi'$, it is bijective so $\phi$ must be injective. As $\phi$ is a covering map, it must be surjective so it is in fact a homeomorphism. Thus, $ B_\epsilon(0)\setminus\{0\} $ is simply connected, an obvious contradiction.
\end{proof}
Thus, if we only have one Picard point (or none) then, for any $\epsilon > 0$, we get a ramification point in the disk of radius $\epsilon$ about the exponential singularity so that the exponential singularity must be an accumulation point of ramification points.

Given a transalgebraic function $f$ on $\Sigma$, its differential $d f$ is not usually a meromorphic one-form. However, it turns out that $d \log f$ always is, which is the content of the next lemma.

\begin{proposition}[{\cite[Lemma~2.15]{PM19}}]\label{p:dlogf}
	Let $f \in \mc{T}(\Sigma)$. The logarithmic derivative $d \log f$ is a meromorphic differential on $\Sigma$ with integer residues. Conversely, if $f$ is function on $ \Sigma$ which is non-zero holomorphic outside a finite set and is such that $d \log f $ is meromorphic with integer residues at poles, then $f \in \mc{T}(\Sigma)$.
\end{proposition}

It turns out that transalgebraic functions on compact Riemann surfaces have a very simple form.

\begin{theorem}[{\cite[Theorem~2.17]{PM19}}]\label{t:tt}
	The space of transalgebraic functions on a compact Riemann surface $ \Sigma$ is equal to the space of functions of the form
	\begin{equation}\label{eq:fM01}
		f(z) = M_0(z) e^{M_1(z)}\,,
	\end{equation}
	where $M_0$ and $ M_1$ are meromorphic functions on $ \Sigma$ and $M_0 \neq 0$.
\end{theorem}

The choice of $M_0$ and $M_1$ in \eqref{eq:fM01} is not quite unique: we can add to $M_1$ a (local) constant $c$, and multiply $M_0$ by $e^{-c}$ without changing $f$.

For more on transalgebraic functions and their underlying geometry (including the so-called log-Riemann surfaces), see \cite{BP15,BP15a,BP15b,PM19}.

\subsection{Transalgebraic spectral curves}

Looking back at the definition of spectral curves \cref{d:AlgSC+Adm}, $x$ and $y$ were only required to be holomorphic at all but finitely many points. In particular, they may be transalgebraic functions on $\Sigma$.

\begin{definition}\label{TransAlgSC}
	Let $\mathcal{S} = (\Sigma, x, y, B)$ be a spectral curve. We say that it is \emph{transalgebraic} if $x$ and $y$ are transalgebraic functions on $\Sigma$ such that $x y$ is a meromorphic function on $\Sigma$.
\end{definition}

Because of the requirement that $x y$ is meromorphic, the one-form $\omega_{0,1} = y dx$ is meromorphic on $\Sigma$, since by \cref{p:dlogf} we know that $\frac{dx}{x}$ is always meromorphic and by assumption $xy$ is meromorphic. Interestingly, all correlators $\omega_{g,n}$ produced by the topological recursion will still be meromorphic.

For topological recursion, we would like to consider $x$ as a branched covering $x: \Sigma \to \mathbb{P}^1$. Even though $x$ is transalgebraic on $\Sigma$, it can still be thought of as a branched covering \cite{BP13}. However, if $x$ has exponential singularities, the covering will not be of finite degree. Nevertheless, it makes sense, and we can define its ramification locus as follows.

\begin{definition}
	Let $ x $ be a transalgebraic function on $\Sigma$.  If $x $ is not meromorphic, we define its \emph{degree} to equal $ \infty$. We consider all the exponential singularities $a$ of $ x$ as ramification points, with ramification order $ r_a = \infty$. We write $R_\infty $ for the collection of exponential singularities, $R_0$ for the collection of finite ($r_a<\infty$) ramification points, and note $ R = R_0 \cup R_\infty $. Following \cite{BP15b}, we will sometimes refer to the elements of $R_\infty$ as infinite ramification points.
\end{definition}

We now focus on compact transalgebraic spectral curves, for which $x$ is a transalgebraic function on a compact Riemann surface $\Sigma$. In this case, by Theorem \ref{t:tt}, we can write 
\begin{equation}
	x(z) = M_0(z) e^{M_1(z) }
\end{equation}
for some meromorphic functions $M_0$ and $M_1$ on $\Sigma$ with $M_0 \neq 0$. We also define $M_2(z) \coloneqq x(z) y(z)$, which is another meromorphic function on $\Sigma$. We will use the $M_0, M_1$ and $M_2$ notation throughout the paper. 
%
%
%

\begin{example}\label{ex:atlantes}
	The typical example of a compact transalgebraic spectral curve is \cref{ex:rspin} but with $\Sigma = \P^1$. In other words, we consider the spectral curve $\mathcal{S}_\infty=(\Sigma,x,y,B)$ with $\Sigma = \P^1$, $x = z e^{-z ^r}$, $y = e^{z^r}$, and $B = \frac{d z_1 dz_2}{(z_1-z_2)^2}$, where $r \in \mathbb{Z}_{\geq 1}$. In this case, $M_0(z) = z$, $M_1(z) = - z^r$, and $M_2(z) = x(z) y(z) = z$. $R_0$ contains the $r$ finite ramification points at the solutions of $z^r =\frac{1}{r} $, while $R_\infty$ has a single point at $\infty$.
\end{example}

In topological recursion we are interested in the behaviour of $x$ near its ramification points. Around a finite ramification point in $ R_0$, the behaviour of $x$ is, locally at least, identical to meromorphic curves. So let us focus on the exponential singularities $ a \in R_\infty$, which correspond to the poles of $M_1(z)$.

\begin{lemma}
Let $a \in R_\infty$. Suppose that $M_1$ has a pole of order $m_1$ at $a$, and that $M_0$ has either a zero of order $m_0$ (for $m_0 \geq 0$) or a pole of order $|m_0|$ (for $m_0 < 0$) at $a$. Then there exists a local coordinate $ \zeta$  near $a$ such that either
	\begin{equation}
		x(\zeta)=e^{\zeta^{-m_1}}\,,
	\end{equation}
	if $m_0=0$, or
	\begin{equation}
	 	x(\zeta) = \zeta^{m_0} e^{-\frac{m_0}{m_1}\zeta^{-m_1}}\,,
	\end{equation}
	if $m_0\neq 0$.
\end{lemma}
\begin{proof}
	If $m_0=0$, this is obvious as near $a$ $\log(x)$ is a well defined meromorphic function with a pole of order $m_1$ at $a$, so there exists a local coordinate such that $\zeta^{-m_1}=\log(x(\zeta))$. If $m_0\neq 0$ we may take $z$ as a coordinate such that, near $a$, $M_0(z)=z^{m_0}$. Letting $\zeta=\sum_{n\geq 1}a_nz^n$ where $a_1\neq 0$ observe 
	\begin{equation}
		\begin{split}
			& M_1(z)+\frac{m_0}{m_1}\zeta^{-m_1}-m_0\log\left(\zeta/z\right) = 0, \\
			\Rightarrow & M_1(z) + \frac{m_0}{m_1}\frac{1}{(a_1z)^{m_1}}\left(1+\sum_{n=1}^{\infty}\frac{a_{n+1}}{a_1}z^n\right)^{-m_1} - m_0\log(a_1) - m_0\log\left(1+\sum_{n=1}^{\infty}\frac{a_{n+1}}{a_1}z^n\right) = 0,
		\end{split}
	\end{equation}
	where we assumed the branch of the logarithm was chosen so that $\log(\zeta/z) = \log(a_1)+\log(a_1^{-1}\zeta/z)$. Then, expanding $M_1(z) = \sum_{n=-m_1}^{\infty} A_n z^n$ with $A_{-m_1}\neq 0$ one finds $a_1 = [-m_0/(m_1A_1)]^{1/m_1}$. The rest of the $a_n$ may then be solved for recursively in terms of the $A_n$.
\end{proof}

However, unlike in the case of finite ramification points we see that there are infinitely many different choices of $\zeta$ corresponding to the branch choices for the $m_1$th root and the logarithm. In other words, $\mathfrak{f}_a(z)$ is countably infinite. Moreover, unlike the finite case, even in the local coordinate $\zeta$, the local deck transformations have no simple expression in terms of elementary functions (except when $m_0=0$). One can derive series expansions around a ramification point of the form (in terms of the local coordinate $\zeta$)
\begin{equation}\label{e:deckexp}
	\sum_{n\geq 0}s_n\zeta^{nm_1+1},
\end{equation} 
where $s_0$ is an $m_1$th root of unity and $s_1$ is $s_0\log(s_0^{m_0})/m_0$ for some branch choice of the logarithm and $m_0\neq 0$ (the explicit expansion is given below for $m_0=0$). Here, different choices of the $m_1$th root of unity and different branches of the logarithm will yield different local deck transformations. The radius of convergence of these series will depend on the choice of logarithm; as we will see shortly, there is no open set on which all such expansions converge.

When $m_0=0$ we may explicitly solve for the deck transformations. Indexing by $k\in\mathbb{Z}$ and $m=0,1,\dots,m_1-1$ and then denoting the deck transformations as $\sigma_a^{k,m}$ we find
\begin{equation}\label{e:m_0=0dt}
	\sigma_a^{k,m}(z) = \frac{ \theta^m \zeta } { \left( 1+2\pi i k \zeta^{m_1} \right)^{1/m_1} } \stackrel{\zeta \to 0}{=} \theta^m\zeta\sum_{n=0}^{\infty}\binom{-1/m_1}{n}(2\pi i k)^n\zeta^{m_1n},
\end{equation}
where the radius of convergence is $|\zeta|<|2\pi k|^{-1}$ and $\vartheta = \exp(2\pi i/m_1)$ is a primitive $m_1$th root of unity.

To proceed with an examination of these deck transformations when $m_0\neq 0$ we fix some notation. By \cref{e:deckexp}, each local deck transformation is uniquely defined by the first and second coefficient in its expansion about $a$ (more abstractly, this is due to the unique lifting property; see \cite{BP15b}). The first coefficient is an $m_1$th root of unity which we fix as $s_0=\vartheta^m$. The second coefficient is $s_0\log(s_0^{m_0})/m_0$. If we fix a choice of $\log$ with a branch cut chosen along an irrational angle in the complex plane (in particular, it must not exclude any power of $\theta$), then the choices of $s_1$ are in one-to-one correspondence with the integers. We denote the local deck transformation with first coefficient $\theta^m$ and second coefficient $\theta^m(2\pi i m/m_1-2\pi i k/m_0)$ as $\sigma_a^{k,m}$.

To proceed, we first find solutions for the partial inverses of $x$ in terms of the Lambert $W$ function
\begin{equation}
	x = \zeta^{m_0}e^{-\frac{m_0}{m_1}\zeta^{-m_1}} \Rightarrow x^{-m_1/m_0}=\zeta^{-m_1}e^{\zeta^{-m_1}} \Rightarrow \zeta = \left[W_k\left(x^{-m_1/m_0}\right)\right]^{-1/m_1},
\end{equation}
where $W_k$ is the $k$th branch of the Lambert $W$ function defined by the relation
\begin{equation}
	z=we^w \Leftrightarrow \exists k\in\mathbb{Z} \ni w=W_k(z). 
\end{equation}
Normally, $W_0$ denotes the principal branch that is real-valued on the non-negative half of the real line and $W_{-1}$ is the branch that is real-valued on the interval $[-1/e,0]$. Otherwise, there is no standard convention in the literature regarding the choices of branches of the Lambert $W$ function; one such choice is given in \cite{CGHJK}, which we will use in the following.

Our deck transformations then take the form
\begin{equation}
	\sigma_a^{k,m}(\zeta) = \theta^{m'} \left[W_{k'}\left(\phi^{l}\zeta^{-m_1}e^{\zeta^{-m_1}}\right)\right]^{-1/m_1},
\end{equation}
where $\phi=\exp(-2\pi i m_1/m_0)$ and $m',k',l$ are integers, which we will determine in what follows. We then note the following expansion of the Lambert $W$ function from \cite{CGHJK}, which is valid for all non-zero $k$ when $\log(|z|)$ is sufficiently large
\begin{multline}\label{e:LWexp}
	W_k(z) = \log(z) + 2\pi ik -\log(\log(z)+2\pi i k)\\
	+ \sum_{a=0}^{\infty}(-1)^a\sum_{b=1}^{\infty}\frac{1}{b!} \Stircyc{a+b}{a+1} (\log(z)+2\pi i k)^{-a-b}\log^{[b]}(\log(z)+2\pi i k),
\end{multline}
where $\Stircyc{n_1}{n_2}$ denotes an unsigned Stirling number of the first kind and $\log^{[b]}$ denotes the logarithm composed with itself $b$ times; all the logarithms in the above expansion are the principal branch of the logarithm. The above expansion is also valid when $k=0$ provided $|z|$ is sufficiently large; for small $z$ the principal branch satisfies $W_0(z) \sim z$. Using the above expansion we find, for $\zeta$ a small positive number
\begin{equation}
	\sigma_a^{k,m}(\zeta) = \vartheta^{m'}\zeta-\frac{\vartheta^{m'}}{m_1}\left(2\pi i k'-2\pi i \frac{m_1}{m_0}l\right)\zeta^{m_1+1}+\mathcal{O}(\zeta^{2m_1+1}).
\end{equation}
This gives us that $m'=m$ and that $k'$ and $l$ should satisfy
\begin{equation}
	\frac{m_1}{m_0}l-k'=m-\frac{m_1}{m_0}k,
\end{equation}
where $l$ must be chosen so that $-\frac{2\pi im_1l}{m_0}\in(-\pi,\pi)$. Recalling that $m=0,\dots,m_1-1$ we see that $k = \mathcal{O}(k')$ as $k\to\infty$. This leads us to the following lemma, which considers the asymptotic behaviour as the chosen branch of the logarithm becomes `large' in some sense.

\begin{lemma}\label{l:deckasym}
	For $m_0\neq 0$ (the corresponding $m_0=0$ formula is obvious from \eqref{e:m_0=0dt})
	\begin{equation}
		\begin{split}
			\sigma_a^{m,k}(\zeta)& = \theta^m( 2\pi i m_1 k/m_0 )^{-1/m_1} \left( 1+\mc{O}\left( \frac{\log(|k|)}{k} \right) \right), \quad |k|\to\infty,\\
			\frac{d\sigma_a^{m,k}}{d\zeta}(\zeta)& = \theta^m\zeta^{-m_1-1}( 2\pi i m_1 k/m_0 )^{-1/m_1-1}\left( 1+\mc{O}\left( \frac{\log(|k|)}{k} \right) \right), \quad |k|\to\infty .
		\end{split}
	\end{equation}
\end{lemma}

\begin{proof}
	This is a direct consequence of the fact that $k\sim\frac{m_0}{m_1}k',\hspace{1mm}|k|\to\infty$, and the expansion \eqref{e:LWexp}.
\end{proof}


\subsection{Transalgebraic spectral curves as limits}
\label{sec:TransAlgSCasLimits}

Compact transalgebraic spectral curves naturally arise as limits of sequences of compact meromorphic spectral curves. The guiding light behind our definition of topological recursion for transalgebraic spectral curves is that it should commute with taking such limits. Schematically, if $\mathcal{S}_N$ is a sequence of compact meromorphic spectral curves such that $\lim_{N \to \infty} \mathcal{S}_N$ is a compact transalgebraic spectral curve, and the $\omega^N_{g,n}[\mathcal{S}_N]$ are the correlators constructed from usual topological recursion on $\mathcal{S}_N$, we want the correlators $\omega_{g,n}[\lim_{N \to \infty} \mathcal{S}_N]$ associated to the transalgebraic spectral curve $\lim_{N \to \infty} \mathcal{S}_N$ to satisfy
\begin{equation}\label{eq:limit}
	\lim_{N \to \infty} \left( \omega_{g,n}^N[\mathcal{S}_N]  \right)= \omega_{g,n} \left[ \lim_{N \to \infty} \mathcal{S}_N \right].
\end{equation}
Therefore, we should study such sequences of spectral curves. Considering such limits will also allow us to extend the notion of admissibility from \cref{d:AlgSCAdm} to exponential singularities.

Therefore, consider such a sequence of compact meromorphic spectral curves $\mathcal{S}_N = (\Sigma, x_N, y_N, B)$, such that $x_N \to x$ and $y_N  \to y$ as $N \to \infty$, where $x$ and $y$ are transalgebraic functions on $\Sigma$ with $x y$ meromorphic.  Explicitly, we will consider the sequence
\begin{equation}
	x_N=M_0\left(1+ (\tau - 1) \frac{M_1}{N}\right)^{-N} \left(1 + \tau \frac{M_1}{N}\right)^N\,, \qquad y_N=\frac{M_2}{x_N}\,,
\end{equation}
which converges compactly to 
\begin{equation}
	x = M_0 e^{M_1}, \qquad y = \frac{M_2}{x},
\end{equation}
away from the poles of $M_1$. 

\begin{remark}
	We introduce the parameter $ \tau$ for two reasons. First, we will see in Theorem \ref{t:main} that the limiting correlators do not depend on the choice of $\tau$, which is evidence that our definition of topological recursion on transalgebraic spectral curves is the correct one for the limiting curve and not an artefact of the particular sequence chosen. Secondly, when constructing quantum curves, we will see that we get a priori different quantum curves for each choice of $\tau$. However, at least in the cases of interest in this paper, we will see that this $\tau$ dependence can be naturally transformed away.
\end{remark}

For the spectral curves $\mc{S}_N$, we divide the ramification points of $x_N$, denoted collectively as $R^N$, into two sets of ramification points:
\begin{enumerate}
	\item $R^N_\infty=\{M_1=-\frac{N}{\tau}\}\cup\{M_1 = \frac{N}{1-\tau} \} \cup \{M_1=\infty\}$ consists of the ramification points colliding at essential singularities of $x$;
	\item $R^N_0=R^N\setminus R^N_\infty$ consists of those ramification points not colliding at essential singularities of $x$. 
\end{enumerate}

Let us consider what admissibility means for ramification points of transalgebraic spectral curves. The notion of admissibility at the ramification points in $R_0$ is clear: it should be the same as for algebraic spectral curves, that of \cref{d:AlgSCAdm}. At the points in $R_\infty$ we need a new definition based on the notion of admissibility for points in $R_\infty^N$. 

We distinguish two cases for an exponential singularity $a\in R_\infty$ depending on whether $M_2=xy$ has a pole at $a$ or not. Let $a \in R_\infty$, which means that it is a pole of $M_1$, and suppose that $x y$ has a pole at $a$. Then, for finite $N$, we have $s_a\leq -1$ and therefore $\mathcal{S}_N$ is admissible at $a$ by \cref{d:AlgSCAdm} and the correlators will be regular at $a$. Moreover, for sufficiently large $N$, $M_2$ will be regular and non-zero at the other points $a'$ in $R_\infty^N$ that collide at the essential singularity $a$, and hence $s_{a'}=1$ in \cref{d:AlgSCAdm} at all such points so $\mathcal{S}_N$ is admissible at all such points. Thus, if $x y$ has a pole at $a$, each spectral curve in the sequence is admissible, and so it makes intuitive sense that the limit of these curves should be admissible.

There appears to be significant challenges in defining the topological recursion in the case where $x y$ does not have a pole at an exponential singularity of $x$. As it appears in no cases of interest, it is not done here\footnote{Curiously, via limiting arguments, it seems that the modularity condition for admissibility should be $ m_0\pmod{m_1} = \pm 1 \pmod{s_a}$ where $m_0$ and $m_1$ are the order of $M_0$ and $M_1$, respectively, at $a$, which very naturally generalises \cref{d:AlgSCAdm}.}. Therefore, we will define admissibility at infinite ramification points as follows.

\begin{definition}\label{d:tadmis}
	Let $\mathcal{S}$ be a compact transalgebraic spectral curve. We say that it is \emph{admissible} if both of the following conditions are satisfied:
	\begin{enumerate}
		\item it is admissible in the sense of \cref{d:AlgSCAdm} at the finite ramification points in $R_0$;
		\item $xy = M_2$ has poles at the infinite ramification points in $R_\infty$.
	\end{enumerate}
\end{definition}

\section{Topological recursion on transalgebraic spectral curves}

\label{s:TRtransalgebraic} 

We are now ready to define topological recursion on transalgebraic spectral curves. Recall that we want topological recursion to commute with limits of sequences of meromorphic spectral curves, as stated schematically in \eqref{eq:limit}. If $\mathcal{S}_N$ is a sequence of compact meromorphic spectral curves such that $\lim_{N \to \infty} \mathcal{S}_N$ is a compact transalgebraic spectral curve, and the $\omega_{g,n}^N[\mathcal{S}_N]$ are the correlators constructed from usual topological recursion on $\mathcal{S}_N$, we want the correlators $\omega_{g,n}[\lim_{N \to \infty} \mathcal{S}_N]$ associated to the transalgebraic spectral curve $\lim_{N \to \infty} \mathcal{S}_N$ to satisfy
\begin{equation}
	\lim_{N \to \infty} \left( \omega_{g,n}^N[\mathcal{S}_N] \right)= \omega_{g,n} \left[ \lim_{N \to \infty} \mathcal{S}_N \right].
\end{equation}

However, it is not straightforward to define topological recursion in this limit. The usual formulation of topological recursion considers residues at ramification points, and to define the integrand one needs to sum over local deck transformations at those ramification points. In other words, one needs to take the pullback of the pushforward of a one-form under the map $x$. To include infinite ramification points in the topological recursion, one would need to take the pullback of the pushforward of a one-form under the local map $x = \zeta^{m_0} e^{-\frac{m_0}{m_1}\zeta^{-m_1}}$. 
Using \cref{l:deckasym}, one can see that for a $1$-form $\eta$ that is holomorphic at all essential singularities of $x$, $x^*x_*\eta$ is well-defined in the sense that the sum over deck transformations is convergent. However, for the topological recursion, we want to look at forms with poles at the essential singularities. In this case the sum in $x^*x_*\eta$ is not absolutely convergent, but there is a natural way to define the principal value.\footnote{In particular, we sum as follows: we first sum over the index $m$, then sum over the sign of $k$ if $k\neq 0$, and then finally sum from $k=1,\dots,\infty$ (at some point adding in the $k=0$ term). Then, one can see the sum from $k=1$ to $k=\infty$ \emph{is absolutely convergent} if the $1$-form $\eta$ has, at $a\in R_\infty$, a pole of order no more than $\Erd_x(a)$ for each $a\in R_\infty$.} However, even after summing, the resulting differential may not have an isolated singularity at the essential singularities of $x$. Thus, defining the residue at these points becomes highly changing. One possible approach is to pushforward the entire TR integrand to the $x$ plane, where it should be meromorphic as a function of $x$, but then it is not clear what residues in the $x$ plane one should be taking as essential singularities do not have well-defined branchpoints, although $x=0,\infty$ (the Picard points) are the most compelling candidates. Even presuming one does all this in a reasonable manner, proving the $N\to\infty$ limit commutes in the sense of \eqref{eq:limit} remains a daunting challenge.

Instead, our approach consists in first rewriting the topological recursion in a different way, which trades out the sum over the deck transformations of $x$ for a sum over ramification points, coinciding points, and deck transformations of $y$. We present this rewriting for compact meromorphic spectral curves in the next section, and then generalise it to transalgebraic spectral curves.\footnote{This rewriting of topological recursion for compact meromorphic spectral curves is inspired by private notes of Nitin K. Chidambaram.}

\subsection{Rewriting topological recursion}

\label{s:rewriting}

Let us review some of the notation required for this rewriting that was introduced in \cref{notation}. Let $C=\{t,t_1,\dots,t_i\}\subset\Sigma$ and $C'=C\setminus\{t\}$ be sets of cardinality $i+1$ and $i$, respectively. For a symmetric $i$-differential $\eta$ we set, by definition
\begin{equation}
	\Res_{C=t}\eta(t_1,\dots,t_i)=\Res_{C'=t}\eta(t_1,\dots,t_i)=\Res_{t_1=t} \cdots \Res_{t_i=t} \eta(t_1,\dots,t_i)\,.
\end{equation}
Then, for a set $ C \subset \Sigma $, we denoted by $ t^C $ one arbitrarily chosen element in this set. Finally, for the purposes of taking many residues in a compact notation, we defined 
\begin{equation}
	\Res_{\substack{t_l=a_l\\l=1,\dots,n}} = \Res_{t_1=a_1} \cdots \Res_{t_n=a_n}\,.
\end{equation}
With this notation (see \cref{notation} for more details), we may proceed to the theorem of this section.

\begin{theorem}\label{t:rewrite}
	Let $\mc{S}$ be a compact meromorphic admissible spectral curve\footnote{To be precise, we also need to assume here that $\mc{S}$ can be ``fully globalised'', in the language of \cite{BBCKS23}. What this means is that we can replace in topological recursion the sums over local deck transformations $\mf{f}'_a(t)$ at the ramification points with a sum over the whole fibre $\mf{f}'(t)$, and perform manipulations along the lines of \cite[Theorem 5]{BE13}. All spectral curves considered in the present paper are fully globalisable, according to the conditions determined in \cite{BBCKS23}. \label{f:fg}} and write $Y(t) = y^{-1} (y(t))$. Then the correlators of topological recursion satisfy the alternative recursive formula
	\begin{equation}
		\begin{split}
			\omega_{g,n+1}(z_0,z_{\llbracket n \rrbracket}) 
			&= \Res_{t \in R} \sum_{m=2}^{{\deg} (x)}\int_{*}^t B(z_0, \cdot ) \sum_{C_1,\dots,C_j\vdash t_{\llbracket m-1 \rrbracket } } \!\!\! \frac{(-1)^{1-\delta_{j,m-1}}}{j!} \!\! \Res_{\substack{t^{C_l} \in R \cup z_{\llbracket n \rrbracket} \cup Y(t)\\l=1,\dots,j}} \,\, \Res_{\substack{C_l=t^{C_l}\\l=1,\dots,j}} 
			\\
			& \quad \left( \prod_{l=1}^{j}  \frac{1}{x(t)-x(t^{C_l})} \prod_{t_0\in C_l\setminus\{t^{C_l}\}} \frac{1}{x(t_0)-x(t^{C_l})} \right)\frac{\mc{W}_{g,n,m}(t,t_{\llbracket m-1 \rrbracket } \,| \, z_{\llbracket n \rrbracket})}{\prod_{l=1}^{m-1}(y(t)-y(t_l))}\,,
		\end{split}
	\end{equation}
	where we slightly abuse notation by writing that the $C_l$ partition the dummy integration variables outside the integral.
\end{theorem}

\begin{proof}
Starting with topological recursion from \cref{BE-TR}, we first replace the sums over local deck transformations $\mf{f}'_a(t)$ by a sum over the whole fibre $\mf{f}'(t)$, as in \cite[Theorem 5]{BE13} (see \cref{f:fg} and \cite{BBCKS23}). Then, we perform the rewriting:
	\begin{multline}
		\sum_{\emptyset \neq Z \subseteq \mf{f}'(t)} K_{|Z | +1}(z_0,t, Z) \mc{W}_{g,n,|Z | +1}(t, Z \, | \, z_{\llbracket n \rrbracket}) = \\ 
		\left( \int_*^t B(z_0, \cdot) \right) \!\! \sum_{m\geq 2} \sum_{\emptyset \neq \{\zeta_1,\dots,\zeta_{m-1}\} \subseteq \mf{f}'(t)} \Res_{\substack{t_l = \zeta_l \\ l=1,\dots,m-1}} \frac{\mc{W}_{g,n,m}(t,t_{\llbracket m-1 \rrbracket} \, | \, z_{\llbracket n\rrbracket})}{\prod_{l=1}^{m-1}(x(t_l)-x(t))(y(t)-y(t_l))}\,.
	\end{multline}
In our new writing, the summand is well-defined when two or more of the elements of $Z=\{\zeta_1,\dots,\zeta_{m-1}\}$ coincide, it will just result in higher order poles at such a $ t_l = \zeta_l$. Therefore, instead of summing over subsets of $ \mf{f}'(t)$, we will sum over tuples of size at most $ \deg x - 1$, but then we need to subtract tuples with repeating terms. This gives us two main terms: the original sum plus the added terms where two or more $t_l$ coincide and the subtracted terms where two or more $t_l $ coincide. We first examine the first term, for fixed size of the tuple $m-1$,
	\begin{equation}
		\begin{split}
			\sum_{\zeta_1,\dots,\zeta_{m-1} \in \mf{f}'(t)} & \frac{1}{(m-1)!} \Res_{\substack{t_l = \zeta_l \\ l = 1, \dots, m-1 }} \frac{\mc{W}_{g,n,m} (t, t_{\llbracket m-1 \rrbracket } \, | \, z_{\llbracket n\rrbracket} ) }{\prod_{l=1}^{m-1} (x(t_l)-x(t)) (y(t)-y(t_l))}
			\\
			= & \frac{1}{(m-1)!} \Res_{\substack{t_l \in R \cup z_{\llbracket n\rrbracket} \cup Y(t) \\ l = 1, \dots, m-1}} \frac{\mc{W}_{g,n,m} (t, t_{\llbracket m-1 \rrbracket } \, | \, z_{\llbracket n\rrbracket})}{\prod_{l=1}^{m-1} (x(t)-x(t_l)) (y(t)-y(t_l)) }.
		\end{split}
	\end{equation} 
All we have done here is used the fact that $\Sigma$ is a compact Riemann surface so the sum of all the residues of any differential must be zero. This gives a factor $ (-1)^{m_1}$, which we reabsorb by flipping the $ m-1$ factors $ x(t) - x(t_l)$. That we only pick up residues at the listed points is because the $\omega_{g,n}$ only have poles at coinciding points and ramification points.

We now wish to apply the same logic to the terms with the coinciding points. To this end, we want to know what happens when $j\leq m-1 $ of the same $t_l$ are specialised to the same sheet. So we consider
\begin{equation}
	\begin{split}
		& \sum_{\zeta \in \mf{f}'(t)} \Res_{\substack{t_l = \zeta \\ l=1, \dots, j}} \frac{\mc{W}_{g,n,m}(t,t_{\llbracket m-1 \rrbracket } \, | \, z_{\llbracket n \rrbracket} ) }{\prod_{l=1}^{m-1} (x(t_l)-x(t)) (y(t)-y(t_l))}
			\\
		= & \sum_{\zeta \in \mf{f}'(t)} \Res_{\substack{t_l = \zeta \\ l=1, \dots, j-1}} \frac{\mc{W}_{g,n,m}(t, t_1, \dots, t_{j-1}, \zeta, t_{j+1}, \dots, t_{m-1} \, | \, z_{\llbracket n \rrbracket} ) }{dx(t) (y(t)-y(\zeta )) \prod_{\substack{l=1 \\ l \neq j}}^{m-1} (x(t_l)-x(t)) (y(t)-y(t_l))}
			\\
		= & \sum_{\zeta \in \mf{f}'(t)} \Res_{t_j = \zeta} \Res_{\substack{t_l = t_j \\ l=1, \dots, j-1}}\frac{\mc{W}_{g,n,m}(t, t_{\llbracket m-1 \rrbracket } \, | \, z_{\llbracket n \rrbracket} ) }{(x(t_j)-x(t)) (y(t)-y(t_j)) \prod_{\substack{l=1 \\ l \neq j}}^{m-1}(x(t_l)-x(t)) (y(t)-y(t_l))}
			\\
		= & \Res_{t_j \in R, z_{\llbracket n \rrbracket}, Y(t)} \Res_{\substack{t_l = t_j \\ l=1, \dots, j-1}} \frac{\mc{W}_{g,n,m}(t, t_{\llbracket m-1 \rrbracket } | z_{\llbracket n \rrbracket} ) }{(x(t)-x(t_j)) (y(t)-y(t_j)) \prod_{\substack{l=1 \\ l \neq j}}^{m-1}(x(t_l)-x(t))(y(t)-y(t_l))}.
	\end{split}
\end{equation}
	
Hence, the subtracted terms with the coinciding deck transformations may be written as
\begin{equation}
	\begin{split}
		& - \sum_{j=1}^{m-2} \frac{1}{j!} \sum_{\zeta_1, \dots, \zeta_j \in \mf{f}'(t)} \sum_{C_1, \dots, C_j \vdash t_{\llbracket m-1 \rrbracket } } \Res_{\substack{C_l = \zeta_l \\ l = 1, \dots, j}} \frac{\mc{W}_{g,n,m} (t, t_{\llbracket m-1 \rrbracket } | z_{\llbracket n \rrbracket} ) }{\prod_{l=1}^{m-1} (x(t_l)-x(t)) (y(t)-y(t_l))} \\
		= & - \sum_{j=1}^{m-2} \frac{1}{j!} \sum_{C_1, \dots, C_j \vdash t_{\llbracket m-1 \rrbracket } } \Res_{\substack{t^{C_l}\in R\cup z_{\llbracket n \rrbracket}\cup Y(t) \\ l = 1, \dots, j}} \Res_{\substack{C_l = t^{C_l} \\ l=1, \dots, j}}\\
		& \hspace{10mm} \left( \prod_{l=1}^{j} \frac{1}{x(t)-x(t^{C_l})} \prod_{t_0\in C_l \setminus \{ t^{C_l} \}} \frac{1}{x(t_0)-x(t^{C_l})} \right) \frac{\mc{W}_{g,n,m} (t, t_{\llbracket m-1 \rrbracket } | z_{\llbracket n \rrbracket} ) }{\prod_{l=1}^{m-1} (y(t)-y(t_l))}\,,
	\end{split}
\end{equation}
and the desired result then follows.
\end{proof}

\subsection{Topological recursion on transalgebraic spectral curves}

With the topological recursion for compact meromorphic spectral curves rewritten as in \cref{t:rewrite}, we are in a position of topological recursion for compact transalgebraic spectral curves. We will present the definition of the topological recursion for transalgebraic spectral curves straight away, and then spend the rest of the section arguing and demonstrating why it works. 

\begin{definition}\label{d:main}
	Let $ \mc{S} = \left( \Sigma, x, y, B \right)$ be a compact transalgebraic admissible spectral curve, with $ x = M_0 \exp(M_1) $, $y = M_2/x$ and $ M_0, M_1, M_2 $ meromorphic functions on $\Sigma$. Fix $\tau \in \mathbb{C}$ and define the sequence of spectral curves $\mc{S}_N = \left( \Sigma, x_N, y_N, B \right)$, where
	 \begin{equation}
	 	x_N = M_0 \left( 1 + (\tau - 1) \frac{M_1}{N} \right)^{-N} \left( 1 + \tau \frac{M_1}{N} \right)^{N}\,, \quad y _N = M_2/x_N\,.
	\end{equation}
	Then, if $\omega_{g,n}^N$ are the correlators constructed  by topological recursion for the spectral curve $\mc{S}_N$, we \emph{define} the correlators of the spectral curve $\mc{S}$ as the $N\to\infty$ limit of the $\omega_{g,n}^N$.
\end{definition}

This defines nothing if the limit depends on $\tau$ or does not yield well-defined meromorphic correlators. The main result of this section is the following theorem, which shows that these issues do not occur, and, therefore, that the above definition makes sense.

\begin{theorem}\label{t:main}
	Let $ \mc{S} = \left( \Sigma, x, y, B \right) $ be a compact transalgebraic admissible spectral curve. Then the $ \omega_{g,n} $ constructed from \cref{d:main} are well-defined meromorphic differentials on $\Sigma^n$ and do not depend on the choice of $ \tau $.
\end{theorem}

\begin{proof}
	Our strategy will be to first prove the that $ \omega_{g,n} $ are well-defined for $ \tau = 0 $, and then show that the limit is independent of $ \tau $. The proof is divided into eight steps.
	
	\emph{First step: start the induction and flip contours.}\par
	We proceed inductively in the $\tau=0$ case on $-\chi_{g,n}=2g+n-2$. For $-\chi_{g,n}=-1,0$ (corresponding to $\omega_{0,1}$ and $\omega_{0,2}$) the result holds trivially, so we may proceed to the induction step. For finite $N$ we may use \cref{t:rewrite} to write
	\begin{equation}\label{e:finiteNtrans}
	\begin{split}
		\omega^N_{g,n+1} (z_0, z_{\llbracket n \rrbracket})
		&=
		\Res_{t=R^N} \sum_{m=2}^{\deg (x_N)} \left( \int_{*}^tB(z_0,\cdot) \right) \sum_{C_1, \dots, C_j \vdash t_{\llbracket m-1 \rrbracket }} \frac{(-1)^{1-\delta_{j,m-1}}}{j!} \Res_{\substack{t^{C_l} \in R^N \cup z_{\llbracket n \rrbracket} \cup \mc{Y}(t) \\ l = 1, \dots, j}} \Res_{\substack{C_l = t^{C_l} \\ l=1, \dots, j}}
		\\
		&\quad \left( \prod_{l=1}^{j} \frac{x_N(t)}{x_N(t) - x_N(t^{C_l})} \prod_{t_0 \in C_l \setminus \{t^{C_l}\}} \frac{x_N(t^{C_l})}{x_N(t_0) - x_N(t^{C_l})} \right) \frac{\mc{W}_{g,n,m}^N (t, t_{\llbracket m-1 \rrbracket } \, | \, z_{\llbracket n \rrbracket}) }{ \prod_{l=1}^{m-1} ( M_2(t) - M_2(t_l) )}\,,
	\end{split}
	\end{equation}
	where $\mc{Y}(t)=(xy)^{-1}\big((xy)(\{t\})\big)$. This is slightly different from \cref{t:rewrite}, so a couple of remarks are in order so it is clear how we get here:
	\begin{itemize}
		\item in the denominator of the integrand in the original topological recursion we rewrote $ y_N (t) - y_N (\sigma(t)) = (M_2(t)-M_2(\sigma(t))) / x_N(t)$ so we ended up with $ M_2 = xy = x_Ny_N $ in the denominator and $ x_N (t) $ in the numerator, where $\sigma(t)\in \mf{f}'(t)$ is a deck transformation;
		\item as $ x_N (t) = x_N (\sigma(t)) $ for every deck transformation, we can choose which deck transformation we take the argument of $ x_N $ to be at; in particular, we take $j$ of them to just be $t$ and the other $ m - 1 - j $ to be precisely those deck transformations that gives us $t^{C_l}$;
		\item when we flipped the contour, we then had to pick out residues at $\mc{Y}(t)$ rather than $Y(t)$.
	\end{itemize}
	
	\emph{Second step: $R_\infty^N $ does not contribute.}\par
	We now observe that the residues at $ t^{C_l} \in R_\infty^N $ vanish for sufficiently large $N$. Namely, for $ t^{C_l} \in R_\infty^N$ that satisfy $ M_1(t^{C_l}) = N $, $ x_N(t^{C_l}) $ has a pole of order $ N $. As $ x_N (t^{C_l}) $ appears in the denominator one time with no corresponding $ x_N(t^{C_l}) $ in the numerator, the overall integrand in the variable $t^{C_l}$ gains a zero of order $N$. We claim the rest of the integrand has a pole of at worst uniformly bounded order in $N$. At these ramification points $\omega_{0,1}^N$ has simple poles, and, for sufficiently large $N$, $M_2$ will be regular and non-zero, which means that $s_a=1$ at these points. From Lemma \ref{PoleOrder} we then know that the $\omega_{g,n}^N$ have poles of order no more than $2g$ and so the $\mc{W}_{g,n,m}^{N}$ have poles of bounded order in $N$. Similarly, $M_2$ is meromorphic and constant in $N$. For $ x_N(t^{C_l}) / (x_N(t_0)-x_N(t^{C_l})) $, the $x_N$ appears in both the denominator and the numerator. 
	Finally, we need to examine taking the residues at $ C_l = t^{C_l} $. This will be a residue of a pole of order no more than three (two from a potential $\omega_{0,2}$ contribution plus one for the difference of the $x_N$ in the denominator). Thus, this residue may be replaced by multiplication by $(t_0-t^{C_l})^3$ and twice differentiating by $t_0$, for each $t_0 \in C_l \setminus \{t^{C_l}\}$, before taking the limit as $ t^{C_l} \to t_0$. By the quotient rule for differentiation, we will have the same total power of derivatives of $x_N$ in the numerator and denominator, just in different combinations and orders of differentiation. Thus, at the residues at points where $M_1 = N$ we may drop the residue in $t^{C_l}$.

	Now we examine the residues in $t^{C_l}$ where the point at which the residue is taken satisfies $M_1 = \infty$. As before, when we take the residue at $C_l=t^{C_l}$, this will correspond to derivatives. Here though, the pole counting is a little more subtle so we do it explicitly. In particular, observe
	\begin{equation}\label{e:xexp1}
	\begin{split}
		\frac{x_N(t^{C_l})\omega_{0,2}(t_0,t^{C_l})(t_0-t^{C_l})^3}{(x_N(t_0)-x_N(t^{C_l}))dt_0 dt^{C_l}}
		&=
		\frac{x_N(t^{C_l})}{x_N'(t^{C_l})} - (t_0-t^{C_l}) \frac{x_N(t^{C_l})x_N''(t^{C_l})}{x'_N(t^{C_l})^2}
		\\
		&+ (t_0 - t^{C_l})^2 \left(\frac{x_N(t^{C_l})x_N''(t^{C_l})^2}{4x_N'(t^{C_l})^3} - \frac{x_N(t^{C_l}) x_N'''(t^{C_l})}{x_N'(t^{C_l})^2} \right) + \mc{O} \left((t_0 - t^{C_l})^3 \right) \,.
	\end{split}
	\end{equation}

	In the constant term and the $t_0 - t^{C_l}$ term, there is no pole at $t^{C_l}$ equalling a pole of $M_1$. However, the $(t_0-t^{C_l})^2$ term has a simple pole here. On the other hand, the $\omega_{g,n}^N$ are regular at these points (this is because we are in the $s_a\leq -1$ case at these points; see \cref{r:free}) and $M_2(t_0)$, which has at least a simple pole by admissibility, appears in the denominator. Thus, in terms with an $\omega_{0,2} (t_0,t^{C_l})$ we do not have contributions from these points. For terms without this factor, the pole at $t_0=t^{C_l}$ is simple. Thus, observing
	\begin{equation}\label{e:xexp2}
		\frac{x_N(t^{C_l})(t_0-t^{C_l})}{x_N(t_0)-x_N(t^{C_l})} = \frac{x_N(t^{C_l})}{x_N'(t^{C_l})} + \mc{O} \left(t_0 - t^{C_l}\right)\,,
	\end{equation}
	we see the same argument still holds. In summary, we may replace the residues in each $t^{C_l}$ at all the points in $R_N$ with just those at $R_N^0$.

	\emph{Third step: integrand well-defined.}\par
	This shows that the integrand in $t$ is well defined in the limit: indeed, we may commute the limit in $ N \to \infty $ with the residues (integrals) in the $ t_0 \in C_l$ and $ t^{C_l} $ using dominated convergence and use the induction assumption that $\omega_{g,n}^N \to \omega_{g,n} $. Note that, although it may appear that the sum over $m$ becomes infinite in the limit, for any fixed $g$ and $n$ only finitely many terms are non-zero so commuting the limit with this sum is entirely trivial.
	
	\emph{Fourth step: integral well-defined at $ R_0$.}\par
	However, we want the integral, not just the integrand, to be well-defined in the limit. To this end, we note that the contributions from the residues at $t = R_0^N $ go to the contributions at $t = R_0 $ in the limit by pulling the limit in $N$ inside each integral using dominated convergence, as before, and applying the induction assumption. However, this simple argument will not work for the residues at $ t = R_\infty $ as these points can collide in the limit. 

	\emph{Fifth step: work in coordinate $w$ for integrals at $ R_\infty$.}\par
	To deduce that the integral must be well-defined in the limit, we will pushforward to work in the $M_1$ plane where all elements of $R_\infty^N$ fall at $M_1 = N, \infty$. To move to the origin, let $ w=1/M_1 $. For a deck transformation $\sigma$ of $M_2$, i.e., $\sigma(t) \in \mc{Y} (t)$, we define a corresponding transformation $\nu$ through $\nu(w) = \nu (1/M_1(t)) = 1 / M_1(\sigma(t)) $. Although $\nu$ may depend on $t$, and so is multi-valued, we sum over $\nu$ at every step; this is well-defined. In completing this sum, as from now on we will suppress this detail, note that the sum over $\nu$ must include both the sum over $\sigma$ (from all elements of $\mc{Y}(t)$) and partial inverse of $M_1$ (from the fact we pushforward in $M_1$). Let $\nu_1, \dotsc, \nu_r$ be all such non-trivial $\nu$ and note that, for general $N$, $\nu_p ( w = 1/N ) \neq 1/N$ for all $p = 1, \dotsc, r$. We claim that the integrand in $t$ of topological recursion, pushed forward under $ M_1 $ and then written in the coordinate $w$, takes the following form (where $ \exp_N(z) = (1 - z/N)^{-N}$):
	
	\begin{equation}\label{e:integrandexp}
		\frac{\mc{N}_N^d (w \, | \, z_0, z_{\llbracket n \rrbracket}) \exp_N(w^{-1})^d + \mc{N}_N^{d-1} (w \, | \, z_0, z_{\llbracket n \rrbracket}) \exp_N (w^{-1})^{d-1} + \dotsb + \mc{N}_N^0 (w \, | \, z_0, z_{\llbracket n \rrbracket})}{\mc{D}^d_N (w \, | \, z_0, z_{\llbracket n \rrbracket}) \exp_N (w^{-1})^{d} + \mc{D}_N^{d-1} (w \, | \, z_0, z_{\llbracket n \rrbracket}) \exp_N(w^{-1})^{d-1} + \dotsb +\mc{D}^N_0(w \, | \, z_0, z_{\llbracket n \rrbracket})}\,,
	\end{equation}
	where each of the $\mc{N}_N^k$ and $\mc{D}_N^k$ are meromorphic functions\footnote{Due to the presence of $\int_*^t B(z_0,\cdot)$ in the integrand, this is not strictly true for non-zero genus. However, all we want to do is integrate, so we may slightly misuse terminology.} with the order of their zeros and poles at $ w = 1/N$ bounded uniformly in $N$ and it is assumed $\mc{D}^d_N(w \, | \, z_0, z_{\llbracket n \rrbracket})$ is not identically zero.\par
	To get this expansion, first note we can write every derivative of $x_N$ as $\exp_N (M_1) = \exp_N (1/w)$ times a sequence of meromorphic functions with the order of their poles and zeros bounded uniformly in $N$. Then, when we take the residues at $ t_0 = t^{C_l}$, we get expansion like \eqref{e:xexp1} and \eqref{e:xexp2}, which we may put over a common denominator.\par
	Then, we will claim that when we take the residues at $t^{C_l} = R_N^0, z_{\llbracket n \rrbracket}, \mc{Y}(t) $ the total number of factors of derivatives of $x_N(t)$ in the denominator is greater than or equal to those in the numerator in each term; putting everything over a common denominator and pushing forward results in an expression of the form \eqref{e:integrandexp}. The fact that we get this same power behaviour will be demonstrated in the proceeding (sixth) step and for now can be taken as a claim. Like any fraction, such an expression is non-unique as we may multiply the numerator and the denominator by the same factor without changing the total value; however, the ratio $\mc{N}_N^d (w \, | \, z_0, z_{\llbracket n \rrbracket}) / \mc{D}^d_N (w \, | \, z_0, z_{\llbracket n \rrbracket})$ is unique and all we will eventually care about.
	
	\emph{Sixth step: properties of $ \mc{N}/\mc{D}$.}\par
	We now need to verify a couple of important properties of $ \mc{N}_N^d (w \, | \, z_0, z_{\llbracket n \rrbracket}) / \mc{D}^d_N (w \, | \, z_0, z_{\llbracket n \rrbracket}) $. First, we claim it has no pole at $w = 0$. Second, we claim that it does not have essential singularity in the limit and its only potential pole near $w = 0$ is the one at $w = 1/N$. Note that the only place an essential singularity could come from is the $\exp_N ( 1 / \nu_p(w) ) $. Finally, along the way, it will become clear that, as we have stated, the degree of the numerator and denominator in $exp_N(1/w)$ must be the same and that the ratio $\mc{N}_N^d (w \, | \, z_0, z_{\llbracket n \rrbracket}) / \mc{D}^d_N (w \, | \, z_0, z_{\llbracket n \rrbracket})$ is well-defined.

	The proof for these claims is a bit more involved. We will examine the individual terms that we put over a common denominator before pushing forward to the $w$-plane; by examining the ratio of the coefficient of the highest power of $x_N$ in the numerator to the one in the denominator, we can deduce the behaviour of the coefficients in the fraction put over a common denominator (before pushing forward, $x_N$ is $\exp_N (M_1)$ times a meromorphic function $M_0$, so looking at leading powers of $x_N$ is the same as looking at leading powers of $\exp_N (M_1)$). As there are no poles at $t \in R_\infty$, there are none at $w = 0$, and the lack of $\exp_N ( M_1 ( \sigma(t)))$ factors in the leading order will give us the no essential singularities result. To this end we first perform the residues at $C_l = t^{C_l}$ in the integrand in $t$ and will be left with an expression of the form
	\begin{equation}\label{e:evalt_cl}
		\Res_{\substack{t^{C_l} \in R_N^0 \cup z_{\llbracket n \rrbracket} \cup \mathcal{Y}(t) \\ l = 1, \dotsc, j}} \frac{ x_N(t) }{ x_N(t)-x_N (t^{C_l}) } \frac{f_N(t, t^{C_l}, t_{\llbracket m-1 \rrbracket } \setminus C_l \, | \, z_{\llbracket n \rrbracket}) }{ (M_2(t) - M_2(t^{C_l}))^{|C_l|} }\,,
	\end{equation}
	where $f_N$ is a differential in all its arguments except $t$. Furthermore, $f_N$ is meromorphic in $t$ and $t^{C_l}$ and remains so in the limit (note that in \eqref{e:xexp1} and \eqref{e:xexp2} the derivatives in $x_N$ appear in the same power in the numerator and denominator so we may cancel out the factor of $\exp_N (M_1)$), and there is no pole in $t$ or $t^{C_l}$ at the poles of $M_1$ (we established this before in the second step to show that the residues at $t^{C_l}=R_\infty$ do not contribute).

	First we examine the residues
	\begin{equation}
		\Res_{\substack{t^{C_l} \in R_N^0 \cup z_{\llbracket n \rrbracket} \\ l = 1, \dotsc, j}} \frac{ x_N(t) }{ x_N(t)-x_N (t^{C_l}) } \frac{f_N(t, t^{C_l}, t_{\llbracket m-1 \rrbracket } \setminus C_l \, | \, z_{\llbracket n \rrbracket}) }{ (M_2(t) - M_2(t^{C_l}))^{|C_l|} }\,.
	\end{equation}
	Here, the residues in the $ t^{C_l} $ are taken at points that do not depend on $t$. Thus, due to the pole of $M_2$ at the poles of $M_1$ guaranteed by admissibility, these residues will all contribute some sort of zero to the ratio of leading order coefficients in powers of $x_N$. Finally, it is trivial that these residues will never contribute $\exp_N(M_1(\sigma(t)))$ as nothing depends on $\sigma(t)$.
	
	However, a pathology can occur here. In the limit we can have two or more different elements of $R_0^N$ collide; if this happens, the independent contribution of each to the final integrand will not be well-defined and only the sum over the residues at these colliding points is well-defined in the limit. Let us examine this case and assure ourselves that this presents no issues for the well-definedness of the ratio $ \mc{N}_N^d (w \, | \, z_0, z_{\llbracket n \rrbracket}) / \mc{D}^d_N (w \, | \, z_0, z_{\llbracket n \rrbracket}) $ in the limit. First note that these colliding points cannot be poles of $x_N$ as the location of the poles of $ x_N $ in $ R_0^N $ do not depend on $N$ as they are just the poles of $M_0$ and the poles of $M_0$ will never collide with the zeros of $ (1-M_1/N)^{N+1}dx_N = (1-M_1/N)dM_0 + M_0dM_1 $. Thus, the general scenario we must examine is when we have $ a^{1}_N,\dots,a^k_N \in R_0^N $ which are all zeros of $dx_N$ and all collide in the limit. We then examine the following expression in which we computed the residue in terms of derivatives
	\begin{equation}
		\sum_{i=1}^{k} \lim\limits_{ t^{C_l} \to a^i_N } \frac{1}{M!}\frac{d^M}{d(t^{C_l})^M} \frac{ x_N(t) }{ x_N(t)-x_N (t^{C_l}) } \frac{f_N(t, t^{C_l}, t_{\llbracket m-1 \rrbracket } \setminus C_l \, | \, z_{\llbracket n \rrbracket}) }{ (M_2(t) - M_2(t^{C_l}))^{|C_l|} }\,,
	\end{equation}
	where $ M \in \mathbb{Z}_{\geq 0} $ is chosen large enough so all the limits in $t^{C_l}$ are finite. By using the product rule we may write this expression as
	\begin{equation}
		\frac{1}{M!} \sum_{h=1}^{M} \sum_{i=1}^{k} \frac{x_N(t)}{( x_N(t)-x_N(a^i_N) )^h} F_N^h(t, a^i_N, t_{\llbracket m-1 \rrbracket } \setminus C_l \, | \, z_{\llbracket n \rrbracket})\,,
	\end{equation}
	where the $ F_N^h $ are meromorphic functions in $t$ such that the orders of all its zeros and poles (in $t$) are uniformly bounded in $N$. Denoting this uniform bound by $N_0$, we see for sufficiently large $N$ the prefactors of the $ F_N^h $ in the above expression must have zeros and poles of order larger than $N_0$ (except possibly for $ h = 1 $ if $a_N^i$ is a zero of $x_N$ for some $i$ and all $N$ sufficiently large). As the sum over $h$ must be well defined in the limit, we can conclude that each individual term in the sum over $n$ is well-defined in the limit. Therefore, the $ h = 1 $ expression
	\begin{multline}
		\frac{1}{M!} \sum_{i=1}^{k} \frac{x_N(t)}{ x_N(t)-x_N(a^i_N) } F_N^1(t, a^i_N, t_{\llbracket m-1 \rrbracket } \setminus C_l \, | \, z_{\llbracket n \rrbracket}) \\ 
		= \frac{x_N(t)}{M!} \prod_{i=1}^{k} \left[x_N(t)-x_N(a_N^i)\right]^{-1} \sum_{i=1}^{k} F_N^1(t, a^i_N, t_{\llbracket m-1 \rrbracket } \setminus C_l \, | \, z_{\llbracket n \rrbracket}) \prod_{\substack{ j=1 \\ j\neq i }}^{k} \left[x_N(t)-x_N(a_N^j)\right]\,,
	\end{multline}
	is well-defined in the limit. Observing that this is degree $k$ in $x_N(t)$ in both the numerator and the denominator we see that this is precisely the desired leading order in $x_N$ expression that will contribute to the ratio $ \mc{N}_N^d (w \, | \, z_0, z_{\llbracket n \rrbracket}) / \mc{D}^d_N (w \, | \, z_0, z_{\llbracket n \rrbracket}) $, whereas the terms other than $ j = 1$ will contribute only to terms lower order in $\exp_N(1/w)^d$.

	More involved are the contributions from the residues at $\mc{Y}(t)$. Here, we divide these into three sub-cases: 
	\begin{itemize}
		\item[(i)] we may take the residues at $ \sigma(t) \in \mc{Y}'(t)$ that do not preserve $M_1$, these correspond to the $\nu_1, \dotsc, \nu_r$ and are where the limit of $\mc{N}_N^d (w \, | \, z_0, z_{\llbracket n \rrbracket}) / \mc{D}^d_N (w \, | \, z_0, z_{\llbracket n \rrbracket}) $ may have potential essential singularities;
		\item[(ii)] we may take the residues at $\sigma(t) \in \mc{Y}'(t)$ that preserve $M_1$, i.e., $M_1\circ\sigma = M_1$; 
		\item[(iii)] we may take the residue at $t$ itself.
	\end{itemize}
	Starting with sub-case (i), take an element $\sigma(t) \in \mc{Y}'(t) $, fix an $l$, and inspect the following residue
	\begin{equation}
		\Res_{t^{C_l} \in \sigma(t)} \frac{ x_N(t) }{ x_N(t)-x_N (t^{C_l}) } \frac{f_N(t, t^{C_l}, t_{ \llbracket m-1 \rrbracket } \setminus C_l \, | \, z_{\llbracket n \rrbracket}) }{ (M_2(t) - M_2(t^{C_l}))^{|C_l|} }\,,
	\end{equation}
	where $f_N$ is as before. Here we have a pole of order $| C_l |$ at $\sigma(t)$ in the variable $t^{C_l}$. We can therefore calculate the residue with the formula
	\begin{equation}
		\lim_{t^{C_l} \to \sigma(t)} \frac{(-1)^{| C_l |}}{| C_l |!}\frac{ d^{| C_l| -1}}{ d(t^{C_l})^{| C_l | -1}} \frac{x_N(t) f_N(t, t^{C_l}, t_{\llbracket m-1 \rrbracket } \setminus C_l \, | \, z_{\llbracket n \rrbracket})}{x_N(t) - x_N ( t^{C_l} )} \left( \frac{t^{C_l} \sigma(t) }{ M_2 (t^{C_l}) - M_2(t)}\right)^{| C_l |} \,.
	\end{equation}
	If we take the derivatives of the $1 / (x_N(t) - x_N(t^{C_l}))$, we will end up with subleading terms in powers of $x_N$; these therefore do not concern our analysis. All we care about when we take derivatives of $f$, is that derivatives cannot create poles. Finally, we have the expansion
	\begin{equation}
		\left( \frac{t^{C_l} \sigma(t) }{ M_2 (t^{C_l}) - M_2(\sigma(t)) } \right)^{| C_l |} = \frac{1}{M_2' (\sigma(t))^{| C_l | }} \sum_{k=0}^{\infty} S_k (\sigma(t)) (t^{C_l} - \sigma(t))^k \,,
	\end{equation}
	where $S_k$ has a pole of order at most $k$ at elements of $R_\infty$ (poles of $M_1$). The pre-factor has a zero of at least order $| C_l |$ (as $M_2$ has a pole at all elements of $R_\infty$), and the only $S_k$ that can contribute are those with $k \leq | C_l |$. Thus, for the leading terms in $x_N$, we will never get poles. Furthermore, from the above discussion, we see the $\exp_N (1 / \nu_p(w))$ will never enter the leading order power in $ \exp_N (1/w) $.

	Now, we move on to sub-case (ii): the deck transformations that preserve $ M_1 $; let $\sigma(t)$ be such a deck transformation and examine the expression
	\begin{equation}
		\lim_{t^{C_l} \to \sigma(t)} \frac{(-1)^{| C_l |}}{| C_l |!} \frac{ d^{| C_l |-1}}{ dt_{C_l}^{| C_l | -1}} \frac{ x_N (t) f_N (t, t^{C_l}, t_{\llbracket i-1 \rrbracket} \setminus C_l \, | \, z_{\llbracket n \rrbracket})}{ x_N(t) - x_N (t^{C_l}) } \left( \frac{t^{C_l} \sigma(t)}{M_2 (t^{C_l}) - M_2(t)}\right)^{| C_l |} \,,
	\end{equation}
	which is the same as the prior case as nothing in the steps changes up to this point. The only thing that changes in analysing this expression is that when we take derivatives of the $x_N(t) / (x_N(t) - x_N (t^{C_l}) )$ factor we do not end up with only subleading terms as $ x_N(\sigma(t)) $ now has a factor of $ \exp(M_1(t)) $. If we take $k$ derivatives of this factor, we get a pole of order at most $k$ in the ratio of the coefficients of the leading powers of $x_N$. Thus, we still cannot get a pole as we have the factor of $ M_2' ( \sigma(t) )^{- | C_l |}$, as before.

	Finally, we examine sub-case (iii), where we take the residue at $t^{C_l}=t$:
	\begin{equation}\label{eq:resatt}
		\Res_{t^{C_l}=t} \frac{ x_N(t) }{ x_N(t) - x_N (t^{C_l}) } \frac{f_N(t, t^{C_l}, t_{\llbracket i-1 \rrbracket } \setminus C_l \, | \, z_{\llbracket n \rrbracket} ) }{(M_2 (t) - M_2 ( t^{C_l} ) )^{| C_l |}}.
	\end{equation}
	Here, we may have a pole of order at most $ | C_l |+3$. In particular, we get a pole of order one from the $ x_N(t) / (x_N(t) - x_N (t^{C_l}))$ factor, a pole of order $| C_l |$ from the difference of the $M_2$ in the denominator, and a potential double pole in $ f_N $ at $ t^{C_l} = t $ due to possible presence of an $ \omega_{0,2}(t, t^{C_l})$. Here, however, the $M_2' (t)^{- | C_l |}$ has at least a zero of order $2 | C_l |$. Using identical arguments with the expansions of the individual factors, this case will not create an undesired pole in $\mc{N}_N^d(w \, | \, z_0, z_{\llbracket n \rrbracket})/\mc{D}^d_N(w \, | \, z_0, z_{\llbracket n \rrbracket})$.

	\emph{Seventh step: integral well-defined.}\par
	With these properties established, we can prove the $\omega_{g,n}^N$ are well-defined in the limit. First note
	\begin{equation}
	\begin{split}
		\Res_{w = 1/N} \frac{\mc{N}_N^d (w \, | \, z_0, z_{\llbracket n \rrbracket}) \exp_N (1/w)^d + \dotsb}{\mc{D}^d_N (w \, | \, z_0, z_{\llbracket n \rrbracket} ) \exp_N (1/w)^d + \dotsb}
		&=
		\Res_{w = 1/N} \frac{\mc{N}_N^d (w \, | \, z_0, z_{\llbracket n \rrbracket}) }{\mc{D}^d_N(w \, | \, z_0, z_{\llbracket n \rrbracket} ) } \left( 1 + \mc{O}(w-1/N)^N \right)
		\\
		&=
		\Res_{w = 1/N} \frac{\mc{N}_N^d (w \, | \, z_0, z_{\llbracket n \rrbracket}) }{\mc{D}^d_N(w \, | \, z_0, z_{\llbracket n \rrbracket} ) } \,.
	\end{split}
	\end{equation}
	
	As we have established $\mc{N}_N^d (w \, | \, z_0, z_{\llbracket n \rrbracket}) / \mc{D}^d_N (w \, | \, z_0, z_{\llbracket n \rrbracket})$ has no pole at $w=0$, we may change the residue at $w = 1/N$ to a contour integral about a small circle around $w = 0$. Then, using dominated convergence to bring the limit as $ N \to \infty $ inside the contour, we conclude
	\begin{equation}
		\lim_{N \to \infty} \Res_{w = 1/N} \frac{\mc{N}_N^d (w \, | \, z_0, z_{\llbracket n \rrbracket}) \exp_N (1/w)^d + \dotsb}{\mc{D}^d_N (w \, | \, z_0, z_{\llbracket n \rrbracket}) \exp_N (1/w)^d + \dotsb} = \Res_{w = 0} \frac{\mc{N}_\infty^d(w \, | \, z_0, z_{\llbracket n \rrbracket}) }{\mc{D}^d_\infty (w \, | \, z_0, z_{\llbracket n \rrbracket})}.
	\end{equation}

\vspace{11pt}
	
	\emph{Eighth step: independence of $\tau$.}\par
	Finally, we claim that all choices of $\tau$ yield the same result in the $N \to \infty$. To establish this, we first prove $ \partial_{\tau}^m \omega_{g,n+1}^N (z_0, z_{\llbracket n \rrbracket}) |_{\tau = 0} $ exists and goes to zero for every value of $ m \in \mathbb{Z}_{\geq 1}$ as $N \to \infty$ proceeding inductively on $-\chi_{g,n} = 2g + n - 2$.  
	
	First note that the result is straightforward for $\omega_{0,1}^N$ and trivial for $\omega_{0,2}$. So we may proceed directly to the induction step and assume the result holds for all prior correlators. We first argue that we may commute derivatives in $\tau$ with all residues in \eqref{e:finiteNtrans}. To do this, we transform all residues into contour integrals; even if the point at which the residue is being taken depends on $\tau$, the contour may be taken to be locally constant in $\tau$. Then we may commute the derivatives in $\tau$ with the $\tau$-independent contour integrals and, as the derivatives in $\tau$ cannot create new poles, we may switch all contour integrals back to the same residues.

	In an identical manner to the second step of the proof, we wish to argue the residues at $ t^{C_l} = R^N_\infty $ do not contribute. For the points in $R^N_\infty$ that satisfy $ M_1 = N/(1-\tau) $, $ x_N(t^{C_l}) $ has a pole of order $N$ and the argument proceeds identically to the argument in the second step. Similarly, for the points where $ M_1 = \infty $ the expansions \eqref{e:xexp1} and \eqref{e:xexp2} are the same as before and the identical argument works for general choices of $\tau$. 
	
	The only new thing to check is that the points in $R^N_\infty$ that satisfy $ M_1 = -N/\tau $ do not contribute. Here $x_N$ has a zero of order $N$ rather than a pole of order $N$. Denoting the collection of these points as $V_N$, by \eqref{e:evalt_cl} we look at the expression (with the $\tau$ dependence suitably inserted)
	\begin{equation}\label{e:bigzeroxN}
		\Res_{\substack{t^{C_l} \in V_N}} \frac{ x_N(t) }{ x_N(t)-x_N (t^{C_l}) } \frac{f^\tau_N(t, t^{C_l}, t_{\llbracket m-1 \rrbracket } \setminus C_l \, | \, z_{\llbracket n \rrbracket}) }{ (M_2(t) - M_2(t^{C_l}))^{|C_l|} }\,,
	\end{equation} 
	where, as before, $f^\tau_N$ will only have poles of order bounded uniformly in $N$. It will be important to observe, that as the correlators are regular at the poles of $M_1$ and $M_2$ has poles at the poles of $M_1$ that $f_N^\tau$ has no pole at the poles of $M_1$ for all $\tau$. Then we note, for any $ a_N \in V_N $,
	\begin{equation}
		\frac{x_N(t)}{x_N(t)-x_N(t^{C_l})}=1+\mathcal{O}((t-t^{C_l})^N),
	\end{equation}
	so the expression in \eqref{e:bigzeroxN} is in fact equal to
	\begin{equation}
		\Res_{\substack{t^{C_l} \in V_N}} \frac{f^\tau_N(t, t^{C_l}, t_{\llbracket m-1 \rrbracket } \setminus C_l \, | \, z_{\llbracket n \rrbracket}) }{ (M_2(t) - M_2(t^{C_l}))^{|C_l|} } 
		\eqcolon 
		F^\tau_N(t, t^{C_l}, t_{\llbracket m-1 \rrbracket } \setminus C_l \, | \, z_{\llbracket n \rrbracket}).
	\end{equation}
	We now claim that $ [ \partial^m F^\tau_N ]_{\tau=0} \equiv 0 $ for all $ m \in \mathbb{Z}_{\geq 0} $, which would mean, at least locally near $\tau=0$, that these points do not contribute. To prove this we wish to commute the $\tau$ derivatives with the residue in the above expression. There is a slight subtlety, as all points in $V_N$ collide at poles of $M_1$ in the limit $ \tau \to 0 $. However, as the $ f_N^\tau $ are regular at the poles of $M_1$, near $ \tau = 0 $ we may draw a contour around each of the poles of $M_1$ that includes all points in $V_N$ (but no other poles of the integrand) and commute the $\tau$ derivatives with this contour.
	
	Next, we note that $ f_N^0 = f_N $ has no pole at the poles of $M_1$ so $ [ \partial^m f^\tau_N ]_{\tau=0} $ is in fact regular at the poles of $M_1$ for all $ m \in \mathbb{Z}_{\geq 0} $. Then, $ [ \partial^m F^\tau_N ]_{\tau=0} $ just involves taking residues of $ [ \partial^m f^\tau_N ]_{\tau=0} $ at the poles of $M_1$ and therefore vanishes.
	
	As argued previously, we can commute the $\tau$ derivatives with the residues and obtain
	\begin{equation}
		\begin{split}
			\partial_\tau^k\omega^N_{g,n+1} (z_0, z_{\llbracket n \rrbracket})
			&=
			\Res_{t\in R^N} \sum_{m=2}^{\deg (x_N)} \!\!\! \left( \int_{*}^t B(z_0,\cdot) \right) \!\!\! \sum_{C_1, \dots, C_j \vdash t_{ \llbracket m-1 \rrbracket}} \!\!\!\!\!\! \frac{(-1)^{1-\delta_{j,m-1}}}{j!} \!\!\! \Res_{\substack{t^{C_l} \in R^N_0 \cup z_{\llbracket n \rrbracket} \cup \mc{Y}(t) \\ l = 1, \dots, j}} \Res_{\substack{C_l = t^{C_l} \\ l=1, \dots, j}}
			\\
			&\quad \partial_\tau^k\left( \prod_{l=1}^{j} \frac{x_N(t)}{x_N(t) - x_N(t^{C_l})} \! \prod_{t_0 \in C_l \setminus \{t^{C_l}\}} \!\! \frac{x_N(t^{C_l})}{x_N(t_0) - x_N(t^{C_l})} \right) \frac{\mc{W}_{g,n,m}^N (t, t_{\llbracket m-1 \rrbracket} \, | \, z_{\llbracket n \rrbracket}) }{ \prod_{l=1}^{m-1} ( M_2(t) - M_2(t_l) )}\,,
		\end{split}
	\end{equation}
	The expression that the derivative in $\tau$ hits in the above equation depends on $\tau$ through the $x_N$ and the $\omega_{g',n'}^N$. $ \del_\tau^i x_N $ goes to zero in the limit $N \to \infty$ and $\partial_\tau^i\omega_{g',n'}^N$ goes to zero by the induction assumption. Thus, the entire second line (\textit{i.e.}, the part inside all the residues) is converging to zero as $N \to \infty$. As all the residues in the variables $C_l$ may just be converted into integrals\footnote{If two or more ramification points in $R^N_0$ collide in the limit, then we will have to write one contour integral around the point in $R_0$ they collide at.} we get that the entire integrand in $t$ is converging to zero. Analogously to proving the existence of the limit when $\tau=0$, we must now argue the integral itself goes to zero. 
	
	For the residues at $t = R^N_0$ the fact that the integrand is going to zero is clear as we may take the $N \to \infty$ limit inside the contour integral; so we concentrate on the residues at $t = R^N_\infty$. Here, for generic $\tau$, we will have residues at solutions of $M_1(z) = N / (1-\tau), -N/\tau$. When we set $\tau = 0$, we end up with no poles at $M_1(z) = \infty$, even after taking derivatives, as we cannot create poles by taking derivatives. For our purposes, we may therefore neglect the residues at $M_1(z) = -N / \tau$, as they will drop out in the end. 

	Thus, at these points, we need to take $\tau$ derivatives of the analogous expression to \eqref{e:integrandexp} where the $\mc{N}$ and $\mc{D}$ coefficients acquire the suitable $\tau$ dependence and $\exp_N (w^{-1}) = (1 + (\tau - 1) / (Nw) )^{-N} (1 + \tau / (Nw) )^N$ is appropriately modified\footnote{$\tau$ derivatives commute with the pushforward in $M_1$ as $M_1$ does not depend on $\tau$.}. After taking $m$ $\tau$ derivatives, the ratio of the new leading order coefficients will have no pole at $w = \tau = 0$ by the quotient rule and the fact that the $\tau$ derivative can only decrease the order of poles at $w = 0$. We may then conclude that the same argument with the $N \to \infty$ limit holds 
	\begin{equation}
		\begin{split}
			\Res_{w = (1-\tau)/N} & \partial_\tau^m \frac{\mc{N}_N^{d,\tau} (w \, | \, z_0, z_{\llbracket n \rrbracket}) \exp_N (1/w)^d + \dotsb}{\mc{D}^{d,\tau}_N (w \, | \, z_0, z_{\llbracket n \rrbracket} ) \exp_N (1/w)^d + \dotsb}
      \\
			&=
			\Res_{w = (1-\tau)/N} \partial_\tau^m \frac{\mc{N}_N^{d,\tau} (w \, | \, z_0, z_{\llbracket n \rrbracket}) }{\mc{D}^{d,\tau}_N(w \, | \, z_0, z_{\llbracket n \rrbracket} ) } \left( 1 + \mc{O}(w-(1-\tau)/N)^N \right)
			\\
			&=
			\Res_{w = (1-\tau)/N} \partial_\tau^m \frac{\mc{N}_N^{d,\tau} (w \, | \, z_0, z_{\llbracket n \rrbracket}) }{\mc{D}^{d,\tau}_N(w \, | \, z_0, z_{\llbracket n \rrbracket} ) } \,,
		\end{split}
	\end{equation}
	but this time, after taking the derivatives in $\tau$, the ratio of the leading order coefficients is converging to zero. 
	
	Finally, we argue that this result, namely that $ \partial_{\tau}^m \omega_{g,n+1}^N (z_0, z_{\llbracket n \rrbracket}) |_{\tau = 0} $ exists and goes to zero in the limit, perhaps unsurprisingly, actually establishes the theorem. Note that we have the following expansion for sufficiently small $\tau$ and generic choices of $z_1, \dotsc, z_n \in \Sigma$
	\begin{equation}
		\omega_{g,n+1}^N (z_0, z_{\llbracket n \rrbracket}) = \sum_{m=0}^{\infty} \frac{\tau^m}{m!} \partial_{\tau}^m \omega_{g,n+1}^N(z_0, z_{\llbracket n \rrbracket}) |_{\tau = 0} \,.
	\end{equation}
	Denote the radius of convergence of this sum as $ \rho_N(z_0, z_{\llbracket n \rrbracket}) $. We claim that $ \rho_N \to \infty $ provided $ z_i \notin R \quad \forall i=0,\dots,n $. To prove this claim we examine the singularity structure of $ \omega_{g,n+1}^N (z_0, z_{\llbracket n \rrbracket})=\omega_{g,n+1}^N (z_0, z_{\llbracket n \rrbracket};\tau) $. This is straightforward as these $\omega_{g,n+1}^N$ only have poles at ramification points.
	
	The ramification points of $x_N$ can be put into three categories: solutions of $M_1(z) = (1-\tau)/N$; solutions of $M_1(z) = -\tau/N$; poles of $M_1$; poles of $M_0$; zeros of $dx_N$ that converge to elements of $R_0$. For the first two cases we clearly see that, for fixed $z$, these singularities go to infinity in the $\tau$ plane. The next two cases can never create singularities in the $\tau$ plane as, by assumption, $z_i \notin R$. The fifth and final case is dealt with by computing
	\begin{equation}
		\left( 1+(\tau-1)\frac{M_1}{N} \right)^{N+1} \left( 1+\tau\frac{M_1}{N} \right)^{-N+1} dx_N = \left( 1+(\tau-1)\frac{M_1}{N} \right) \left( 1+\tau\frac{M_1}{N} \right) dM_0 + M_0dM_1 \, ,
	\end{equation}
	and noting that, for a fixed point on $\Sigma$ that is not a zero of this differential in the limit, the only zero this has in the $\tau$ plane shoots off to infinity in the limit $N \to \infty$. 
	
	As $\rho_N$ is the distance between zero and the nearest singularity in the $\tau$ plane we indeed have that $\rho_N \to \infty$. Thus, given any fixed $\tau$ we are inside the radius of convergence, \textit{i.e.}, $|\tau|<\rho_N$, for all $N$ sufficiently large so we may commute the limit in $N$ with the infinite sum. This proves the theorem.
\end{proof}

Now that we know  that correlators for transalgebraic spectral curves are well-defined, we can easily prove a number of corollaries.

\begin{corollary}
Let $\mathcal{S}$ be a compact transalgebraic admissible spectral curve. For $2g+n-2 \geq 1$, the correlators $ \omega_{g,n}$ constructed by topological recursion on $\mathcal{S}$ satisfy the following properties:
	\begin{itemize}
		\item Symmetry: the $ \omega_{g,n}$ are symmetric in all of their $ n $ variables.
		\item Pole structure: the $\omega_{g,n}$ have poles only at the ramification points of $ x $.
		\item Residueless: the $\omega_{g,n}$ have vanishing residue at all points.
		\item Homogeneity: rescaling $ \omega_{0,1}$ by a constant $ c \in \C^*$ to $ c\omega_{0,1}$ results in a rescaling $\omega_{g,n} \to c^{2-2g-n} \omega_{g,n}$.
	\end{itemize}
\end{corollary}

\begin{proof}
	These properties are well-known for ordinary topological recursion, and were proved in \cite{EO07,BE13,BBCCN18}. They carry over as they hold for each curve in our sequence of spectral curves. 
\end{proof}

We also give a direct formula for topological recursion on transalgebraic spectral curves, in a wide variety of cases, without using a sequence of correlators and taking limits.

\begin{lemma}\label{l:formula}
Let $\mathcal{S}$ be a compact transalgebraic admissible spectral curve.
	If $M_1$ is a well-defined function of $M_2$ we may use the following formula to recursively compute the correlators of topological recursion.
	\begin{equation}
	\begin{split}
		\omega_{g,n+1} (z_0, z_{\llbracket n \rrbracket})
		&=
		\Res_{t \in R} \sum_{m = 2}^{\deg (x)}\left(\int_{*}^tB(z_0,\cdot)\right) \sum_{C_1, \dotsc, C_j \vdash  t_{\llbracket m-1 \rrbracket } } \frac{(-1)^{1-\delta_{j,m-1}} }{j!} \Res_{\substack{t^{C_l} \in R_0 \cup Z\cup \mc{Y}(t) \\ l = 1, \dotsc, j}} \Res_{\substack{C_l = t^{C_l} \\ l = 1, \dotsc, j}}
		\\
		&
		\left( \prod_{l=1}^{j} \frac{x(t)}{x(t) - x(t^{C_l})} \prod_{t_0 \in C_l \setminus \{ t^{C_l} \} } \frac{x(t)}{x(t_0) - x(t^{C_l})}\right) \frac{\mc{W}_{g,n,m}(t, t_{ \llbracket m-1 \rrbracket } \, | \, z_{\llbracket n \rrbracket})}{ \prod_{l=1}^{i-1} ((xy)(t) - (xy)(t_l))} \,,
	\end{split}
	\end{equation}
	where $\mc{Y}(t) = M_2^{-1} \big(M_2 (t) \big)$ and the residues at the infinite ramification points $R_\infty$ are defined as
	\begin{equation}
		\Res_{t \in R_\infty} \leftrightarrow \frac{1}{(2g-1)!} \lim_{w \to 0^+} \frac{d^{2g-1}}{dw^{2g-1}} {M_1}_*\,,
	\end{equation}
	where the expression on the right is to be interpreted as follows: we take the pushforward under the map $M_1$ and define $w = 1/M_1$ so the infinite ramification points are all located at $w = 0$; the formula is then the standard one for a pole of order $2g$ at $w = 0$ except we take the limit as $w \to 0$ along the positive real axis.
\end{lemma}

\begin{proof}
	Adopting the notation of the proof of \cref{t:main}, as $\mc{N}_N^d ( w \, | \, z_0, z_{\llbracket n \rrbracket}) / \mc{D}^d_N ( w \, | \, z_0, z_{\llbracket n \rrbracket})$ has a pole of order at most $2g$ at $w = 1/N$, we will have that $\mc{N}_\infty^d (w \, | \, z_0, z_{\llbracket n \rrbracket}) / \mc{D}^d_\infty (w \, | \, z_0, z_{\llbracket n \rrbracket})$ will have a pole of order no more than $2g$ at $w = 0$. Ergo, we can compute the topological recursion in the limit (note here, by assumption, there are no $\nu$). By definition, this is
	\begin{equation}
	\begin{split}
		&\frac{1}{(2g-1)!} \lim_{w \to 0^+} \frac{d^{2g-1}}{dw^{2g-1}} \frac{\mc{N}_\infty^d (w \, | \, z_0, z_{\llbracket n \rrbracket}) \exp(d/w) + \dotsb}{\mc{D}^d_\infty (w \, | \, z_0, z_{\llbracket n \rrbracket} ) \exp(d/w) + \dotsb}
		\\
		=&\frac{1}{(2g-1)!} \lim_{w \to 0^+} \frac{d^{2g-1}}{dw^{2g-1}} \frac{\mc{N}_\infty^d (w \, | \, z_0, z_{\llbracket n \rrbracket}) }{\mc{D}^d_\infty (w \, | \, z_0, z_{\llbracket n \rrbracket} )} \left( 1 + \mc{O}(\exp(-w^{-1}) ) \right)
		\\
		=&\frac{1}{(2g-1)!} \lim_{w \to 0^+} \frac{d^{2g-1}}{dw^{2g-1}} \frac{\mc{N}_\infty^d (w \,| \, z_0, z_{\llbracket n \rrbracket}) }{\mc{D}^d_\infty ( w \, | \, z_0, z_{\llbracket n \rrbracket} )} = \Res_{w = 0} \frac{\mc{N}_\infty^d (w \, | \, z_0, z_{\llbracket n \rrbracket}) }{\mc{D}^d_\infty (w \, | \, z_0, z_{\llbracket n \rrbracket})}.
		\qedhere
	\end{split}
	\end{equation}
\end{proof}

\begin{remark}
	We believe that this formula works even when $M_1$ is not a well-defined function of $M_2$, but it is tricky to establish due to the possibility of the $\exp(1/\nu(w))$ contributing to the leading order coefficients in the limit. 
	If one could prove that the $\exp(1/\nu(w))$ do not contribute to the leading order in the limit, the lemma would hold even when $M_1$ is not a well-defined function of $M_2$.
\end{remark}

\subsection{Essential singularities only contribute for \texorpdfstring{$n=1$}{n=1}}
\label{s:EssSingn=1}

As highlighted in \cref{re:compact}, given transalgebraic functions $x$ and $y$ with exponential singularities on a compact Riemann surface $\Sigma$, with $x y$ meromorphic, one can define two distinct spectral curves:
\begin{enumerate}
	\item A compact transalgebraic spectral curve, where the Riemann surface is taken to be $\Sigma$ itself;
	\item A non-compact meromorphic spectral curve, where the Riemann surface is taken to be $\Sigma \setminus R_\infty$ (which ignores the essential singularities of $x$).
\end{enumerate}
In this section, we show that topological recursion on the compact transalgebraic spectral curve differs from topological recursion on the non-compact meromorphic spectral curve at only finitely many steps. More explicitly, what this means is that for all but finitely many $ (g,n)$, the $ \omega_{g,n}$ defined by topological recursion on transalgebraic spectral curves (\cref{d:main}) may also be calculated using the recursive step of standard topological recursion (\cref{BE-TR}) ignoring essential singularities.

We will prove this by analysing the $ N$-dependence of the $ \omega_{g,n}^N$; we show that $ \lim_{N \to \infty} \omega_{g,n}^N$ has no principal part at $ R_\infty $ for all but finitely many $ (g,n)$.

We begin by examining the loop equations of \cref{AbsLoopEqns}. We wish to prove that the $ i $th loop equation (corresponding to $ \mathcal{E}_{g,n,i} $) can be written with all the same sheets rather than all different ones. First we define the non-regularised $ (0,2) $ correlator
\begin{equation}
	\bar{\omega}_{0,2}(z_1,z_2) \coloneqq \omega_{0,2}(z_1,z_2) - \frac{ dx(z_1) dx(z_2)}{(x(z_1)-x(z_2))^2} = - \sum_{\zeta \in \mf{f}' (z_1)} \omega_{0,2} (\zeta, z_2 ) + \textup{regular} \,.
\end{equation}
	
Then we have the following result.

\begin{proposition}\label{p:lleq}
	On a meromorphic spectral curve,  
	\begin{equation}
		\sum_{\substack{Z \subset \mf{f}_a(z) \\ |Z | = i } } \mc{E}_{g,n, i} ( Z \, | \, z_{\llbracket n \rrbracket} ) 
		= (-1)^{i-1}(i-1)! \sum_{z' \in \mf{f}_a(z)} \bar{\mathcal{E}}_{g,n,i}(z',\dotsc,z' \, | \, z_{\llbracket n \rrbracket }) + \mc{O} ( z^{r_a - 1 - 2g} dz^i)
	\end{equation}
	for a local coordinate $ z $ near a ramification point $ a $ with $ s_a = 1$, where the bar indicates we replace any occurrence of $ \omega_{0,2} $ with $ \bar{\omega}_{0,2} $.\par
	For an admissible transalgebraic spectral curve $ \mc{S}$, for fixed $g$, either side of the expression is holomorphic of arbitrarily high vanishing order on $ \mc{S}_N$ near points $ a \in R_\infty^N$ as $N \to \infty$.
\end{proposition}
 
\begin{proof}
	We will repeatedly apply the linear loop equations on the meromorphic spectral curves converging to our transalgebraic one. The linear loop equation, the $ i =1 $ case of \cref{AbsLoopEqns}, always holds up to $ \mc{O}( x^0(z)dx(z)) = \mc{O} ( z^{r_a - 1} dz )$. By assumption, $ s_a = 1$, so by \cref{PoleOrder}, the pole order of $ \omega_{g,n}$ at $ a $ is bounded by $ 2g$.\par
  Writing $ \zeta_k $ for the local Galois conjugates of a $ \zeta_1$,
\begin{multline}
	\sum_{k_1} \sum_{k_2 \neq k_1} \cdots \sum_{k_i \neq k_1,\dots,k_{i-1}} \mathcal{E}_{g,n,i}(\zeta_{k_1},\dotsc,\zeta_{k_i}(z) \, | \, z_{ \llbracket n \rrbracket })\\
	\sim -\sum_{k_1} \sum_{k_2 \neq k_1} \cdots \sum_{k_{i-1} \neq k_1,\dots,k_{i-2}} \sum_{j=1}^{i-1} \mathcal{E}_{g,n,i}(\zeta_{k_1},\dotsc, \zeta_{k_{i-1}},\overline{\zeta_{k_j}} \, | \, z_{ \llbracket n \rrbracket })\,,
\end{multline}
where the bar over the entry indicates we should replace every instance of $ \omega_{0,2} $ evaluated at this entry with $ \bar{\omega}_{0,2} $ and $ \sim$ means up to $ \mc{O}(z^{r_a -1 - 2g} dz^i)$. We then note the summation is entirely symmetric in $ k_1,\dots,k_{i-1} $ (we are summing over all permutations of $ i-1 $ indices where no two indices are the same) so all terms in the sum over $ j $ are equal to the term with $ j=1 $ and therefore the above is equal to
\begin{multline}
	-(i-1) \sum_{k_1} \sum_{k_2 \neq k_1} \cdots \sum_{k_{i-1} \neq k_1,\dots,k_{i-2}} \mathcal{E}_{g,n,i}(\zeta_{k_1},\dotsc,\zeta_{k_{i-1}},\overline{\zeta_{k_1} } \, | \, z_{ \llbracket n \rrbracket })\\
	\sim (i-1) \sum_{k_1} \sum_{k_2 \neq k_1} \cdots \sum_{k_{i-2}\neq k_1,\dots,k_{i-3}} \sum_{j=1}^{i-2} \mathcal{E}_{g,n,i}(\zeta_{k_1}, \dotsc, \zeta_{k_{i-2}}, \overline{\zeta_{k_{j}} }, \overline{\zeta_{k_1} } \, | \, z_{ \llbracket n \rrbracket }).
\end{multline}
By the same argument we may remove the sum over $ j $ by picking up a factor of $ i-2 $. Repeating this argument a further $ i-3 $ times yields the desired first result.\par
For the second, we use that for any $ a \in R_\infty^N$, by admissibility, $ s_a = 1$, so we may apply the first result, and moreover the order of the loop equation from \cref{AbsLoopEqns} is $ \mc{O} ( z^{r_a-i} dz^i)$, as $ s_a = 1$ and $ i \leq r_a$.
Then $ r_a = \mc{O} (N)$, so indeed the vanishing order of either side grows arbitrarily large as $ N \to \infty $.
\end{proof}

	

Now we use our rewritten loop equations to derive the desired $ N $-dependence of the $ \omega_{g,n}^N $. Fixing an essential singularity $ a $ of $ x $, we define a local coordinate near this essential singularity through $ \zeta^{-m_1} = M_1 $; using this notation, the essential singularity corresponds to $ \zeta (a) = 0 $. Setting $ \tau = 0 $ in $ x_N $, let $ \vartheta $ be a primitive $ m_1 $th root of unity and define the coordinate $ t $ such that $ t^{-N} = x_N $ so that $ t (\zeta) = M_0( \zeta )^{-1/N} (1 - \frac{1}{\zeta^{m_1} N}) $ where the branch of the $ N $th root is chosen such that $ M_0( \zeta = \vartheta^m N^{-1/m_1} ) $ does not lie on the cut for any values of $ m $ and $ N $ and the limit value of $ t $ ($ \lim_{N \to \infty} t $ is a constant function) is not a pole of any of the correlators. Our claim is then

\begin{lemma} \label{NDependenceCorrelators}
	With the above conditions, the principal part of $ \omega_{g,n+1}^N (t, z_{\llbracket n \rrbracket} ) $ at $\zeta (t) = \theta^m N^{-1/m_1}$ is given by
	\begin{equation}
		\Res_{\zeta =\vartheta^m N^{-1/m_1}} \Big( \int_{\theta^m N^{-1/m_1}}^\zeta \!\!\!\omega_{0,2}(t, \mathord{\cdot} ) \Big) \omega_{g,n+1}^N ( \zeta, z_{ \llbracket n \rrbracket } ) = N^{1 - n + \frac{m_2}{m_1} \chi_{g,n+1}} \sum_{l=1-2g}^{-1} w^{l,m}_{g,n} (z_{\llbracket n \rrbracket } ) d\xi_l^m(t) \,,
	\end{equation}
	where
	\begin{equation}
		d\xi_l^m(t') = \Res_{\zeta =\vartheta^m N^{-1/m_1}} \Big( \int_{\theta^m N^{-1/m_1}}^\zeta \!\!\!\omega_{0,2}(t', \mathord{\cdot} ) \Big) t(\zeta )^{l} dt ( \zeta ) ,\qquad l<0 \,,
	\end{equation}
	$ \chi_{g,n} = 2-2g-n $ is the Euler characteristic, and $ w^{l,m}_{g,n} = \mathcal{O}(N^0) $.
\end{lemma}

We require a lemma for the proof.

\begin{lemma} \label{02BarNBehaviour}
	With the notation as above,
	\begin{equation}
		\bar{\omega}_{0,2}^N (t,t) = \frac{dt \, du}{ (t - u)^2 } - \frac{dt^N du^N}{(t^N - u^N)^2} \bigg|_{u = t} + \mc{O}(t^0dt^2) = \frac{(N - 1) (N + 1) (5N - 6)}{24 N} \frac{dt^2}{t^2} + \mc{O} (t^0 dt^2 )\,.
	\end{equation}
\end{lemma}

\begin{proof}
	The first equality holds by general invariance of the principal part  of $ \frac{dz dw}{( z - w )^2} $ under change of local coordinates.\par
	The second equality is a direct calculation, using geometric series.
	\begin{equation*}
	\begin{split}
		\frac{dt \, du}{ (t - u)^2 } - \frac{dt^N du^N}{(t^N - u^N)^2}
		&=
		\bigg( \Big( \sum_{m = 0}^{N - 1} t^{N - m - 1} u^m \Big)^2 - N^2 t^{N-1} u^{N-1} \bigg) \frac{dt \, du}{ (t^N - u^N)^2 }
		\\
		&=
		\bigg( N (N - 1) t^{N - 1} u^{N - 1} - \! \sum_{k = 0}^{N - 2} \! ( k + 1) \Big( t^{2N - 2 - k} u^k + t^k u^{2N - 2 - k} \Big) \bigg) \frac{dt \, du}{ (t^N - u^N)^2 }
		\\
		&=
		 \sum_{k = 0}^{N - 2} ( k + 1) \Big( t^{2N - 2 - k} u^k + t^k u^{2N - 2 - k}- 2 t^{N - 1} u^{N - 1} \Big) \frac{dt \, du}{ (t^N - u^N)^2 }
		\\
		&=
		\sum_{k = 0}^{N - 2} ( k + 1) \sum_{l = 0}^{N - 2 - k}  \frac{ ( t^{2N - 3 - k - l} u^{k + l} - t^{k + l} u^{2N - 3 - k - l} )dt \, du}{ (t - u) \big( \sum_{m = 0}^{N - 1} t^{N - 1 - m} u^m \big)^2 }
		\\
		&=
		\sum_{j = 0}^{N - 2}  \sum_{k = 0}^j ( k + 1) \frac{ ( t^{2N - 3 - j} u^j - t^j u^{2N - 3 - j} ) dt \, du}{ (t - u) \big( \sum_{m = 0}^{N - 1} t^{N - 1 - m} u^m \big)^2 }
		\\
		&=
		\sum_{j = 0}^{N - 2}  \frac{ (j + 1) (j + 2)}{2} \frac{( t^{2N - 3 - j} u^j - t^j u^{2N - 3 - j} ) dt \, du}{ (t - u) \big( \sum_{m = 0}^{N - 1} t^{N - 1 - m} u^m \big)^2 }
		\\
		&=
		\sum_{j = 0}^{N - 2}  \frac{ (j + 1) (j + 2)}{2} t^j u^j \sum_{i = 0}^{2N - 4 - 2j}  \frac{t^{2N - 4 - 2j - i} u^i dt \, du}{ \big( \sum_{m = 0}^{N - 1} t^{N - 1 - m} u^m \big)^2 }\,.
	\end{split}
	\end{equation*}
	At this point, we may set $ u = t$, and obtain
	\begin{equation*}
	\begin{split}
		\bar{\omega}_{0,2}^N (t,t) 
		&=
		\sum_{j = 0}^{N - 2}  \frac{ (j + 1) (j + 2)}{2} (2N - 3 - 2j)  \frac{t^{2N - 4} dt^2}{ \big( N t^{N - 1} \big)^2 } + \mc{O} ( t^0 dt^2)
		\\
		&=
		\frac{ (N - 1) (N + 1) (5N - 6)}{24 N} \frac{dt^2}{t^2} + \mc{O} ( t^0 dt^2) \,. \qedhere
	\end{split}
	\end{equation*}
\end{proof}

\begin{proof}[Proof of \cref{NDependenceCorrelators}]
	The principal part is given by the left-hand side because of the projection property of \cref{AbsLoopEqns}.\par
	To prove the equality, we use induction on the negative of the Euler characteristic $ -\chi_{g,n} = 2g - 2 + n $. We start with $ \omega_{0,1} $. However, we modify $ \omega_{0,1} $ locally near each ramification point $ z = \theta^mN^{1/m_1} $ by subtracting $ -N^{-1} (xy) (\theta^mN^{1/m_1})d\log(x_N) $; as this is a pure function of $ x $, this modification does not affect topological recursion in any way, and it ensures $ \omega_{0,1} $ satisfies the local linear loop equations at the ramification points colliding at $ a $. Examining the expansion coefficients in $ t $ of $ M_2 $ around $ t( z = \theta^m N^{1/m_1} ) = 0 $,
	\begin{equation*}
		M_2(t) = N^{\frac{m_2}{m_1}}\theta^{mm_2} \sum_{l=0}^{\infty} y_l\xi_l(t) \,,
	\end{equation*}
	where the $ y_l $, to leading order in $ N $, do not depend on $ N $ or $ m $. As $ \omega_{0,1}(t) = -N(M_2(t)-M_2(t=0))dt/t $, the claimed result holds in this case. The claim for $ \omega_{0,2} $ also holds:
	\begin{equation*}
		\omega_{0,2}(t,z_1) = \sum_{l=1}^{\infty} w^{l,m}_{0,2}(z_1)\xi_l^m(t), \qquad z\to\theta^mN^{1/m_1} \,.
	\end{equation*}
	
	However, we will be particularly concerned with not $ \omega_{0,2} $ but $ \bar{\omega}_{0,2} $. There are two cases we need to examine when there is such a term in $\mathcal{E}_{g,n,i}(\zeta_{k_{1}}, \dotsc, \zeta_{k_i} \, | \, z_{\llbracket n \rrbracket })$. The first is when a term has a factor of 
	\begin{equation*}
		\bar{\omega}_{0,2}(\zeta_{k_a} ,z_b) = \omega_{0,2}(\zeta_{k_a}, z_b) - \frac{ dx_N(z) dx_N(z_b)}{(x_N (z) - x_N (z_b))}.
	\end{equation*}
	Here, at the ramification points of interest, $ t = 0 $, we note that the second term has a zero of order $ N-1 $ and so will not need to be taken into account in examining the loop equations. The first term is just $\omega_{0,2}$ and so follows the claim.

	The second case is when there is a factor of $ \bar{\omega}_{0,2}(\zeta_{k_a} , \zeta_{k_a} ) $. In this case, by \cref{02BarNBehaviour}, $ \bar{\omega}_{0,2} ( \zeta_{k_a} , \zeta_{k_a} ) = \mc{O} (N^2) \frac{dt^2}{t^2} $.

	Now we are equipped to perform the induction step. Let $ \alpha $ be a primitive $ N $th root of unity and examine the loop equation from \cref{p:lleq}
	\begin{equation*}
		\sum_{r=1}^{N}\mathcal{E}_{g,n,i}(\alpha^r t, \dotsc, \alpha^r t \, | \, z_{\llbracket n \rrbracket }) 
		= 
		\sum_{\mu \vdash \llbracket i \rrbracket } \sum_{\substack{\sqcup_{k=1}^{l(\mu)}N_k = \llbracket n \rrbracket \\ \sum_{k=1}^{l(\mu)}g_k = g+l(\mu)-n}}\prod_{k=1}^{l(\mu)}\omega_{g_k,|\mu_k|+|N_k|}(\alpha^r t, \dotsc, \alpha^r t, z_{N_k}).
	\end{equation*}
	Let us first examine the terms that contain a factor of $ \omega_{g,n+1} $. These are
	\begin{equation*}
		i\sum_{r=1}^{N} \omega_{g,n+1}(\alpha^r t, z_{ \llbracket n \rrbracket } )\omega_{0,1}(\alpha^r t)^{i-1}.
	\end{equation*}
	Each factor of $ \omega_{0,1} $ gives an $ N $ dependence of $ N^{1+\frac{m_2}{m_1}} $ and the sum over $ r $ gives an additional (possible) factor of $ N $. Thus the highest order $N$ dependence the coefficient of $ \omega_{g,n+1}(\alpha^r t,z_{ \llbracket n \rrbracket }) $ can have is $ N^{1+(i-1)(1+\frac{m_2}{m_1})} $.

	Now let us examine the highest order $N$ dependence another term may have using the induction assumption. Examining the expansion of $ \bar{\omega}_{0,2}(\alpha^rt,\alpha^rt) $ and the induction assumption, we see that the highest order occurs when the partition $\mu$ consists only of singletons and pairs where each pair has genus zero and none of the $z_{ \llbracket n \rrbracket }$ to yield a $ \bar{\omega}_{0,2}(\alpha^r t,\alpha^r t) $ factor. We will calculate the largest possible $N$ dependence of such a term. Let $s$ be the number of singletons and $d$ the number of pairs. The pairs give us $d$ factors $ \bar{\omega}_{0,2}(\alpha^r t,\alpha^r t) $, which give an $N$ dependence of $N^{2d}$. The singletons give a total $N$ dependence, by the induction assumption, of $ N^{2s-s-n+\frac{m_2}{m_1}(i-1+\chi_{g,n+1})} $. The sum then gives us an additional factor of $ N $, for a total dependence of (using that $2d+s=i$)
	\begin{equation*}
		N^{2d+s-n+\frac{m_2}{m_1}(i-1+\chi_{g,n+1})+1}=N^{i+1-n+\frac{m_2}{m_1}(i-1+\chi_{g,n+1})}.
	\end{equation*}

	Thus, $\omega_{g,n+1}$ indeed has the claimed $N$ dependence and we are done.
\end{proof}

As a direct consequence of \cref{NDependenceCorrelators}, we find the following corollary. 

\begin{corollary} \label{NoPoleAtLargeN}
Let $a$ be a pole of $M_1$. Then all correlators $\omega_{g,n}^N$ with $ 2gm_2 \geq (2-n)(m_1+m_2) $ have vanishing principal part at $ a $ in the limit as $ N \to \infty $; in particular, this includes all correlators with $ n \geq 2 $.
\end{corollary}

\begin{proof}
	Given that $ t = M_0(z)^{-1/N}( 1-z^{m_1}/N ) $, we have the following expansion, where the $ a_l = \mathcal{O}(N^0) $ are order one coefficients that, to leading order in $ N $, do not depend on $ m $:
	\begin{equation*}
		t = \sum_{l=1}^{\infty} \frac{a_l}{\theta^{ml}N^{l/m_1}} ( z-\theta^mN^{1/m_1} )^l.
	\end{equation*}
	Ergo, \cref{NDependenceCorrelators} implies that we have the expansion
	\begin{equation}\label{ExpansionAtSing}
		\omega_{g,n+1}^N(z,z_{ \llbracket n \rrbracket }) 
		=
		N^{1-n+\frac{m_2}{m_1}\chi_{g,n+1}} \sum_{l=2}^{2g} \frac{A_{l,m}(z_{ \llbracket n \rrbracket })}{(z/(\theta^mN^{1/m_1})-1)^l} d\frac{z}{\theta^mN^{1/m_1}} + \mathcal{O}\left((z-\theta^mN^{1/m_1})^0\right)\,,
	\end{equation}
	where the $ A_{l,m} = \mathcal{O}(N^0) $. So if $ 1 - n + \frac{m_2}{m_1} \chi_{g,n+1} - \frac{1}{m_1} < 0$, the limit $ N \to \infty $ vanishes. Changing $ n + 1 \to n $, this is equivalent to the condition in the corollary.
\end{proof}

\begin{corollary}\label{c:regular}
	The correlators $\omega_{g,n}$ with $ 2g - 2 + n > 0$ are regular at essential singularities where $m_2 \geq m_1$; in particular, this includes all essential singularities where $M_1$ has only a simple pole ($m_1=1$).
\end{corollary}
\begin{proof}
	For $n>1$ we have established the correlators may never have poles. For $n=1$ we know the correlators do not have poles if $ (2g-1)m_2 \geq m_1 $ by the previous corollary; this is trivially true if $m_2 \geq m_1$ as $g \geq 1$.
\end{proof}

\begin{proposition} \label{TransAlgTRViaAlgTR}
	Let $\mathcal{S}$ be a compact transalgebraic admissible spectral curve. For $ 2 g m_2 \geq (2 - n) (m_1 + m_2) $, the correlators $ \omega_{g,n}$ defined via topological recursion  on $\mathcal{S}$ (\cref{d:main}) are regular at all essential singularities $a \in R_\infty$. In particular, this includes all correlators with $n \geq 2$.\par
	The correlators $\omega_{g,n}$ satisfying the condition above may be calculated via the topological recursion of \cref{BE-TR} with residues only at the finite ramification points, but where the $ \omega_{g_k,n_k} $ on the right-hand side of \cref{TR-RecursiveStep} are obtained by the topological recursion of \cref{d:main}.
\end{proposition}

\begin{proof}
	The first statement follows immediately from \cref{NoPoleAtLargeN} and the definition of transalgebraic topological recursion as a limit.\par
	For the second statement, note that the individual contributions of ramification points in \cref{TR-RecursiveStep} are continuous as the spectral curve varies without the type of ramification changing.\footnote{To be precise, this is proven for ramification points of arbitrary order in \cite{BBCKS23}.} This is the case for the $ a \in R_0^N$, which converge to $ R_0$. By the projection property, \cref{ProjProp}, $ \omega_{g,n}^N$ are the sums of their principal parts, and by \cref{NoPoleAtLargeN}, the limit of the contributions at elements of $ R_\infty^N $ vanishes. As $ \omega_{g,n}$ is defined as the limit, this proves the second statement.
\end{proof}

\begin{remark}
	\Cref{TransAlgTRViaAlgTR} does \emph{not} mean that for $ 2 g m_2 \geq (2 - n) ( m_1 + m_2 )$, the correlators $ \omega_{g,n}$ calculated via  topological recursion on the transalgebraic spectral curve (\cref{d:main}) are equal to the ones calculated from  topological recursion on the non-compact meromorphic spectral curve that ignores essential singularities. Rather, the contributions from the essential singularities at higher Euler characteristic propagate to all $ (g,n)$ through the recursion at finite ramification points.\par
	However, as the inequality only fails for $ n = 1$ and $ 2g - 1 \leq \frac{m_1}{m_2} $, it does mean that for any given transalgebraic spectral curve, the limit definition of topological recursion only has to be used a finite number of times, and can then be disregarded for the remaining correlator calculations.
\end{remark}

Now we provide a bound on the order of the poles of the correlators at the infinite ramification points.

\begin{proposition}\label{p:bound}
	Let $\mathcal{S}$ be a compact transalgebraic admissible spectral curve. Let $a \in R_\infty $ be an infinite ramification point. Suppose that $M_1$ has a pole of order $m_1$ at $a$, and let $m_2$ be the order of the pole of $ xy$ at $a$. Then $\omega_{g,1}$ has a pole of order no greater than $m_2 (1-2g) + m_1 + 1$ at $a$.
\end{proposition}

\begin{proof}
	Begin with the expansion we found before in \cref{ExpansionAtSing}, which for $ n =0$ reads
	\begin{equation}
		\omega_{g,1}^N(z) 
		=
		N^{1+\frac{m_2}{m_1}(1-2g)} \sum_{l=2}^{2g} \frac{A_{l,m}}{(z/(\theta^m N^{1/m_1})-1)^l} d\frac{z}{\theta^m N^{1/m_1}} + \mathcal{O}\left((z-\theta^m N^{1/m_1})^0\right)\,,
	\end{equation}
	where $ A_{l,m} = \mc{O}(N^0)$. We then change coordinate $ z = \zeta^{-1}$, such that $ \zeta (a) = 0$ and $ \zeta^{m_1} = M_1$, so we obtain
		\begin{equation}
		\omega_{g,1}^N(\zeta ) 
		\sim
		N^{1+\frac{m_2}{m_1}(1-2g) - \frac{1}{m_1}} \sum_{l=2}^{2g} \frac{A_{l,m} \theta^{-m} }{(\theta^{-m} N^{-1/m_1} \zeta^{-1} -1)^l} d\zeta^{-1} \,,
	\end{equation}
	where we ignore the non-polar part, as it does not contribute to the poles for an admissible spectral curve.\par
	For a fixed small $ \zeta$, take $ N $ large enough that $ N^{-1/m_1}\zeta^{-1} $ is small, and use this to expand denominators as geometric series. We see that the coefficient of $ \zeta^{-k} d\zeta^{-1} = - \zeta^{-k-2} d\zeta $ is then $ \mc{O}(N^{1 + \frac{m_2}{m_1}(1-2g) - \frac{k+1}{m_1}}) $, so for this to persist in the limit, we require that
	\begin{equation*}
		\begin{split}
			0 
			&
			\leq 1 + \frac{m_2}{m_1}(1-2g) - \frac{k+1}{m_1}
			\\
			0
			&
			\leq m_1 + m_2(1-2g) - k-1
			\\
			k
			&
			\leq m_1 + m_2 (1-2g) - 1\,,
		\end{split}
	\end{equation*}
	which after a shift of $2$ for $k$ exactly gives the order in the proposition.
\end{proof}

\begin{proposition}\label{p:11}
	For $g = 1$, we have an explicit formula for the contribution of the infinite ramification points.
	\begin{equation}
	\begin{split}
		&\sum_{a \in R_\infty} \Res_{t=a} \left(\int_a^t\omega_{0,2}(z_0,\cdot)\right) \omega_{1,n} (t, z_{\llbracket n-1 \rrbracket})
		\\
		=
		 -\frac{\delta_{n,1}}{24} &\sum_{a \in R_\infty} \Res_{t=a} \left(\int_a^t\omega_{0,2}(z_0,\cdot)\right) d\left( \frac{d}{dM_2(t)}\right) \log(x(t))
		\\
		=
		-\frac{\delta_{n,1}}{24} &\sum_{a \in R_\infty} \Res_{t=a} \left(\int_a^t\omega_{0,2}(z_0,\cdot)\right) dM_2(t) \left(\frac{d}{dM_2(t)} \right)^2 M_1(t) \,,
	\end{split}
	\end{equation}
\end{proposition}
\begin{proof}
	For $n\geq 2$ the result holds by \cref{NoPoleAtLargeN}, so we concentrate on the $n=1$ case. Take a large positive integer $N$ with a corresponding primitive $N$th root of unity $\alpha$ and fix a ramification point of $x_N$ (we use $\tau = 0$), which we denote $a$, such that $M_1(a) = N$. Fixing a local coordinate $t^{-N}=x_N$ we write the quadratic local loop equation for $\omega^N_{1,1}$ at $a$\footnote{When we consider the local loop equations about $a$ we need $\omega^N_{0,1}$ to be regular at $a$. This is accomplished by taking the local $\omega^N_{0,1}$ to be $\left[ M_2(\alpha^i t)-M_2(a) \right] d\log(x_N(t))$, which differs from the original by a pure function of $x_N$ and therefore does not change the other correlators. See the proof of \cref{NDependenceCorrelators} for further explanation.}
	\begin{equation}
		\frac{1}{2!} \frac{dx_N(t)}{x_N(t)} \sum_{\substack{i,j=1 \\ i\neq j}}^{N} 2 \left[ M_2(\alpha^i t)-M_2(a) \right] \omega_{1,1}(\alpha^j t) + \frac{1}{2!} \sum_{\substack{i,j=1 \\ i\neq j}}^{N} \omega_{0,2}(\alpha^i t,\alpha^j t) = \mathcal{O}\left(\frac{dx_N(t)^2}{x_N(t)}\right).
	\end{equation}
	By \cref{PoleOrder} (with $s_a=1$) $\omega^N_{1,1}$ has only a double pole at $a$; denote the coefficient of this double pole as $A$. Therefore,
	\begin{equation}
		\left[ M_2(\alpha^i t)-M_2(a) \right] \omega_{1,1}(\alpha^j t) = - \alpha^{i-j} M_2'(a) \frac{A}{N} \frac{dx_N(t)}{x_N(t)} + \mathcal{O}(dt),
	\end{equation}
	where we assume that $N$ is chosen so large that $M_2'(a) \neq 0$. Furthermore, from \cite[Lemma~A.5]{BBCCN18} we have
	\begin{equation}
		\frac{1}{2!} \sum_{\substack{i,j=1 \\ i\neq j}}^{N} \omega_{0,2}(\alpha^i t,\alpha^j t) = -\frac{N^2-1}{24N}\left(\frac{dx_N(t)}{x_N(t)}\right)^2.
	\end{equation}
	Putting these two results together we obtain
	\begin{equation}
		A = \frac{1}{M_2'(a)}\frac{N^2-1}{24N},
	\end{equation}
	and we can then repackage this result in a more suggestive form (the base point $*$ is arbitrary)
	\begin{equation}
		\Res_{t=a} \left(\int_*^t\omega_{0,2}(z_0,\cdot)\right) \omega^N_{1,1}(t) = - \frac{N^2-1}{24N^2} \Res_{t=a} \left(\int_*^t\omega_{0,2}(z_0,\cdot)\right) d \frac{d}{dM_2(t)} \log(x_N(t)).
	\end{equation}
	This holds for every $a$ with $M_1(a)=N$. Ergo, we can sum both sides over all such $a$. Moreover, near the elements of $R_\infty$ (poles of $M_1$) the integrands will only have poles at such $a$s; we can therefore replace this sum over residues by integration over a contour $\Gamma$ that is the disjoint union of small circles around each element of $R_\infty$
	\begin{equation}
		\int_{\Gamma(t)} \left(\int_*^t\omega_{0,2}(z_0,\cdot)\right) \omega^N_{1,1}(t) = - \frac{N^2-1}{24N^2} \int_{\Gamma(t)} \left(\int_*^t\omega_{0,2}(z_0,\cdot)\right) d \frac{d}{dM_2(t)} \log(x_N(t)).
	\end{equation}
	Noting that the second Bernoulli number is $B_2 = 1/6 $ so $(2^{1-2}-1)B_2/2! = -1/24$, the claimed result holds upon taking the $N\to\infty$ limit of the above expression.
\end{proof}

Let us show this explicitly in an example.

\begin{example}\label{ex:qOrbifoldrAtlantes}
	Let $z$ be an affine coordinate on $\P^1$, $r,q \in \mathbb{Z}_{\geq 1}$, and consider the spectral curve
	\begin{equation}
		\mathcal{S} = \left( \P^1, x(z)=ze^{-z^{qr}}, y(z)=z^{q-1}e^{z^{qr}} \right),
	\end{equation}
	which we will call the $q$-orbifold $r$-atlantes Hurwitz curve.
	
	Here we will use \cref{l:formula} to calculate the contribution from the essential singularity at infinity to the correlator $\omega_{1,1}$. Given our spectral curve we have $M_1(z) = -z^{qr}$, $M_2(z) = z^q$, and $\mathcal{W}_{1,0,2}(t, t_1 \, |\, \emptyset) = \omega_{0,2}(t, t_1)$. Letting $\theta$ be a primitive $q$th root of unity we see that this contribution will be
	\begin{equation}
		\Res_{t=\infty} \left( \int_\infty^t\omega_{0,2}(z_0,\cdot) \right) \omega_{1,1}(t) = \Res_{t=\infty} \frac{dz_0}{z_0-t} \Res_{t_1=R_0,\mathcal{Y}(t)} \frac{te^{-t^{qr}}}{te^{-t^{qr}}-t_1e^{-t_1^{qr}}} \frac{\omega_{0,2}(t,t_1)}{t^q-t_1^q}.
	\end{equation}
	The residues at $t_1 = R_0$ will drop out, as the integrand has no poles here. For the residues at $t_1 = \mathcal{Y}(t)$ we must be careful to distinguish between the trivial and non-trivial sheets of $M_2$, as the pole structure of the integrand is different in these two cases. First, we look at the non-trivial sheets, where there is only a simple pole
	\begin{equation}
		\sum_{m=2}^{q-1} \Res_{t_1=\theta^m t} \frac{t{\rm e}^{-t^{qr}}}{te^{-t^{qr}}-t_1e^{-t_1^{qr}}} \frac{\omega_{0,2}(t,t_1)}{t^q-t_1^q} = \sum_{m=2}^{q-1} \frac{1}{1-\theta^m} \frac{\theta^m}{-q} \frac{dt}{t^2(1-\theta^m)^2},
	\end{equation}
	which we see has no pole at $t=\infty$ and so will not contribute to the final result. Next, we examine the residue at $t_1=t$. The calculation was done on SageMath \cite{sagemath} and we just present the result here:
	\begin{equation}
		\begin{split}
			&\Res_{t_1=t} \frac{te^{-t^{qr}}}{t_1e^{-t_1^{qr}}-te^{-t^{qr}}} \frac{\omega_{0,2}(t,t_1)}{t_1^q-t^q} = -\frac{qr(r-1)t^{qr-q-1}{\rm d}t}{24} + \mathcal{O}(t^{-2})dt.
		\end{split}
	\end{equation}
	Then, multiplying by $\int_\infty^t\omega_{0,2}(z_0,\cdot)$ and taking the residue at infinity we obtain
	\begin{equation}
		\Res_{t=\infty} \left( \int_\infty^t\omega_{0,2}(z_0,\cdot) \right) \omega_{1,1}(t) = -\frac{rdz_0^{q(r-1)}}{24},
	\end{equation}
	which is in agreement with \cref{p:11}. Note that we did not have to use the re-definition of the residue at $t=R_\infty$ in \cref{l:formula} as the integrand is meromorphic. This is generic to calculations of $\omega_{1,1}$, but will not hold for more complicated correlators.
\end{example}

\section{Quantum curves}

\label{s:QC}

One of the main motivations for introducing topological recursion on transalgebraic spectral curves is to make a sense of a conundrum related to sequences of meromorphic spectral curves, which arises in the context of quantum curves. To understand  this, we introduce the notion of quantum curves, and the topological recursion/quantum curve correspondence.

\subsection{The topological recursion/quantum curve correspondence}

Topological recursion originally appeared in the context of matrix models, where the correlators $ \omega_{g,n} $ are  generating functions for expectation values of the traces of the matrices under consideration \cite{E04,EO07}. But the trace is only one of the most natural basis-independent objects one can form from a matrix; another is, of course, the determinant. Traces and determinants are intimately connected, and, fundamentally, this relation is what gives rise to the topological recursion/quantum curve correspondence.

In a matrix model, the expectation values of the determinants satisfy certain differential equations; roughly speaking, the solution of these differential equations is the \emph{wave function} $\psi$ and the operator that kills it is the \emph{quantum curve}. Because of the well known relation $ \det \exp = \exp \Tr $, it is intuitively clear that the wave function should involve the exponential of the $ \omega_{g,n} $. The connection between the differential equation satisfied by the wave function $\psi$ and the topological recursion satisfied by the correlators $\omega_{g,n}$ is made explicit in the topological recursion/quantum curve correspondence.

Let us now be a little more precise. Let $ \mathcal{S} = \left( \Sigma,x, y, B\right) $ be a meromorphic spectral curve. The functions $x$ and $y$ satisfy a relation $P(x,y)=0$. If the spectral curve is compact, then $P$ will be polynomial, but in general it may not be.

Define the wave function $\psi$ associated to the spectral curve $\mathcal{S}$ as
\begin{equation}\label{e:wfuncconj}
	\psi(x(z)) = \exp\left[ \sum_{n=1}^{\infty} \sum_{g=0}^{\infty} \frac{\hslash^{2g+n-2}}{n!} \int^z\cdots\int^z \left( \omega_{g,n}  - \delta_{g,0} \delta_{n,2} \frac{dx(z_1) dx(z_2)}{(x(z_1) - x(z_2))^2} \right) \right] \,,
\end{equation}
which is an exponential of the correlators $\omega_{g,n}$ constructed from topological recursion. Here $ \hslash $ is a formal expansion parameter, there are $n$ integrations in each term, and it is conventional to write $ \psi $ as a function of $x(z)$, rather than $z$, even though it is not globally well-defined as such.\footnote{This convention is the natural one as the way one obtains the expectation values of the traces from the $ \omega_{g,n} $ is through formal expansion in $x$ where the expectation values of the traces are read off from the expansion coefficients.} The exact nature of the integration should be defined carefully (see, for example, \cite{BE17}).

The statement of the topological recursion/quantum curve correspondence is that there should exist an operator $\hat{P}(\hat{x}, \hat{y}, \hslash )$ such that 
\begin{equation}
	\hat{P} \psi = 0\,,
\end{equation}
where $ \hat{x} = x \cdot $ and $ \hat{y} = \hslash \frac{d}{dx}$. Furthermore, this $ \hat{P}$ should be a quantisation of $P$, in the sense that $ \hat{P} (x, y, 0) = P(x,y)$. If such a $\hat{P}$ exists, we call it a ``quantum curve''. 

Of course, there is no unique quantisation of $P$, due to non-commutativity of $ \hat{x}$ and $ \hat{y}$. Moreover, we may allow corrections of order $\hslash$ in the operator $\hat{P}$. In our context, we define quantisation as follows.

\begin{definition}\label{d:Quant}
	Let $\mathcal{S}$ be a meromorphic spectral curve, with $x$ and $y$ satisfying the relation $ P(x,y) = 0 $. We say that $\hat{P}(\hat{x},\hat{y};\hslash) $  is a quantisation of $P(x,y)$ if we have the following expansion for some $ m\in\mathbb{N}\cup\{\infty\} $:
		\begin{equation*}
			\hat{P}(\hat{x},\hat{y};\hslash) = P(\hat{x},\hat{y}) + \sum_{i=1}^{m} \hslash^i\hat{P}_i(\hat{x},\hat{y}),
		\end{equation*}
		where $P(\hat{x},\hat{y})$ is taken to be normally ordered (in each term all the $\hat{x}$ are put to the left of the $\hat{y}$) and the $ \hat{P}_i $ are  normal ordered polynomials of degree at most $ \deg P - 1 $. We say that the quantisation is \emph{simple} if $ m < \infty $.
\end{definition}

We can now state the topological recursion/quantum curve correspondence.\footnote{This conjecture is sometimes referred to as the Gukov-Sulkowski conjecture in the literature \cite{GS11}; however, the result has been well-known in the context of matrix models \cite{M90} long before the topological recursion was introduced, and was already being considered more generally in \cite{BE09} in the context of the then recently discovered topological recursion before \cite{GS11}.}

\begin{conjecture}\label{con:GS}
Let $\mathcal{S}$ be a meromorphic spectral curve, with $x$ and $y$ satisfying the relation $ P(x,y) = 0 $. Let $\psi(x(z))$ be the wave function \eqref{e:wfuncconj} associated to $\mathcal{S}$, with the $\omega_{g,n}$ constructed from topological recursion. Then there exists a quantisation $\hat{P}(\hat{x},\hat{y}; \hslash)$ of $P(x,y)$ such that
\begin{equation}
\hat{P}(\hat{x},\hat{y}; \hslash) \psi(x(z)) = 0.
\end{equation}
We call $\hat{P}(\hat{x},\hat{y}; \hslash) $ a quantum curve.
\end{conjecture}
%
%

As stated here, the conjecture is imprecise. To start with, it requires a proper definition of integration in the wave function \eqref{e:wfuncconj} (see \cite{BE17}). Furthermore, the wave function \eqref{e:wfuncconj} is the ``perturbative wave function'', and as stated the conjecture is only expected to hold when the spectral curve is genus zero. For higher genus spectral curves, non-perturbative corrections should be added to \eqref{e:wfuncconj}. Nevertheless, the statement can be made precise, and the conjecture has been proved for a wide class of compact meromorphic genus zero spectral curves with arbitrary ramification in \cite{BE17}, as well as for every compact meromorphic spectral curves with only simple ramification in \cite{EG23,MO22,EGMO21}.

\subsection{Quantum curves for transalgebraic spectral curves}

The conjecture has also been proved for a number of non-compact meromorphic genus zero spectral curves, such as the spectral curve from \cref{ex:rspin} \cite{MSS13}. However, in contrast to the compact cases mentioned above, in these non-compact cases the existence of the quantum curve is proved from the enumerative geometric interpretation of the correlators (which is $r$-completed cycles Hurwitz theory in the case of \cref{ex:rspin}), and not directly from topological recursion. 

Part of the motivation for the current paper was to prove the existence of quantum curves for such cases directly from topological recursion. Our idea is simple: as the transalgebraic spectral curve is obtained as the $N \to \infty$ limit of a sequence of compact meromorphic spectral curves, if quantum curves are known to exist for the spectral curves at finite $N$, then we simply need to take the $N \to \infty$ limit of these quantum curves to get the quantum curve for topological recursion on the transalgebraic spectral curve.

Therefore, to achieve this program we need to consider transalgebraic curves for which the finite $N$ spectral curves are known to satisfy the topological recursion/quantum curve correspondence. In this respect, we will use the results of \cite{BE17}. Using this methodology, we will prove the topological recursion/quantum curve correspondence for a large class of compact meromorphic spectral curves, which are called ``regular''. Moreover, the proof is constructive, as it provides an explicit way of calculating the quantum curve.

To understand the regularity condition on spectral curves, we need to introduce the Newton polygon of a plane curve.

\begin{definition}\label{d:Npoly}
  The \emph{Newton polygon} $\Delta$ of $P$ is the convex hull of the exponents in $P$, i.e., the convex hull in $ \R^2 $ of $ A \coloneqq \{(i,j)\in\mathbb{N}^2 \, | \, \alpha_{i,j} \neq 0\}$.
\end{definition}

Regularity for compact meromorphic spectral curves is defined as follows.\footnote{Regular is called ``admissible'' in \cite{BE17}.}

\begin{definition}[Definition 2.7, \cite{BE17}]
	Let $\mathcal{S}$ be a compact meromorphic spectral curve. Then $x$ and $y$ satisfy a polynomial equation $P(x,y) = 0$. We say that $\mathcal{S}$ is \emph{regular} if $P(x,y) = 0$ is smooth as an affine curve and its Newton polygon has no integral interior point.\footnote{As every Newton polygon with an interior contains a non-integral interior point it is common to state this condition without the word ``integral''.}
\end{definition}
In particular, all regular spectral curves have genus zero by Baker's formula \cite{B1895}, which states that the number of interior points of the Newton polygon is greater or equal than the genus of the curve. In fact, compact meromorphic spectral curves that are regular can be classified \cite{BE17}. A compact meromorphic spectral curve is regular if and only if it falls into one of the following cases:
\begin{itemize}
	\item $P(x,y)$ is linear in $x$, i.e., $P(x,y) = x E_1 (y) - E_2 (y)$, where $E_1, E_2$ are polynomials.
	\item $P(x,y)$ has Newton polygon $\Delta$ given by the convex hull of $ \{ (0,0), (0,2), (2,0) \}$.
	\item $P(x,y)$ is obtained from one of the two previous cases via a transformation $ (x,y) \to (x^a y^b, x^c y^d)$ with $ad - bc = 1$ and a rescaling by powers of $x$ and $y$ to get an irreducible polynomial equation.
\end{itemize}
For all regular spectral curves, the quantum curve associated to the corresponding wave function is constructed in \cite{BE17}.

We would like to extend the notion of regularity to transalgebraic spectral curves. In the spirit of defining the transalgebraic in terms of limits of the algebraic, it would seem most natural to define transalgebraic curves $\mathcal{S}$ as regular precisely when the considered sequence of meromorphic curves that converge to $\mathcal{S}$ are regular. The following lemma precisely characterises when this is the case.

 \begin{lemma}\label{l:class}
	Let $\mathcal{S} = (\Sigma, x , y, B)$ be a compact transalgebraic spectral curve, with the notation of \cref{sec:TransAlgSCasLimits}. Then the curves $\mathcal{S}_N = (\Sigma, x_N, y_N, B)$ with
	\begin{equation}
		x_N = M_0 \left( 1 + (\tau - 1) \frac{M_1}{N} \right)^{-N} \left( 1 + \tau \frac{M_1}{N} \right)^{N}\,, \qquad y _N = M_2/x_N\,
	\end{equation}
	are regular for all $N$ if and only if $\Sigma \cong \mathbb{P}^1$ and $xy \in \mathop{\mathrm{Aut}}(\mathbb{P}^1)$.
\end{lemma}

\begin{proof}
	If $x_Ny_N=M_2$ is a M\"obius transformation, then we may define an affine coordinate $z$ as $z=x_Ny_N$. As $x_N$ is meromorphic on $\P^1$, this means that we can write
	\begin{equation}
		x_N = f_N(z)
	\end{equation}
	for some rational function $f_N$. Clearing denominators and using $z=x_Ny_N$ it follows that
	\begin{equation}
		E_N^{(1)} (z) x_N - E_N^{(2)} (z) = 0
	\end{equation}
	for some polynomials $E_N^{(1)}$ and $E_N^{(2)}$. But this is a regular curve, as it is obtained from the curve
	\begin{equation}
		E_N^{(1)}(y_N) x_N - E_N^{(2)}( y_N) = 0
	\end{equation}
	via the transformation $(x_N, y_N) \mapsto (x_N, x_N y_N)$.
	
	Alternatively, assume $\mathcal{S}_N$ is regular for every $N$. By the classification of regular compact meromorphic spectral curves, $\mathcal{S}_N$ will be a transformation of either  a curve $P_N( x_N,y_N ) = 0$ that is linear in $x_N$, or a curve $P_N( x_N,y_N ) = 0$ with Newton polygon $\Delta_N$ given by the convex hull of $ \{ (0,0), (0,2), (2,0) \}$. First consider the later case. Up to suitable rescaling by overall powers of $x_N$ and $y_N$ the most general curve of this form is
	\begin{equation}
		P_N(x_N,y_N)=k_0 + k_1x_N^ay_N^b + k_2x_N^cy_N^d + k_3x_N^{a+c}y_N^{b+d} + k_4x_N^{2a}y_N^{2b} + k_5x_N^{2c}y_N^{2d},
	\end{equation}
	for some constants $k_0,\dots,k_5 \in \mathbb{C}$ and $ad-bc=1$. However, if we take a sequence of points $ p_N \in \mc{S}_N$ converging to a point $ p \in \mc{S}$, then we can interpret the $P_N$ in the local rings at $ p_N$, $ \C \llbracket x - x(p_N), y - y(p_N) \rrbracket$, and they converge to a non-polynomial equation in the local ring at $p$, $ \C \llbracket x - x(p), y - y(p) \rrbracket $, locally cutting out $ \mc{S}$. So for sufficiently large $N$, $\mathcal{S}_N$ will never take this form as the number of terms of $P_N$ has to grow unboundedly.
	
	Ergo, we focus on the former case; when $P_N( x_N,y_N ) = 0$ is a transformation of a curve linear in $x_N$. Thus, for some polynomials $E_N^{(1)}$ and $E_N^{(2)}$ and $ad-bc=1$,
	\begin{equation}
		P_N( x_N,y_N ) = x_N^ay_N^bE_N^{(1)}(x_N^cy_N^d) + E_N^{(2)}(x_N^cy_N^d) = 0.
	\end{equation}
	Letting $E_N = E_N^{(2)} / E_N^{(1)}$ and choosing an affine coordinate $w$ on $\mathbb{P}^1$ we have
	\begin{equation}
		(x_N^ay_N^b)(w) = E_N((x_N^cy_N^d)(w)).
	\end{equation}
	We now wish to count the number of sheets of these two equal functions. Using $y_N=M_2/x_N$ and denoting the degree (as a branched covering) of the rational function $E_N$ as $D_N$, we have
	\begin{equation}
		|a-b|\deg(x_N) + |b|\deg(M_2) \geq \deg(x_N^ay_N^b) = \deg(E_N(x_N^cy_N^d)) \geq D_N\left[|c-d|\deg(x_N)-d\deg(M_2)\right].
	\end{equation}
	By assumption, $\mathcal{S}_N$ is transalgebraic in the limit so $E_N$ must not be meromorphic in the limit. Thus, $D_N\to\infty$ so the only way this inequality can hold for arbitrarily large $N$ is if $|c-d|=0$ so $c=d$. However, we now have that $(a-b)c=1$ so $c=\pm 1$ and $a=b\pm 1$. This gives us
	\begin{equation}
		x=[xy]^{-b}[E_N([xy]^{\pm 1})]^{\pm 1},
	\end{equation}
	so $x$ is a well-defined function of $xy$ and as $y=xy/x$, $y$ is also a well-defined function of $xy$. Therefore $xy$ is in fact a valid coordinate everywhere on the curve so we have that the function $xy:\Sigma\cong\mathbb{P}^1\to\mathbb{P}^1$ can be taken to be injective. Ergo, as $xy$ is also meromorphic and therefore rational, $xy \in \mathop{\mathrm{Aut}}(\mathbb{P}^1)$.
\end{proof}

The preceding lemma, then, justifies the succeeding definition.
 
\begin{definition}\label{d:regulartrans}
	Let $\mathcal{S} = (\Sigma, x , y, B)$ be a compact transalgebraic spectral curve. We say that $\mathcal{S}$ is \emph{regular} if $\Sigma\cong\mathbb{P}^1$ and $xy \in \mathop{\mathrm{Aut}}(\mathbb{P}^1)$.
\end{definition}

\begin{remark}
	The reader may wonder whether our regularity condition is merely an artefact of our considered sequence. However, this is almost certainly not the case. Any sequence of curves converging to a transalgebraic curve will have more than six terms, and the equality $\deg(x_N^ay_N^b) = \deg(E_N)\deg(x_N^cy_N^d)$ should always enforce that $\deg(x_N^cy_N^d)$ will be finite in the limit so $c=d=1$.
\end{remark}

In contrast to the meromorphic case, where admissibility and regularity were very much independent conditions, in the transalgebraic case there is a nice classification of all regular curves that are also admissible.

\begin{proposition}\label{p:TransRegImpAdmiss}
	Let $\mathcal{S} = (\Sigma, x , y, B)$ be a regular transalgebraic spectral curve. Then $\mathcal{S}$ is admissible if and only if $M_1$ is a polynomial in $xy$.
\end{proposition}
\begin{proof}
	We first prove necessity. Let $z=xy$ be an affine coordinate. $M_1$ is meromorphic and we are in genus zero so it is rational. If it were not polynomial, then it would have a pole at some point $z = p \neq \infty$. As $x$ will then have an essential singularity at this point, and $xy=M_2$ does not have a pole at this point, the curve may not be admissible by definition \ref{d:tadmis}.
	
	Now we prove sufficiency. In light of the argument for necessity, it is clear that the curve will satisfy definition \ref{d:tadmis} at all essential singularities of $x$ so we must only concern ourselves with the finite ramification points where admissibility is defined in definition \ref{d:AlgSCAdm}. Again letting $z=xy$ be an affine coordinate, we see $\omega_{0,1}(z)=z \, d \log (x(z))$. Let $a\in R_0$ be a ramification point and examine two cases: when $x(a) \in \{0,\infty\}$, or when $x(a) \in \mathbb{C}^{\times}$. In the first case we have that $s_a=1$ and in the second case we have that $s_a=r_a$+1. In either case admissibility holds.
\end{proof}

Given a compact transalgebraic admissible regular spectral curve $\mathcal{S}$, the strategy to construct its quantum curve is then clear. It proceeds in two steps:
\begin{enumerate}
\item For all $N$, we construct the wave function $\psi_N$ from the correlators $\omega_{g,n}^N$ obtained via the usual topological recursion on $\mathcal{S}_N$. From \cite{BE17}, we can construct an associated quantum curve $\hat{P}_N$ such that
\begin{equation}
\hat{P}_N \psi_N = 0.
\end{equation}
\item The $N \to \infty$ limit of the correlators $\omega_{g,n}^N$ gives the correlators $\omega_{g,n}$ associated to the transalgebraic spectral curve $\mathcal{S}$, and hence the $N  \to \infty$ limit of the wave-function $\psi_N$ gives the wave function $\psi$ associated to $\mathcal{S}$. Its quantum curve is thus obtained by taking the $N \to \infty$ limit of $\hat{P}_N$:
\begin{equation}
\lim_{N \to \infty}\left( \hat{P}_N \psi_N \right)= \left( \lim_{N \to \infty} \hat{P}_N \right) \left( \lim_{N \to \infty} \psi_N \right) = \hat{P} \psi = 0.
\end{equation}
\end{enumerate}
This strategy is studied in detail in \cref{QCForTransAlgTR} -- see \cref{t:TR/QCpole} and \cref{t:TR/QCzero}. In particular, we apply this procedure explicitly to calculate the quantum curve for the spectral curve of \cref{ex:atlantes} in the next section.

\subsection{A particular example}

Let us recall the spectral curve from \cref{ex:atlantes}:
\begin{equation}
\mathcal{S}_\infty =  \left(\Sigma=\P^1, \quad x(z)=z e^{-z^r}, \quad y(z)=e^{z^r},  \quad B=\frac{dz_1 dz_2}{(z_1-z_2)^2} \right),
\end{equation}
where $r \in\mathbb{Z}_{\geq 1} $ is a fixed integer. The functions $x$ and $y$ satisfy the relation
\begin{equation}
P(x,y) = y - e^{x^r y^r} = 0.
\end{equation}
We note that this transalgebraic curve is regular, since $M_2(z) = x(z) y(z) = z$. Thus, all spectral curves $\mathcal{S}_N$ in the sequence are regular, and the results of \cite{BE17} apply for finite $N$. As a result, we obtain the quantum curve associated to $\mathcal{S}_\infty$:

\begin{proposition}\label{p:qc}
Let $\mathcal{S}_\infty$ be the compact transalgebraic spectral curve from \cref{ex:atlantes}. We use the results and notations from \cref{QCForTransAlgTR}. Let $\psi_\infty(x;0)$ be the wave function associated to $\mathcal{S}_\infty$ and constructed from the correlators $\omega_{g,n}^\infty$ with the integration base point $z=0$. Then $\psi_\infty(x;0)$ satisfies the quantum curve differential equation
	\begin{equation}
		\label{e:OurCurve}
		\left( \hat{y} - e^{(\hat{x}\hat{y})^r}  \right) \psi_\infty(x;0) = 0.
	\end{equation}
\end{proposition}

\begin{proof}
As $\mathcal{S}_\infty$ is regular, by \cref{l:class} we know that for all $N$ $\mathcal{S}_N$ is regular, and we can use the results of \cref{QCForTransAlgTR}.

	For all $\tau \in \mathbb{C}$, we write the equation $P(x,y) = 0$ satisfied by the functions $x$ and $y$ as follows, following \cref{r:pres}:
	\begin{equation}
		P(x,y) 
		= y e^{(\tau-1)(xy)^r} - e^{\tau(xy)^r} 
		= \sum_{m=0}^{\infty} \frac{(\tau-1)^m}{m!} x^{rm} y^{rm+1} - \sum_{m=0}^{\infty} \frac{\tau^m}{m!}x^{rm}y^{rm}.
	\end{equation}
	Using the notation of \cref{QCForTransAlgTR} we find that $ \lfloor \alpha_m \rfloor = m-1+\delta_{m,0} $, $ q_m(x) = 0 $ if $m \neq 0,1 \pmod{r}$, $ q_{rm}(x) = -\frac{\tau^mx^{rm}}{m!} $, and $ q_{rm+1}(x) = \frac{(\tau-1)^mx^{rm}}{m!} $. Choosing the base point $b=\{z=0\}$, we find the following coefficients
	\begin{equation}
	\begin{split}
		H_1 &= \hslash\left( \frac{d}{dx} - \frac{1}{x} \right), \quad H_i = \hslash\left( x\frac{d}{dx}-1 \right),\\
		F_1 &= \hslash\frac{d}{dx}, \quad F_i=\hslash x\frac{d}{dx}, \quad G_i=0.
	\end{split}
	\end{equation}
	As $z=0$ is a zero of $x$ which is not in the ramification locus, we can apply  \cref{t:TR/QCzero}. We get the following quantum curve, where $ \hat{x} = x $ and $ \hat{y} = \hslash\frac{d}{dx} $:
	\begin{equation}
	\begin{split}
		\hat{P}(\hat{x},\hat{y};\hslash) 
		&=
		-1+ \frac{1}{x} \sum_{m=0}^{\infty} \frac{(\tau-1)^m \hslash^{rm+1}}{m!} \left( x\frac{d}{dx}-1 \right)^{rm} x \frac{d}{dx}-\frac{1}{x} \sum_{m=1}^{\infty} \frac{\tau^m \hslash^{rm}}{m!} \left( x\frac{d}{dx} - 1 \right)^{rm-1} x^2 \frac{d}{dx}
		\\
		&=
		\sum_{m=0}^{\infty} \frac{(\tau-1)^m \hslash^{rm+1}}{m!} \left( x \frac{d}{dx} \right)^{rm} \frac{d}{dx}  - \sum_{m=0}^{\infty} \frac{\tau^m \hslash^{rm}}{m!} \left( x\frac{d}{dx} \right)^{rm} 
		\\
		&=
		e^{(\tau-1) (\hslash x \frac{d}{dx})^r} \hslash \frac{d}{dx}  - e^{\tau ( \hslash x \frac{d}{dx} )^r}\,,
	\end{split}
	\end{equation}
	where we used $ x^{-1}  \big ( x \frac{d}{dx} - 1) x = x \frac{d}{dx}$. Finally, for any value of $\tau$ we can multiply on the left by an invertible operator to show that $\hat{P} (\hat{x}, \hat{y}; \hslash ) \psi_\infty(x;0) = 0$ if and only if
	\begin{equation}
	\left( \hat{y} - e^{(\hat{x}\hat{y})^r} \right) \psi_\infty(x;0)  = 0,
	\end{equation}
	which completes the proof.
\end{proof}

\begin{remark}
	Note that we started with a quantum curve that depended on $ \tau $ and ended up with a result that had no $ \tau $ dependence. This is because all the quantum curves for different $ \tau $ were related by multiplication on the left by an invertible operator and multiplying on the left by an invertible operator does not change the solution of the corresponding differential equation. It is unclear to the authors whether this holds in general, or is unique to the case considered.
\end{remark}

\section{Topological recursion for Atlantes Hurwitz numbers}
\label{TRAtlantes}

The astute reader may recognise the quantum curve \eqref{e:OurCurve}: it appeared in the work of \cite{ALS16}, where it is proved that it annihilates the wave function for Atlantes Hurwitz numbers. In fact, to quote \cite{ALS16}: ``We have an example where the dequantization of the quantum curve doesn't give a spectral curve suitable for the corresponding topological recursion.'' They conclude that ``the dequantization of $\hat{y} - e^{\hat{x}^r \hat{y}^r}$ cannot be the spectral curve for the atlantes Hurwitz numbers, suitable for the construction of the topological recursion.'' 

What do they mean by that? By ``dequantization'' of $\hat{y} - e^{\hat{x}^r \hat{y}^r}$, they mean the relation $P(x,y) = y - e^{x^r y^r}=0$. They then assume that this relation is associated with the spectral curve of \cref{ex:rspin}, namely the non-compact meromorphic spectral curve
\begin{equation}
	\mathcal{S} = \left(\Sigma=\mathbb{C}, \quad x(z)=z e^{-z^r}, \quad y(z)=e^{z^r},\quad  B=\frac{dz_1 dz_2}{(z_1-z_2)^2} \right).
\end{equation}
As it was already conjectured, with substantial evidence, in \cite{SSZ15} (later proved in \cite{DKPS23}) that topological recursion on this spectral curve produces correlators $\omega_{g,n}$ that are generating functions for $r$-completed cycles Hurwitz numbers, which are \emph{not} Atlantes Hurwitz numbers, they conclude that Atlantes Hurwitz numbers provide an example of enumerative invariants satisfying a quantum curve relation that does not arise from topological recursion.

However, this is not the end of the story. The key realisation of the present paper is that there are in fact two different spectral curves with an enumerative interpretation that share the same relation $P(x,y) = y - e^{x^r y^r}=0$:
\begin{enumerate}
	\item The non-compact meromorphic spectral curve $\mathcal{S} = \left( \mathbb{C}, x(z)=z e^{-z^r}, y(z)=e^{z^r},  B=\frac{dz_1 dz_2}{(z_1-z_2)^2} \right) $;
	\item The compact transalgebraic spectral curve $\mathcal{S}_\infty = \left( \P^1, x(z)=z e^{-z^r}, y(z)=e^{z^r},  B=\frac{dz_1 dz_2}{(z_1-z_2)^2} \right) $.
\end{enumerate}
While both spectral curves share the same functions $x$, $y$, and bidifferential $B$, the Riemann surface over which $x$ and $y$ are defined is different. In $\mathcal{S}$,  the exponential singularity of $x$ at infinity is not included, while it is included in $\mathcal{S}_\infty$. According to our proposal, topological recursion on $\mathcal{S}$ may produce different correlators than topological recursion on $\mathcal{S}_\infty$: indeed, this is exactly what happens for $r \geq 2$ (the correlators happen to be the same for $r=1$ by \cref{c:regular}).

Where does that leave us? On the one hand, we know from \cite{DKPS23} that topological recursion on $\mathcal{S}$ produces generating functions for $r$-completed cycles Hurwitz numbers. Moreover, a quantum curve for $r$-completed cycles Hurwitz numbers has  been obtained in \cite{MSS13}, from the geometry of Hurwitz numbers (not directly from topological recursion). In our notation, their result is:
\begin{equation}	\label{e:TheirCurve}
	\hat{P}(\hat{x},\hat{y};\hslash) = \hat{y} - \hat{x}^{1/2}{\rm e}^{\frac{1}{r+1} \sum_{i=0}^{r} \hat{x}^{-1}(\hat{x}\hat{y})^i\hat{x}(\hat{x}\hat{y})^{r-i}}\hat{x}^{-1/2}.
\end{equation}
While this is a quantisation of the relation $P(x,y) = y - e^{x^r y^r}$, it is clearly not the same as the one that we obtained above in \eqref{e:OurCurve}, as it corresponds to a different choice of ordering of the non-commuting operators $\hat{x}$ and $\hat{y}$.

On the other hand, in the present paper, we defined correlators $\omega_{g,n}^\infty$ for the spectral curve $\mathcal{S}_\infty$ that includes the exponential singularity of $x$. We showed in \cref{p:qc} that the wave function constructed from these correlators is annihilated by the quantum curve
\begin{equation}
	\hat{P}_\infty(\hat{x},\hat{y}; \hslash) = \hat{y} - e^{(\hat{x} \hat{y})^r},
\end{equation}
which happens to be the same as the quantum curve for Atlantes Hurwitz numbers. It is then natural to guess that \emph{the correlators $\omega_{g,n}^\infty$ constructed from topological recursion on the transalgebraic curve $\mathcal{S}_\infty$ are generating functions for Atlantes Hurwitz numbers}. This is what we prove in this section, therefore showing that Atlantes Hurwitz numbers \emph{do} fit within the framework of topological recursion, but only if one considers topological recursion on transalgebraic spectral curves.

We can summarize these relations in \cref{t:tr}.
\begin{table}[!ht]
\caption{Topological recursion for $r$-completed cycles and Atlantes Hurwitz numbers}
\label{t:tr}
\begin{tabular}{|l|l|l|}
\hline
Spectral curve & yields generating functions for & Quantum curve \\
\hline
$\left(\mathbb{C}, z e^{-z^r}, e^{z^r},  \frac{dz_1 dz_2}{(z_1-z_2)^2} \right)$ & $r$-completed cycles Hurwitz & $ \hat{y} - \hat{x}^{1/2}{\rm e}^{\frac{1}{r+1} \sum_{i=0}^{r} \hat{x}^{-1}(\hat{x}\hat{y})^i\hat{x}(\hat{x}\hat{y})^{r-i}}\hat{x}^{-1/2}$ \\
$\left(\P^1, z e^{-z^r}, e^{z^r},  \frac{dz_1 dz_2}{(z_1-z_2)^2} \right)$ & r-Atlantes Hurwitz &  $\hat{y} - e^{(\hat{x} \hat{y})^r}$\\
\hline
\end{tabular}
\end{table}

To prove that the correlators $\omega_{g,n}^\infty$ constructed by topological recursion on $\mathcal{S}_\infty$ compute Atlantes Hurwitz numbers, we first need to define what Atlantes Hurwitz numbers are. Let us now review some of the key results relating Hurwitz numbers and topological recursion.

\subsection{Hurwitz numbers}

Hurwitz numbers are counts of covers of a given Riemann surface with a given ramification behaviour, up to equivalence and weighed by automorphisms. We will always take the target curve to be $\P^1$. Via the monodromy representation, they can be interpreted as counting decompositions of the identity in the symmetric group algebra; this is the point of view that we will take in this paper.

\begin{definition}
	Let $ d \in \N$ and $ C_1, \dotsc, C_k \in Z\C [\mf{S}_d]$. The associated \emph{disconnected Hurwitz number} is
	\begin{equation}
		H^\bullet (C_1, \dotsc, C_k) = \frac{1}{d!} [1] \prod_{j=1}^k C_j\,.
	\end{equation}
	Here $ [1]$ is the dual to the unit of $ \C[\mf{S}_d]$ in the natural basis, i.e. it extracts the coefficient of $ 1$.
\end{definition}

In most modern studies of Hurwitz numbers, one or two of the central elements are chosen as free parameters --- usually indexed by partitions, as $ Z\C[\mf{S}_d]$ has a basis given by sums of conjugacy classes, i.e. cycle types, which are naturally indexed by partitions of $d$. All the other central elements are then chosen to be equal, and this `generic' element determines the type of Hurwitz problem. These kinds of Hurwitz problems are related to the Kadomtsev-Petviashvili (KP) and 2D Toda lattice hierarchies: they can be assembled into generating functions which are \emph{hypergeometric tau-functions} or \emph{Orlov-Scherbin partition function} of these hierarchies, \cite{KMMM95,OS01a,OS01b}. For more on this relation, see \cite{HO15,ALS16}. Moreover, in many cases (i.e. for many generic elements) they satisfy topological recursion, which was first conjectured for simple Hurwitz numbers (generic partition $ (2, 1, \dotsc, 1)$) in \cite{BM08} and proved in \cite{EMS11}, and has since been proved in many individual cases, see e.g. \cite{DLN16,DKPS23,KPS22,ACEH20}.\par
The most general and direct relation between these two points of view is given by the following theorem.

\begin{theorem}[\cite{BDKS22,BDKS20}]\label{BDKSTR}
	Consider two formal power series
	\begin{equation}\label{psiandy}
		\hat{\psi} (\hslash^2,y) \coloneqq \sum_{k=1}^\infty \sum_{m=0}^\infty c_{k,m} y^k \hslash^{2m}\,, \qquad \hat{y}(\hslash^2,z) \coloneqq  \sum_{k=1}^\infty \hat{y}_k(\hslash^2) z^k \coloneqq \sum_{k=1}^\infty \sum_{m=0}^\infty s_{k,m}z^k\hslash^{2m}  \,,
	\end{equation}
	and their associated hypergeometric KP tau-function
	\begin{equation}\label{DefHypGeo}
		Z (\underline{p}) = e^{F(\underline{p})} = \sum_{\nu \in \mc{P}} \exp \Big( \sum_{\square \in \nu} \hat{\psi} (\hslash^2,-\hslash c_\square )\Big) s_\nu (\underline{p} ) s_\nu \big( \big\{ \frac{\hat{y}_k(\hslash^2)}{\hslash} \big\} \big)\,.
	\end{equation}
	Define
	\begin{equation}\label{XfromKP}
	\begin{alignedat}{3}
		\psi (y) 
		&
		\coloneqq \hat{\psi}(0,y)\,,  \qquad
		&
		y(z) 
		&
		\coloneqq \hat{y}(0,z)\,, \qquad
		&
		x(z) 
		&
		\coloneqq \log z - \psi(y(z))\,, 
		\\
		X(z) 
		&
		\coloneqq e^{x(z)}\,,
		&
		D 
		&
		\coloneqq \frac{\del}{\del x} \,, 
		&
		Q 
		&
		\coloneqq z \frac{d x}{d z}
	\end{alignedat}
	\end{equation}
	and write
	\begin{equation}\label{HnDef}
		H_n \coloneqq \sum_{k_1, \dotsc, k_n = 1}^\infty \frac{\del^n F}{\del p_{k_1} \dotsb \del p_{k_n}} \bigg|_{p=0} X_1^{k_1} \dotsb X_n^{k_n}\,.
	\end{equation}
	Then these can be decomposed as
	\begin{equation}\label{HnGenusDecomp}
		H_n = \sum_{g=0}^\infty \hslash^{2g-2+n} H_{g,n}\,,
	\end{equation}
	with $ H_{g,n} $ independent of $ \hslash$, and
	\begin{equation}\label{OSUnstable}
		DH_{0,1}(X(z)) = y(z)\,, \qquad H_{0,2}(X(z_1),X(z_2)) = \log \Big( \frac{z_1^{-1} - z_2^{-1}}{X_1^{-1} - X_2^{-1}} \Big)\,.
	\end{equation}
	If moreover $\frac{d\psi (y)}{dy} \big|_{y = y(z)}$ and $ \frac{d y(z)}{dz}$ have analytic continuations to meromorphic functions in $ z$ and all coefficients of positive powers of $\hslash^2$ in $ \hat{\psi}(\hslash^2,y(z))$ and $ \hat{y}(\hslash^2,z)$ are rational functions of $z$ whose singular points are disjoint from the zeroes of $ dx$, then the $n$-point differentials
	\begin{equation}\label{eq:hurcor}
		\omega_{g,n} \coloneqq d_1 \dotsb d_n H_{g,n} + \delta_{g,0}\delta_{n,2} \frac{dX_1 \, dX_2}{(X_1-X_2)^2}
	\end{equation}
	can be extended analytically to $ (\P^1)^n$ as global rational forms, and the collection of $n$-point differentials satisfies meromorphicity and the linear and quadratic loop equations, i.e. blobbed topological recursion~\cite{BS17}, for the spectral curve $ (\Sigma, X(z), \frac{y(z)}{X(z)}, \frac{dz_1\, dz_2}{(z_1-z_2)^2})$, where $\Sigma$ is $\P^1$ minus the exponential singularities of $X(z)$.
	
	Finally, if $ \hat{\psi} $ and $ \hat{y}$ belong to one of the two families
	\begin{align*}
		&
		\text{Family I}
		&
		\hat{\psi}(\hslash^2,y)
		&
		= \mc{S}(\hslash\del_y ) P_1(y) + \log \Big(\frac{P_2(y)}{P_3(y)}\Big)\,;
		&
		\hat{y}(\hslash^2, z) 
		&
		= \frac{R_1(z)}{R_2(z)}\,,
		\\
		&
		\text{Family II}
		&
		\hat{\psi}(\hslash^2,y)
		&
		= \alpha y\,;
		&
		\hat{y}(\hslash^2, z) 
		&
		= \frac{R_1(z)}{R_2(z)} +\mc{S} (\hslash z \del_z)^{-1} \log \Big(\frac{R_3(z)}{R_4(z)}\Big)\,,
	\end{align*}
	where $ \alpha \in \C^\times$ and the $ P_i$ and $R_j$ are arbitrary polynomials such that $ \psi(y)$ and $ y(z)$ are non-zero, but vanishing at zero, and no singular points of $y$ are mapped to branch points by $x$, then the $n$-point differentials also satisfy the projection property, and hence topological recursion, for the spectral curve above.
\end{theorem}

\Cref{BDKSTR} does not explicitly mention Hurwitz numbers, but these are the coefficients of the power series $H_{g,n}$ of \cref{HnGenusDecomp}. The function $ \hat{\psi}$ encodes the generic ramification profile, the function $ \hat{y}$ a specified (fixed) ramification profile, and the exponents $ k_i$ are make up the final, freely chosen, ramification profile.

\subsubsection{\texorpdfstring{$r$}{r}-completed cycles Hurwitz numbers}

A special case of \cref{BDKSTR} is given by the $r$-completed cycles Hurwitz numbers. It corresponds to the choice
\begin{equation}
\hat{\psi}(\hslash^2,y) = \mathcal{S}(\hslash \partial_y) y^r, \qquad \hat{y}(\hslash^2,z) = z
\end{equation}
in Family I. From the theorem, we see that the $\omega_{g,n}$ defined in \eqref{eq:hurcor} satisfy topological recursion for the meromorphic spectral curve $\mathcal{S} =\left (\mathbb{C}, z e^{-z^r}, e^{z^r}, \frac{dz_1 dz_2}{(z_1-z_2)^2} \right)$, as stated earlier in this section. This is precisely the spectral curve of \cref{ex:rspin}. 

Moreover, as stated earlier a quantum curve for $r$-completed cycles Hurwitz numbers has been obtained in \cite{MSS13} from the geometry of Hurwitz numbers, see \cref{e:TheirCurve}. It is a quantisation of the relation $P(x,y) = y - e^{x^r y^r}$.

\subsubsection{Atlantes Hurwitz numbers}

Another type of Hurwitz numbers that will play a key role in the following is Atlantes Hurwitz numbers. In the context of \cref{BDKSTR}, Atlantes Hurwitz numbers correspond to the case:
\begin{equation}
	\hat{\psi} (\hslash^2, y) = y^r \,, \qquad \hat{y}(\hslash^2, z) = z\,.
\end{equation}

If $ r > 1$, this does not fit in one of the two families, so the projection property for Atlantes Hurwitz numbers does not follow from \cref{BDKSTR}, but the meromorphicity property and the linear and quadratic loop equations do.

But what are Atlantes Hurwitz numbers? The notion of Atlantes Hurwitz numbers was introduced in \cite{ALS16}, to encode the value of power-sum symmetric functions evaluated at the Jucy-Murphy elements.

\begin{definition}
	Let $d\geq 1$ and let $ \mf{S}_d$ be the symmetric group on $d$ elements. The \emph{Jucys-Murphy elements} are defined as
	\begin{equation}
		\mc{J}_k = \sum_{j=1}^{k-1} (j \, k) \in \C [\mf{S}_d]\,.
	\end{equation}
	They generate a maximally commutative subalgebra of $ \C [\mf{S}_d]$, called the \emph{Gelfand-Tsetlin algebra}.
\end{definition}

\begin{proposition}[Jucys correspondence~\cite{Juc74}]
	Let $ \sigma_b$ be the elementary symmetric function. Then
	\begin{equation}\label{SymmJMelements}
		\sigma_b (\mc{J}_2, \dotsc, \mc{J}_d) = \sum_{\substack{\alpha \vdash n \\ \ell (\alpha ) = d -b}} C_\alpha\,.
	\end{equation}
	Hence, any symmetric function evaluated at the Jucys-Murphy elements gives a central element in $ \C [\mf{S}_d]$.
\end{proposition}

\begin{proposition}[{\cite{FH59}}]
	The collection of elements given in \cref{SymmJMelements} generate $Z\C[\mf{S}_d]$.
\end{proposition}

\begin{definition}[{\cite{ALS16}}]
	An $r$-block of Atlantes is $ B^\times_r \coloneqq p_r (\mc{J}_2, \dotsc, \mc{J}_d) \in Z\C[\mf{S}_d]$.
\end{definition}

The name Atlantes comes from the following lemma:

\begin{lemma}[{\cite[Lemma~4.3]{ALS16}}]
	The geometric interpretation of the block of Atlantes is the following: we have $r$ simple ramifications, whose monodromies are given by the transpositions $(x_i \, y)$, $ x_i < y$, $i = 1, \dotsc , r$. Here $y$ is an arbitrary number from $2$ to $d$, which is not fixed in advance, but is the same for all transpositions.
\end{lemma}

\begin{figure}
\centering
\begin{tikzpicture}
	\draw (0,2) -- (1,2);
	\draw (0,1.5) -- (1,1.5);
	\draw (0,1) -- (1,1);
	\draw (0,.5) -- (1,.5);
	\draw (0,0) -- (1,0);
	
	\draw (1,2) -- (2,1.5);
	\draw (1,1.5) -- (2,2);
	\draw (1,1) -- (2,1);
	\draw (1,.5) -- (2,.5);
	\draw (1,0) -- (2,0);
	
	\draw (2,2) -- (3,2);
	\draw (2,1.5) -- (3,1.5);
	\draw (2,1) -- (3,1);
	\draw (2,.5) -- (3,.5);
	\draw (2,0) -- (3,0);
	
	\draw (3,2) -- (4,0);
	\draw (3,1.5) -- (3.1,1.5); \draw (3.4,1.5) -- (3.6,1.5); \draw (3.9,1.5) -- (4,1.5);
	\draw (3,1) -- (3.35,1); \draw (3.65,1) -- (4,1);
	\draw (3,.5) -- (3.1,.5); \draw (3.4,.5) -- (3.6,.5); \draw (3.9,.5) -- (4,.5);
	\draw (3,0) -- (4,2);
	
	\draw (4,2) -- (5,2);
	\draw (4,1.5) -- (5,1.5);
	\draw (4,1) -- (5,1);
	\draw (4,.5) -- (5,.5);
	\draw (4,0) -- (5,0);
	
	\draw (5,2) -- (6,1);
	\draw (5,1.5) -- (5.35,1.5); \draw (5.65,1.5) -- (6,1.5);
	\draw (5,1) -- (6,2);
	\draw (5,.5) -- (6,.5);
	\draw (5,0) -- (6,0);
	
	\draw (6,2) -- (7,2);
	\draw (6,1.5) -- (7,1.5);
	\draw (6,1) -- (7,1);
	\draw (6,.5) -- (7,.5);
	\draw (6,0) -- (7,0);
\end{tikzpicture}
\caption{A $3$-block of Atlantes with degree $d=5$.}
\label{f:Atlas}
\end{figure}

Graphically, we often draw a cover as a couple of parallel horizontal lines (sheets) mapped to one horizontal line. Simple ramifications are drawn as crosses connecting two sheets. In a block of Atlantes, we interpret the sheet $y$ as the sky, the sheets $ x_i$ as part of the earth, and the transposition crosses $ (x_i\, y)$ as Atlas holding the sky. See \cref{f:Atlas}.

\begin{definition}[{\cite{ALS16}}]
	Let $ r \geq 1$. We define the disconnected \emph{ $r$-Atlantes single Hurwitz numbers} as
	\begin{equation}
		h^{\bullet,\times r}_{g,\mu} \coloneqq \frac{1}{d!} [1]C_\mu (B^\times_r)^b\,,
	\end{equation}
	where $ \mu \vdash d$, $ g \in \Z$, and
	\begin{equation}
		b =\frac{2g-2 + d + \ell (\mu)}{r}
	\end{equation}
	is determined by the Riemann-Hurwitz formula.
\end{definition}

We can define a wave function for Atlantes Hurwitz numbers, and show that it satisfies a differential equation.

\begin{proposition}[{\cite[Proposition~7.4]{ALS16}}]\label{p:qcat}
	Let 
	\begin{equation}
		Z^{\times r} (\underline{p}, \hslash) = \exp \Big( \sum_{g,\mu} \frac{\hslash^{2g-2+\ell (\mu) + |\mu |}}{(2g-2 + \ell (\mu ) + |\mu |)!} h^{\times r}_{g,\mu} p_\mu \Big)
	\end{equation}
	be the generating function, and $ \Psi^{\times r} (x, \hslash) = Z^{\times r} (\{ p_k = (\hslash^{-1}x)^k \}, \hslash ) $ be the wave function. Then
	\begin{equation}
		\Psi^{\times r}(x,\hslash) = \sum_{n=0}^\infty \frac{x^n}{n! \hslash^n} \exp \Big( \hslash^r \sum_{j=1}^{n-1} j^r \Big)
	\end{equation}
	satisfies the differential equation
	\begin{equation}
		\big(\hat{y} - e^{(\hat{x} \hat{y})^r} \big) \Psi^{\times r}(x,\hslash ) = 0\,,
	\end{equation}
	where $ \hat{x} = x \cdot $ and $ \hat{y} = \hslash  \frac{d}{d x}$.
\end{proposition}
We of course recognise the quantum curve that we obtained in \cref{p:qc}, that is, \eqref{e:OurCurve}.

\subsection{The projection property for Atlantes Hurwitz numbers}

As Atlantes Hurwitz numbers do not fall within Families I and II of \cref{BDKSTR}, while a different $ \hslash$ deformation of its $ (\psi, y)$ does, the correlators $\omega_{g,n}$ do not satisfy the usual topological recursion on the non-compact meromorphic spectral curve $\mathcal{S} = \left (\mathbb{C}, z e^{-z^r}, e^{z^r}, \frac{dz_1 dz_2}{(z_1-z_2)^2} \right)$. Indeed, topological recursion on $\mathcal{S}$ produces generating functions for $r$-completed cycles Hurwitz numbers, not Atlantes Hurwitz numbers, as stated above. The reason is that the projection property does not follow from \cref{BDKSTR} for Atlantes Hurwitz numbers. However, the meromorphicity property and the linear and quadratic loop equations do.

In this section, we analyse to what extent the projection property holds for Atlantes Hurwitz numbers. This will be needed to establish that topological recursion on the transalgebraic spectral curve $\mathcal{S}_\infty =  \left(\P^1, z e^{-z^r}, e^{z^r}, \frac{dz_1 dz_2}{(z_1-z_2)^2} \right)$ produces generating functions for Atlantes Hurwitz numbers. We do this by analysing the proof of the projection property for Family I, as given in \cite[Sections 3 \& 4]{BDKS20}.

\begin{definition}[{\cite[Definition 3.7]{BDKS20}}]
	Fix a spectral curve $ ( \Sigma, x, y, B)$. The space $\Theta_n$ (or $\Theta$ if $n$ is clear from context) is defined as the linear span of functions of the form $\prod_{i=1} f_i(z_i)$, where each $ f_i(z_i)$ 
	\begin{itemize}
		\item is a rational function on the Riemann sphere;
		\item has poles only at the ramification points $ p_1, \dotsc, p_N$ of $ x$;
		\item its principal part at these ramification points is odd with respect to the corresponding deck transformation.	\end{itemize}
\end{definition}

\begin{proposition}[{\cite[Proposition 3.9]{BDKS20}}]
	The differentials $ \omega_{g,n}$ satisfy the projection property and the linear loop equations if and only if $H_{g,n} \in \Theta_n$ for $2g - 2 + n > 0$.
\end{proposition}

\begin{proposition}\label{NoPolesForn>1}
	For $ \hat{\psi}(\hslash^2,y) = P_1(y) + \log \frac{P_2(y)}{P_3(y)}$ and $ \hat{y} (\hslash^2,z) = \frac{R_1(z)}{R_2(z)}$, where the $ P_i$ and $ R_j$ are polynomials with simple zeroes,\footnote{This holds for more general polynomials $P_i$ and $R_j$ using the results of \cite{BBCKS23}.} the functions $ H_{g,n}$ for $ 2g-2+n >0$ and $ n \geq 2$ lie in $\Theta_n$, while the $ H_{g,1}$ are rational functions on the Riemann sphere, with poles at $ p_1, \dotsc, p_N, \infty$, and principal parts at the $ p_i$ odd with respect to the deck transformation.
\end{proposition}


The proof of this proposition is a fairly straightforward extension of the methods of \cite{BDKS20}. As it contains no new ideas, we give it in \cref{ProofOfAtlantesBDKS}.

\begin{corollary} \label{PoleHg1}
	In the situation of \cref{NoPolesForn>1}, the pole at infinity of $ H_{g,1}$ is given by
	\begin{equation}
		[\hslash^{2g}] \sum_{k=0}^{g-2} \frac{D_1y(z_1)}{(k+2)!}D_1^k \Big( \big( \frac{ 1}{\mc{S} (\hslash \del y)} -1\big) \psi (y) \Big)^{k+2} \Big|_{y = y(z_1)}
		+ \frac{(2^{1-2g}-1)B_{2g}}{(2g)!} \del_y^{2g-1} \psi (y) \Big|_{y = y(z_1)} \,.
	\end{equation}
	The pole order of both terms is $ (\deg (P_1)+1 - 2g) (\deg (R_1)-\deg (R_2 ))$, where $ \deg $ is the polynomial degree.
\end{corollary}

We also give the proof of this corollary in \cref{ProofOfAtlantesBDKS}.

In particular, for $r$-Atlantes, the pole of $ H_{1,1}$ is
\begin{equation}
	\frac{(2^{1-2}-1)B_2}{2!} \del_y y^r \Big|_{y = z_1} = - \frac{r}{24} z_1^{r-1}\,,
\end{equation}
which agrees with our calculation in \cref{ex:qOrbifoldrAtlantes} and the result of \cref{p:11}.
The pole of $ H_{2,1}$ is
\begin{equation}
	- \frac{r(r-1)^2}{24} z^{r-3} + \frac{7r(r-1)(r-2)}{5760} z^{r-3} =   \frac{r(r-1)(226-233r)}{5760} z^{r-3}\,.
\end{equation}


\subsection{Topological recursion for Atlantes Hurwitz numbers}

To summarize the main conclusions of the previous section, namely \cref{NoPolesForn>1}, let us introduce differentials for Atlantes Hurwitz numbers as in \cref{BDKSTR}:
\begin{equation}\label{eq:diffa}
\omega_{g,n}^\infty = d_1 \cdots d_n H_{g,n} + \delta_{g,0} \delta_{n,2} \frac{d X_1 d X_2}{(X_1 - X_2)^2}.
\end{equation}
Then \cref{BDKSTR} and \cref{NoPolesForn>1} state that:
\begin{itemize}
\item The $\omega_{g,n}^\infty$ are meromorphic differentials on $\P^1$.
\item The $\omega_{g,n}^\infty$ satisfy the linear and quadratic loop equations for $\mathcal{S}_\infty = (\P^1, z e^{-z^r}, e^{z^r}, \frac{dz_1 dz_2}{(z_1-z_2)^2})$.
\item The $\omega_{g,n}^\infty$ for $2g-2+n > 0$ and $n \geq 2$ satisfy the projection property, and hence have poles only at the finite ramification points of the function $x = z e^{-z^r}$.
\item The $\omega_{g,1}^\infty$ may have poles at both the finite ramification points and the essential singularity of $x$ at infinity.
\end{itemize}
We also know from \cref{p:qcat} that the wave function $\psi_\infty$ constructed from the $\omega_{g,n}^\infty$ satisfies the differential equation
\begin{equation}\label{eq:qqcc}
( \hat{y} - e^{(\hat{x} \hat{y})^r} ) \psi_\infty = 0.
\end{equation}

With this information we can prove that the $\omega_{g,n}^\infty$ are the correlators constructed from topological recursion (\cref{d:main}) on the transalgebraic spectral curve $\mathcal{S}_\infty = (\P^1, z e^{-z^r}, e^{z^r}, \frac{dz_1 dz_2}{(z_1-z_2)^2})$.

\begin{theorem}\label{t:TRatlantes}
Let $\omega_{g,n}^\infty$ be the differentials for Atlantes Hurwitz numbers as in \eqref{eq:diffa}. Then they satisfy topological recursion (\cref{d:main}) on the transalgebraic spectral curve $\mathcal{S}_\infty = (\P^1, z e^{-z^r}, e^{z^r}, \frac{dz_1 dz_2}{(z_1-z_2)^2})$.
\end{theorem}

\begin{proof}
	On the one hand, let $\omega_{g,n}^\infty$ be the differentials defined by \eqref{eq:diffa}. For $n \geq 2$, since they satisfy the projection property, we know that they satisfy the usual topological recursion formula with residues only at the finite ramification points of $x$. However, the $\omega_{g,1}^\infty$ do not, since they do not satisfy the projection property.

	On the other hand, let $\tilde{\omega}_{g,n}^\infty$ be the correlators defined by topological recursion on the transalgebraic spectral curve $\mathcal{S}_\infty$. By \cref{TransAlgTRViaAlgTR}, we know that the correlators $\tilde{\omega}_{g,n}^\infty$ for $n \geq 2$ satisfy the usual topological recursion formula with residues only at the finite ramification points of $x$, just like the $\omega_{g,n}^\infty$ for $n \geq 2$. However, the $\tilde{\omega}_{g,1}^\infty$ do not, as there are contributions from the essential singularity of $x$ at infinity.

	To establish that $\omega_{g,n}^\infty = \tilde{\omega}_{g,n}^\infty$ for all $g$ and $n$, we proceed by induction, using the fact that the wave function $\psi_\infty$ constructed from the $\omega_{g,n}^\infty$ and the wave function $\tilde{\psi}_\infty$ constructed from the $\tilde{\omega}_{g,n}^\infty$ satisfy the same quantum curve equation  \eqref{eq:qqcc}. The base case is obvious, since $\omega_{0,1}^\infty = \tilde{\omega}_{0,1}^\infty$ and $\omega_{0,2}^\infty = \tilde{\omega}_{0,2}^\infty$. So we proceed with the induction step. Assume that $\omega_{g',n'}^\infty = \tilde{\omega}_{g',n'}^\infty$ for all  $g',n'$ such that $2g'-2+n' < k$. Then we show that $\omega_{g,n}^\infty = \tilde{\omega}_{g,n}^\infty$ for all $g,n$ such that $2g-2+n = k$. For the correlators with $n \geq 2$, this is clear, since they are constructed from the same topological recursion formula for the same lower order correlators. As for the correlators $\omega_{g,1}^\infty$ and $\tilde{\omega}_{g,1}^\infty$, we conclude that they must be equal from the fact that $\psi_\infty$ and $\tilde{\psi}_\infty$ satisfy the same quantum curve equation. Indeed, the quantum curve equation must be satisfied order by order in $\hslash$, and up to a given order $O(\hslash^k)$, only correlators $\omega_{g',n'}^\infty$ with $2g'-2+n' \leq k$ contribute to the equation. Thus, up to $O(\hslash^k)$, we know that all correlators contributing to the equation are the same, except potentially for $\omega_{g,1}^\infty$ and $\tilde{\omega}_{g,1}^\infty$. But as the quantum curve equation is the same for both wave functions, we conclude that $\omega_{g,1}^\infty = \tilde{\omega}_{g,1}^\infty$.
\end{proof}


\subsection{Relation between the Atlantes and \texorpdfstring{$r$}{r}-completed cycles Hurwitz quantum curves}

Recall \cref{t:tr}, which summarizes the relations between spectral curves, quantum curves, and topological recursion for $r$-completed cycles and Atlantes Hurwitz numbers. The spectral curves are very similar, differing only by the choice of Riemann surface. As the functions $x$ and $y$ are the same, they satisfy the same relation $P(x,y) = y - e^{x^r y^r} = 0$. Indeed, the two quantum curves are both quantisations of this relation. In this section we study in more details the relation between these two quantum curves.

Let
\begin{equation}
\hat{P}_\infty(\hat{x},\hat{y}; \hslash) = \hat{y} - e^{(\hat{x} \hat{y})^r}
\label{e:OurCurve2}
\end{equation}
be the quantum curve for Atlantes Hurwitz numbers with wave function $\psi_\infty$, and 
\begin{equation}
	\hat{P}(\hat{x},\hat{y};\hslash) = \hat{y} - \hat{x}^{1/2}{\rm e}^{\frac{1}{r+1} \sum_{i=0}^{r} \hat{x}^{-1}(\hat{x}\hat{y})^i\hat{x}(\hat{x}\hat{y})^{r-i}}\hat{x}^{-1/2}
	\label{e:TheirCurve2}
\end{equation}
be the quantum curve for $r$-completed cycles Hurwitz numbers, with wave-function $\psi$. Clearly, in general, this is not the same quantum curve. However, we can observe an interesting relation between the two results that was first noticed in a more limited form in \cite{C16}. Let $\hat{Y} = \hat{x} \hat{y} $, and assume there exists an operator
\begin{equation}
	\hat{H} = \exp\left( \sum_{n=1}^{r} \hslash^{r-n} h_n \hat{Y}^n \right),
\end{equation}
such that $ \hat{H}\psi_\infty = \psi$.  Then we find, using that $ \hat{x}^{-1} \hat{Y} \hat{x} = \hat{Y} + \hslash$, so $ \hat{x}^{-1} \hat{H} \hat{x}$ commutes with $ \hat{Y}$,
\begin{equation}
	\begin{split}
		0 
		&=
		\hat{H}0 = \hat{H} \big( \hat{Y}- \hat{x} e^{\hat{Y}^r} \big) \psi_\infty
		\\
		&=
		\big( \hat{Y} - \hat{H} \hat{x}  e^{\hat{Y}^r} \hat{H}^{-1} \big) \hat{H} \psi_\infty
		\\
		&=
		\big( \hat{Y} - \hat{x} \hat{x}^{-1} \hat{H} \hat{x} e^{\hat{Y}^r} \hat{H}^{-1} \big) \psi
		\\
		& = \big( \hat{Y} - \hat{x} e^{\hat{Y}^r} \hat{x}^{-1}\hat{H}\hat{x}\hat{H}^{-1} \big) \psi.
	\end{split}
\end{equation}
Ergo, our operator $ \hat{H} $ is the solution to
\begin{equation}
	\begin{split}
		 \hat{x}e^{\hat{Y}^r} \hat{x}^{-1}\hat{H}\hat{x}\hat{H}^{-1} 
		 &=
		 \hat{x}^{3/2} e^{\frac{1}{r+1} \sum_{i=0}^{r} \hat{x}^{-1}\hat{Y}^i\hat{x} \hat{Y}^{r-i}}\hat{x}^{-1/2}
		 \\
		\hat{x}^{-1/2} e^{\hat{Y}^r} e^{ \sum_{n=1}^r \hslash^{r-n} h_n (\hat{Y} + \hslash )^n} e^{ -\sum_{n=1}^r \hslash^{r-n} h_n \hat{Y}^n} \hat{x}^{1/2}
		&=
		e^{\frac{1}{r+1} \sum_{i=0}^{r} (\hat{Y} + \hslash)^i \hat{Y}^{r-i}}
		\\
		e^{(\hat{Y} + \frac{\hslash}{2})^r} e^{ \sum_{n=1}^r \hslash^{r-n} h_n \big( (\hat{Y} + \frac{3\hslash}{2} )^n - (\hat{Y} + \frac{\hslash}{2})^n \big)}
		&=
		e^{\frac{1}{r+1} \sum_{i=0}^{r} (\hat{Y} + \hslash)^i \hat{Y}^{r-i}}
		\\
		\big(\hat{Y} + \frac{\hslash}{2} \big)^r + \sum_{n=1}^r \hslash^{r-n} h_n \Big( \big(\hat{Y} + \frac{3\hslash}{2} \big)^n - \big( \hat{Y} + \frac{\hslash}{2} \big)^n \Big)
		&=
		\frac{1}{r+1} \sum_{i=0}^{r} (\hat{Y} + \hslash)^i \hat{Y}^{r-i}
		\\
		\sum_{n=1}^r h_n \sum_{j=0}^n \binom{n}{j} \Big( \big( \frac{3}{2} \big)^{n-j} - \big( \frac{1}{2} \big)^{n-j} \Big) \hslash^{r-j} \hat{Y}^j
		&=
		\frac{1}{r+1} \sum_{i=0}^{r} \sum_{k=0}^i \binom{i}{k} \hslash^{i-k} \hat{Y}^{k+r-i} - \sum_{j=0}^r \binom{r}{j} \big( \frac{\hslash}{2} \big)^{r-j} \hat{Y}^j
		\\
		\sum_{j=0}^r \sum_{n=j}^r h_n \binom{n}{j} \Big( \big( \frac{3}{2} \big)^{n-j} - \big( \frac{1}{2} \big)^{n-j} \Big) \hslash^{r-j} \hat{Y}^j
		&=
		\sum_{j=0}^r \bigg( \frac{1}{r+1} \sum_{i=r-j}^{r} \binom{i}{r-j} - \binom{r}{j} \big( \frac{1}{2} \big)^{r-j} \bigg) \hslash^{r-j} \hat{Y}^j
		\\
		\sum_{n=j}^r h_n \binom{n}{j} \Big( \big( \frac{3}{2} \big)^{n-j} - \big( \frac{1}{2} \big)^{n-j} \Big)
		&=
		\frac{1}{r+1} \binom{r+1}{j} - \binom{r}{j} \big( \frac{1}{2} \big)^{r-j} \,,
		\quad 0 \leq j \leq r\,.
	\end{split}
\end{equation}
This can be solved recursively from $ j = r$, and the first few numbers are
\begin{align}
	h_r &= 0\,,
	\\
	h_{r-1} &= \frac{1}{24} r = \frac{1}{24} \binom{r}{1} \,,
	\\
	h_{r-2} &= - \frac{1}{16} r(r-1) = - \frac{1}{8} \binom{r}{2} \,.
\end{align}
Such an $ \hat{H} $ therefore exists and is unique (up to an arbitrary multiplicative constant). In particular, $ \hat{H} \neq 1 $ except in the case when $ r = 1 $, where it can be seen that \eqref{e:OurCurve2} and \eqref{e:TheirCurve2} agree.
The degree of the operator is no surprise; a degree $ r-1 $ operator is precisely the degree needed to reduce all contributions from the essential singularity to constants by \cref{p:bound} and the fact that all contributions from the essential singularity vanish for $ r = 1 $ by \cref{c:regular}.

\section{Conclusion and open questions}

\label{s:open}

In this paper, we defined topological recursion for transalgebraic spectral curves via limits of sequences of meromorphic spectral curves. We studied properties of the topological recursion on transalgebraic spectral curves and proved the topological recursion/quantum curve correspondence for a subclass of transalgebraic spectral curves. As a particular example, we used our formalism to show that generating series for Atlantes Hurwitz numbers satisfy topological recursion on a transalgebraic spectral curve. \par
This has only been a first investigation of topological recursion on transalgebraic spectral curves; further research could shed more light on this new extension. In particular, we give a number of open questions below:

\begin{itemize}
	\item For a given transalgebraic spectral curve, the quantum curve that we construct a priori seems to depend on the particular sequence of meromorphic spectral curves considered. In the particular case of Atlantes Hurwitz numbers, we showed that all of the quantum curves constructed this way are equivalent, but we do not know if this property holds in general.
	\item For topological recursion on meromorphic spectral curves, the expansion of the correlators $\omega_{g,n}$ at the ramification points has a nice interpretation in terms of intersection theory of the moduli spaces of curves \cite{Eyn11,DOSS14}. Is there a similar interpretation for the expansion of the correlators associated to a transalgebraic spectral curve at the exponential singularities?
	\item Following up on the previous point, in cases where the correlators $\omega_{g,n}$ constructed from topological recursion on a meromorphic spectral curve have an interpretation in Hurwitz theory, the relation between the expansion of the correlators at punctures of the curve and at ramification points gives rise to `ELSV-type formulae' for Hurwitz numbers. Do similar formulae hold for Atlantes Hurwitz numbers, using topological recursion on transalgebraic spectral curves?
	\item The quantum curve for Atlantes Hurwitz numbers looks simpler than the one for $r$-completed cycles Hurwitz numbers, which has nearly the same spectral curve, but excludes the essential singularity. On the other hand, $r$-completed cycles have natural relations to cohomological field theory, via Chiodo classes~\cite{Chio08,SSZ15,DKPS23}, and to Gromov--Witten theory of curves~\cite{OP06a,OP06}, where the `completion' of the cycles seems related to the boundary of $ \overline{\mc{M}}_{g,n}$. It would be interesting to see if Atlantes Hurwitz numbers admit a similar interpretation.
	\item Using the Airy structure approach pioneered by Kontsevich and Soibelman \cite{KS17,ABCO24}, topological recursion on meromorphic spectral  curves can be reformulated in terms of representation theory of $\mathcal{W}$-algebras \cite{BBCCN18}. It would be interesting to investigate whether topological recursion on transalgebraic spectral curves has a similar reformulation in terms of $\mathcal{W}$-algebras.
\end{itemize}

\section*{Acknowledgements}

We would like to thank Sergey Shadrin for extensive discussions on a previous version of this paper, and the anonymous referees of useful comments.\par
The authors acknowledge support from the National Science and Engineering Research Council of Canada, application ID SAPIN-2018-00029. R.K. acknowledges support from a Postdoctoral Fellowship of the Pacific Institute for the Mathematical Sciences. The research and findings may not reflect those of these institutions. The University of Alberta respectfully acknowledges that we are situated on Treaty 6 territory, traditional lands of First Nations and Métis people.


\appendix
\renewcommand*{\thesection}{\Alph{section}}

\section{Extensions of \texorpdfstring{\cite{BE17}}{BE17} to the transalgebraic case}
	\label{QCForTransAlgTR}

In this appendix we extend the results of \cite{BE17} on quantum curves in a way that make them suitable for the construction of quantum curves for transalgebraic spectral curves. For the sake of maintaining a reasonable level of brevity, this appendix will not be self contained and will frequently reference \cite{BE17}. To avoid confusion, we will refer explicitly to the analogues of what we are doing in \cite{BE17} when possible. However, in \cite{BE17} everything was indexed starting from the degree of the curve which for us may be infinite;\footnote{For us, recall, any curve that does not have a well-defined finite degree is of infinite degree.} therefore, we will have to re-index virtually all objects considered in \cite{BE17}, where infinite degree curves were not a consideration.

Let $ P(x,y)=0 $ be the (trans)algebraic equation corresponding to a compact spectral curve $ \mathcal{S} $, with Newton polygon $ \Delta$ (recall \cref{d:Npoly}), and set the notation

\begin{equation}
	P(x,y) = \sum_{i=0}^{d} q_i (x) y^i = \sum_{(i,j) \in \mathbb{N}^2} \alpha_{i,j} y^i x^j \,,
\end{equation}
where $d$ is the degree of the curve (which may be infinite if the curve is transalgebraic) and the $ q_i$ correspond to reindexed versions of the $ p_m$ in \cite[Remark~2.2]{BE17}. The following definitions will be necessary to explain the construction of quantum curves.
\begin{definition}\label{d:QCpoly}
	For $m=0,\dots,d$ define
	\begin{equation}
		Q_m(x,y) = \sum_{i=1}^{d-m-1} q_{m+i+1}(x) y^i \,.
	\end{equation}
\end{definition}

\begin{definition}
	Given $m=0,\dots,d$ denote
	\begin{equation}
		\alpha_m = \inf \{a \, | \, (a,m) \in \Delta \} \,, \qquad \beta_m = \sup \{a \, | \, (a,m) \in \Delta\}\,.
	\end{equation}
\end{definition}

The $\alpha_m$ and $\beta_m$ correspond directly to \cite[Definition~2.3]{BE17}, whereas the $Q_m$ correspond to reindexed versions of the $P_m$ in \cite[Definition~2.5]{BE17}. 

%

Given that we now have a couple of notations that will be critical to our construction of quantum curves, let us consider an example.

\begin{example}
	Consider the spectral curve
		\begin{equation}
		\mathcal{S} = \left( \P^1, \quad x(z) = z+1/z, \quad y(z) = z^2, \quad B = \frac{dz_1 dz_2}{(z_1-z_2)^2} \right)\,.
	\end{equation} 
	The functions $x$ and $y$ satisfy the degree two polynomial equation:
	\begin{equation*}
		P(x,y) = y^2+(2-x^2)y+1 = 0\,.
	\end{equation*}
	Here the non-zero $q_m$ and $Q_m$ are
	\begin{equation}
		q_0(x) = 1, \quad q_1(x) = 2-x^2, \quad q_2(x) = 1, \quad Q_0(x,y) = y =z^2\,.
	\end{equation}
	The Newton polygon $\Delta $ is the hull of $ \{ (0,0), (0,1), (0,2), (2,1) \}$, which means that this spectral curve is not regular as $(1,1)$ is an interior point. From $\Delta$ we also write down the $\alpha_m$ and $\beta_m$ for illustrative purposes
	\begin{equation}
		\alpha_0 = 0 \, , \quad \alpha_1 = 0 \,, \quad \alpha_2 = 0 \,, \quad \beta_0 = 0 \,, \quad \beta_1 = 2 \,, \quad \beta_2 = 0\,.
	\end{equation}
	In this simple case, the infimum and supremum in the definition of the $\alpha_m$ and $\beta_m$, respectively, are actually achieved by points in $A$; however, in general, this may not be the case and the $\alpha_m$ and $\beta_m$ could take on non-integer values.
\end{example}

We now define analogues of the $C_k$ and $D_k$ that appear in \cite[Equations (5.31) \& (5.34)]{BE17}. Let $x=x(z)$ and $x_i=x(z_i)$ for $z,z_i\in\mathbb{C}$ and $i\in\mathbb{Z}_{\geq 1}$.

\begin{definition}\label{d:diffoperate}
	Let $b\in\mathbb{C}$ be a pole of $ dx$ where all the $\omega_{g,n}$ are holomorphic and $x$ is meromorphic. We define
	\begin{equation}
		E_i = -\lim_{z_1\rightarrow b} \frac{Q_{i-1}(x,y)}{x^{\lfloor\alpha_{i}\rfloor+1}} \,,
		\qquad
		F_i = \hslash\frac{x^{\lfloor\alpha_i\rfloor}}{x^{\lfloor\alpha_{i-1}\rfloor}} \frac{d}{ dx}\,,
	\end{equation}
	where $\lfloor\cdot\rfloor$ is the floor function.
\end{definition}

With these definitions out of the way we can construct quantum curves for compact transalgebraic admissible regular spectral curves.

\begin{theorem}\label{t:TR/QCpole}
Let $\mathcal{S} = (\Sigma, x, y, \frac{dz_1 dz_2}{(z_1-z_2)^2})$ be a compact transalgebraic admissible regular spectral curve. Let $b$ be a pole of ${\rm d}x$ at which the $\omega_{g,n}$ are holomorphic. Then the wave-function
	\begin{equation}
		\psi(z;b) = \exp\left[\sum_{n=1}^{\infty}\sum_{g=0}^{\infty}\frac{\hslash^{2g+n-2}}{n!}\int_b^z\cdots\int_b^z \left( \omega_{g,n}-\delta_{n,2}\delta_{g,0} \frac{dx(z_1) dx(z_2)}{(x(z_1) - x(z_2))^2} \right)\right]
	\end{equation}
	satisfies the differential equation
	\begin{equation}
		\left(\frac{q_0(x)}{x^{\lfloor \alpha_0 \rfloor}} + \sum_{i=1}^{d} F_1 F_2 \cdots F_{i-1}\frac{q_i(x)}{x^{\lfloor \alpha_i \rfloor}}F_i + \hslash \sum_{i=1}^{d-1} E_i F_1 F_2 \cdots F_{i-1}\frac{x^{\lfloor \alpha_i \rfloor}}{x^{\lfloor \alpha_{i-1} \rfloor}}\right) \psi(z;b) = 0 \,.
	\end{equation}
	(Note that $d$ may be infinite if the curve is transalgebraic.)
\end{theorem}

\begin{proof}
	Since $\mathcal{S}$ is regular, it has genus zero, and thus we can take $\Sigma = \P^1$. Let $x=M_0\exp(M_1)$, $y=M_2/x$ with $M_0,M_1,M_2$ rational functions on $\P^1$. Consider the sequence of compact, meromorphic spectral curves:
	\begin{equation*}
		\mathcal{S}^N = \left( \P^1, \quad  x^N=M_0\left(1 + (\tau - 1) \frac{M_1}{N}\right)^{-N} \left(1+\tau \frac{M_1}{N}\right)^N \,, \quad  y^N=\frac{M_2}{x^N}, \quad B =\frac{dz_1 dz_2}{(z_1-z_2)^2}\right) \,.
	\end{equation*} 
	By assumption, those spectral curves are all regular, so we can apply \cite[Lemma 5.14]{BE17}.
	
	As the $\omega_{g,n}^N$ converge to the correlators $\omega_{g,n}$, if the $\omega_{g,n}$ are regular at $b$, then the $\omega_{g,n}^N$ must also be regular for large enough $N$. We then define $E_i^N$, $F_i^N$, and $\psi^N(z;b)$ in the natural fashion. We quickly see that $\psi^N(z;b)\to\psi(z;b)$ as the exponential is continuous, the sum is formal, and the $\omega_{g,n}^N$ are well-defined in the limit so we can bring the limit inside the integrals using dominated convergence. The fact that $F_i^N\to F_i$ is clear, as the Newton polygon will converge. Finally, we must deal with the $E_i^N$. In \cite{BE17} it was argued, using arguments based on an inequality of divisors, that the $C_i$ (which correspond to re-indexed $E_i$) must be finite as $z_1\to b$. The argument will carry over in the limit as $N\to\infty$ as
	\begin{enumerate}
		\item $x$ is meromorphic near $b$ by assumptions on $b$;
		\item as $x^N$ will be uniformly convergent away from $x = \infty $, it will be uniformly convergent, in particular, near $b$;
		\item the required inequalities are non-strict.
	\end{enumerate}
	Finally, as already noted, near $b$, $x^N$ is uniformly convergent so by the Moore--Osgood theorem,
	\begin{equation}
		\lim\limits_{N\rightarrow\infty} \lim\limits_{z_1\rightarrow b} \frac{Q^N_{i-1}(x_1,y_1)}{(x^N-x^N_1)(x^N_1)^{\lfloor\alpha^N_{i}\rfloor}} = \lim\limits_{z_1\rightarrow b} \lim\limits_{N\rightarrow\infty} \frac{Q^N_{i-1}(x_1,y_1)}{(x^N-x^N_1)(x^N_1)^{\lfloor\alpha^N_{i}\rfloor}} \,,
	\end{equation}
	at which point we may just take the limit, concluding that $E_i^N\rightarrow E_i$ and the $E_i$'s are not identically infinity.
\end{proof}
%

This theorem therefore gives us a canonical way of creating a quantum curve for compact transalgebraic admissible spectral curves that are regular. In particular, we do not actually have to construct a quantum curve for each finite $N$ and take the limit; the existence of such a sequence of curves guarantees we can construct the quantum curve directly from the limiting curve. It is important to note that if $d=\infty$, the constructed quantum curve need not be simple, in contrast to the $d<\infty$ case.

\begin{remark}\label{r:pres}
While we may construct the quantum curve directly from the limiting curve as $N \to \infty$, different sequences of curves $\mathcal{S}_N$ may yield different presentations for the limiting curve $\mathcal{S}$. What do we mean by that? The limiting curve is given by a quadruple $(\P^1, x, y, B)$. But to extract the quantum curve, one needs to write down the relation $P(x,y) = 0$ satisfied by the functions $x$ and $y$. This relation can be written in different ways, and they may produce a priori different quantum curves via \cref{t:TR/QCpole}. To make this statement clear, consider the spectral curve $\mathcal{S}$ of \cref{ex:atlantes}, with $M_0 = z$, $M_1 =- z^r$ and $M_2 = z$, and the usual $\tau$-dependent sequence of curves $\mathcal{S}_N$. \Cref{t:TR/QCpole} gives a quantum curve that can be read off directly from the equation $P(x,y)=0$ satisfied by the functions of $\mathcal{S}$, but the way this equation is presented depends on $\tau$. More precisely, the relation $P(x,y) = 0$ used to extract the quantum curve should be written as
\begin{equation}
P(x,y) = y e^{\tau(xy)^r} - e^{(\tau-1)(xy)^r} \,.
\end{equation}
As a result, the quantum curve may a priori depend on $\tau$, that is, on the choice of sequence used to construct the limiting curve.

This is not too surprising as for each choice of $\tau$ the entire function $P$ is different, and yet all such $P$ correspond to the same spectral curve. Thus the fact that the limiting quantum curve may depend on the choice of $\tau$ should not be seen as an artefact of defining things in terms of limits, but an actual degeneracy in the choice of $P$ that exists in the limit.
\end{remark}

As in \cite{BE17}, the choice of integration divisor can be generalised from $D = [z] - [b]$ in an analogous way to the generalisation presented in \cite[Remark~5.15]{BE17}; the key steps of the proof carry over virtually without modification.

However, choosing one's base point to be a pole of ${\rm d}x$ is inconvenient when ${\rm d}x$ has no pole; a case that may arise when $x$ has an essential singularity. In \cite{BE17}, the authors considered the case of the base point $b$ being a zero of $q_d(x(b))=0$, but only for $d=2$. Here, we generalise this choice to the case $\infty > d > 2$ and then use it to construct quantum curves with this base point. We begin this process with a lemma before proving a theorem analogous to \cref{t:TR/QCpole}.

\begin{lemma}
	For $b$ a zero of $q_d(x)$ that is not in the ramification locus of $x$,
	\begin{equation}
		\psi_i(x(b);z;b) = \psi(z;b) \lim_{z_1\rightarrow b} \frac{1}{x(z_1)^{\lfloor \alpha_{d-i} \rfloor}} Q_{d-i-1}(x(z_1),y(z_1)) \,,
	\end{equation}
	where the $\psi_i$ are defined in \cite[Definition~5.9]{BE17}.
\end{lemma}
\begin{proof}
	From \cite[Definition~5.9]{BE17}
	\begin{equation}
		\psi_i(x(b);z;b) = \psi(z;b) \lim_{z_1\rightarrow b} \left( \frac{1}{x(z_1)^{\lfloor \alpha_{d-i}\rfloor}} \big( q_{d}(x(z_1))\xi_i(x(z_1);D)-q_{d-i}(x(z_1)) \big) \right) \,.
	\end{equation}
	Using the notation of \cite{BE17} and substituting in the definition of the $\xi_k$, \cite[Definition~5.6]{BE17},
	\begin{equation}
		q_d(x(z_1))\xi_i(z_1) = (-1)^iq_d(x(z_1)) \sum_{n=0}^{\infty} \sum_{g=0}^{\infty} \frac{\hslash^{2g+n}}{n!} \frac{G^{(i)}_{g,n+1}(z_1)}{dx(z_1)^i} \,,
	\end{equation}
	where the $G^{(i)}_{g,n+1}$ are defined in \cite[Definition~5.3]{BE17}. First we examine the power $\hslash^{0}$. Here we have, where the $U^{(i)}_{0,1}$ are defined in \cite[Definition~4.1]{BE17}
	\begin{equation}
		(-1)^iq_d(x(z_1)) \frac{G^{(i)}_{0,1}(z_1)}{dx(z_1)^i} = (-1)^iq_d(x(z_1)) \frac{U^{(i)}_{0,1}(z_1)}{dx(z_1)^i} = Q_{d-i-1}(x(z_1),y(z_1))+q_{d-i}(x(z_1)) \,.
	\end{equation}
	Note that we then have the inequality of divisors (\cite[Lemma~2.6]{BE17})
	\begin{equation}
		\Div (Q_{d-i-1}(x,y)) \geq \alpha_{d-i} \Div_0(x) - \beta_{d-i} \Div_{\infty}(x) \,.
	\end{equation}
	So we therefore have that the limit
	\begin{equation}
		\lim_{z_1\to b} \frac{1}{x(z_1)^{\lfloor \alpha_{d-i} \rfloor}}Q_{d-i-1}(x(z_1),y(z_1)) \,,
	\end{equation}
	is finite. This is in agreement with the results of \cite[Section 5.3.2]{BE17} for $d=2$ as in that case $Q_0(x,y)= q_d (x) y$. Now we examine the higher order powers of $\hslash$. As $b$ is not in the ramification locus of $x$, each $G^{(i)}_{g,n+1}(z_1)$ is regular at $b$ for $2g+n\geq 1$. Furthermore, it cannot be a zero of $dx$ so for each $i$,
	\begin{equation}
		\frac{G^{(i)}_{g,n+1}(z_1)}{dx(z_1)^i}
	\end{equation}
	is regular at $b$. Ergo, if $b$ is not a zero of $x$, the terms of higher order in $\hslash$ never contribute. Assume then that $b$ is a simple zero of $x$; we claim that still
	\begin{equation}\label{e:qdzero}
		\lim_{z_1\to b} \frac{q_d(x(z_1))}{x(z_1)^{\lfloor \alpha_{d-i}\rfloor}} = 0 \,,
	\end{equation}
	for $i = 1, \dots, d - 1$. As our curve is irreducible, there is some $k=0,\dots,d-1$ with $q_{k}(0) \neq 0$ as we could otherwise cancel out an overall factor of $x$ in $P(x,y)$. Let $k_1$ and $ k_2$ be the minimum and maximum such $k$, respectively. By convexity, $ \alpha_k = 0$ if and only if $ k_1 \leq k \leq k_2$. Then, as the $\alpha_m$ are the smallest point on the convex hull at the power of $y^{m}$, they are strictly increasing for $m \geq k_2$ and strictly decreasing for $m \leq k_1$. Finally, $\alpha_0$ and $\alpha_d$ will be non-negative integers. Furthermore $\alpha_0 \leq \alpha_d$ as, if this inequality did not hold, $(1,k_1)$ would be an interior point of the Newton polygon. Thus, we have $\alpha_d=\lfloor\alpha_d\rfloor>\lfloor\alpha_m\rfloor$ for all $d>m>0$. This establishes \eqref{e:qdzero} as the order of the zero of $q_d(x)$ in $x$ is $\alpha_d$.
	
	So we get that the $\hslash$ corrections vanish and we have the explicit expressions,
	\begin{equation}
		\psi_i(x(b);z;b) = \psi(z;b)\lim\limits_{z_1\rightarrow b}\left(\frac{1}{x(z_1)^{\lfloor \alpha_{d-i}\rfloor}}Q_{d-i-1}(x(z_1),y(z_1))\right) \,,
	\end{equation}
	as claimed.
\end{proof}

This gives a theorem analogous to \cref{t:TR/QCpole} except with this new choice of base point. First, we define the new coefficients $G_i$ and $H_i$
\begin{equation}
	G_{i} = \lim\limits_{z_1\rightarrow b} \frac{1}{x(z_1)^{\lfloor \alpha_{i}\rfloor}} Q_{i-1}(x(z_1),y(z_1)), \quad H_i = \hslash\frac{x^{\lfloor\alpha_i\rfloor}}{x^{\lfloor\alpha_{i-1}\rfloor}} \left( \frac{d}{dx}-\frac{1}{x-x(b)} \right) \,.
\end{equation}
Then, \cite[Theorem~5.11]{BE17} reduces to
\begin{multline}\label{e:qcrecorig}
	\hslash\frac{{\rm d}}{{\rm d}x}\psi_{i-1}(x;z;b) = \frac{x^{\lfloor\alpha_{d-i}\rfloor}}{x^{\lfloor\alpha_{d-i+1}\rfloor}} \psi_i(x;z;b) - \frac{q_{d-i+1}(x)x^{\lfloor\alpha_{d-1}\rfloor}}{q_d(x)x^{\lfloor\alpha_{d-i+1}\rfloor}} \psi_1(x;z;b)\\
	+ \hslash\frac{1}{x-x(b)} \left( \psi_{i-1}(x;z;b)-G_{d-i+1}\psi(z;b) \right) \,.
\end{multline}
We can now derive a quantum curve in the manner of \cite[ Lemma~5.14]{BE17}.

\begin{theorem}\label{t:TR/QCzero}
Let $\mathcal{S} = (\Sigma, x, y, \frac{dz_1 dz_2}{(z_1-z_2)^2})$ be a compact transalgebraic admissible regular spectral curve. Let $b$ be a zero of $q_d(x)$ for $d<\infty$ or, if $d=\infty$, a zero of $x$, with $b$ not in the ramification locus of $x$. Then the wave-function
	\begin{equation}
		\psi(z;b) = \exp\left[\sum_{n=1}^{\infty}\sum_{g=0}^{\infty}\frac{\hslash^{2g+n-2}}{n!}\int_b^z\cdots\int_b^z \left( \omega_{g,n}-\delta_{n,2}\delta_{g,0} \frac{dx(z_1) dx(z_2)}{(x(z_1) - x(z_2))^2} \right)\right]
	\end{equation}
	satisfies the differential equation
	\begin{equation}
		\left( \frac{q_0(x)}{x^{\lfloor\alpha_0\rfloor}} + \sum_{i=1}^{d}H_1 \cdots H_{i-1}\frac{q_i(x)}{x^{\lfloor\alpha_i\rfloor}}F_i + \hslash\sum_{i=1}^{d-1}G_iH_1 \cdots H_{i-1}\frac{x^{\lfloor\alpha_i\rfloor}}{x^{\lfloor\alpha_{i-1}\rfloor}(x-x(b))} \right)\psi(z;b)=0 \,.
	\end{equation}
\end{theorem}
\begin{proof}
	First assume $d<\infty$. Rewriting \eqref{e:qcrecorig},
	\begin{multline}
		\psi_i(x;z;b) = H_{d-i+1}\psi_{i-1}(x;z;b) + \frac{q_{d-i+1}(x)}{x^{\lfloor\alpha_{d-i+1}\rfloor}} F_{d-i+1}\psi(z;b)\\
		+ \hslash\frac{x^{\lfloor\alpha_{d-i+1}\rfloor}}{x^{\lfloor\alpha_{d-i}\rfloor}(x-x(b))} G_{d-i+1}\psi(z;b) \,,
	\end{multline}
	where we used the fact that by \cite[Lemma~5.10]{BE17}) 
	\begin{equation*}
		\psi_1(x;b) = \frac{q_d(x)}{x^{\lfloor\alpha_{d-1}\rfloor}} \hslash\frac{{\rm d}}{{\rm d}x}\psi(z;b) \,.
	\end{equation*}
	We can substitute the $i=d-1$ result into the $i=d$ result to obtain
	\begin{multline}
		\psi_d(x;z;b) = H_{1}\psi_{d-1}(x;z;b) + \frac{q_{1}(x)}{x^{\lfloor\alpha_{1}\rfloor}}F_1\psi(z;b) + \hslash\frac{x^{\lfloor\alpha_{1}\rfloor}}{x^{\lfloor\alpha_{0}\rfloor}(x-x(b))} G_{1}\psi(z;b)\\
		= H_{1}H_{2}\psi_{d-2}(x;z;b) + H_1\frac{q_{2}(x)}{x^{\lfloor\alpha_{2}\rfloor}}F_2\psi(z;b) + \hslash H_1\frac{x^{\lfloor\alpha_{2}\rfloor}}{x^{\lfloor\alpha_{1}\rfloor}(x-x(b))}G_{2}\psi(z;b)\\
		+ \frac{q_{1}(x)}{x^{\lfloor\alpha_{1}\rfloor}}F_1\psi(z;b) + \hslash\frac{x^{\lfloor\alpha_{1}\rfloor}}{x^{\lfloor\alpha_{0}\rfloor}(x-x(b))}G_{1}\psi(z;b) \,.
	\end{multline}
	Applying this iteratively, before finally using the fact that (again by \cite[Lemma~5.10]{BE17}) 
	\begin{equation*}
		\psi_d(x;z;b) = -\frac{q_0(x)}{x^{\lfloor\alpha_0\rfloor}}\psi(z;b) \,,
	\end{equation*}
	 yields the desired result. Taking the limit to get the $d=\infty$ result is completely analogous to the $d=\infty$ case in \cref{t:TR/QCpole}.
\end{proof}

\section{Proofs of \texorpdfstring{\cref{NoPolesForn>1,PoleHg1}}{prop}}\label{ProofOfAtlantesBDKS}

In this appendix, we give the proofs of \cref{NoPolesForn>1,PoleHg1}. All statements and proofs can be directly adapted from \cite[Section~4.2]{BDKS20}; in particular \cite[Lemmata~4.3-10]{BDKS20} will be altered for our purposes. We will give all statements, together with a full proof of the analogue of \cite[Lemma~4.3]{BDKS20}, and will indicate what has to be changed for the other ones. The analogue of \cite[Lemma~4.10]{BDKS20} actually has a different result in this case, so we give details for this as well, in the proof of \cref{PoleHg1}.

We start with the basic structural result for the $H_{g,n}$.

\begin{proposition}[{\cite[Proposition 3.10]{BDKS20}}]\label{ShapeOfHgn}
	In the setup of \cref{BDKSTR}, for $n \geq 3$,
	\begin{equation}
		H_{g,n} = [\hslash^{2g-2+n}] \sum_{\gamma \in \Gamma_n} \prod_{v_i \in \mc{I}_\gamma} \bar{U}_i \!\! \prod_{\{ v_i, v_k\} \in E_\gamma \setminus \mc{K}_\gamma} \!\! w_{i,k} \! \prod_{\{ v_i, v_k\} \in \mc{K}_\gamma} \! \bigg( \bar{U}_iw_{i,k} + \hslash u_k \mc{S}(u_k \hslash Q_kD_k) \frac{z_i}{z_k -z_i}\bigg)  + \textup{const}\,,
	\end{equation}
	where $\Gamma_n$ is the set of simple graphs on $n$ vertices $ v_1, \dotsc , v_n$, $E_\gamma$ is the set of edges of a graph $ \gamma$, $\mc{I}_\gamma$ is the subset of vertices of valency $\geq 2$, and $\mc{K}_\gamma$ is the subset of edges with one end $v_i$ of valency $1$ and another end $v_k$, and where
	\begin{equation}
	\begin{split}
		\bar{U}_i f
		&
		\coloneqq \sum_{s=0}^\infty \sum_{j=1}^\infty D_i^{j-1} \bigg( \frac{L_{s,i}^j}{Q_i} [u_i^s] \frac{e^{u_i \mc{S}(u_i \hslash Q_i D_i) \hat{y}(z_i) - y(z_i)}}{u_i \hslash \mc{S}(u_i \hslash )}f \bigg)\,,
		\\
		w_{k,l}
		&
		\coloneqq e^{\hslash^2 u_k u_l \mc{S} (u_k \hslash Q_k D_k) \mc{S} (u_l \hslash Q_l D_l) \frac{z_k z_l}{(z_k - z_l)^2}} - 1\,,
		\\
		L_{s,i}^j
		&
		\coloneqq \Big( [v^j] ( \del_y + v \psi'(y))^s e^{v \big( \frac{ \mc{S}(v \hslash \del_y)}{\mc{S} (\hslash \del y)} \hat{\psi} (y) - \psi (y)\big)} \Big) \Big|_{y = y(z_i)}\,.
	\end{split}
	\end{equation}
For $n = 2$ and $g > 0$ we have:
	\begin{equation}
		H_{g,2} = [\hslash^{2g}]  \bigg( \bar{U}_1\bar{U}_2 w_{1,2} + \bar{U}_1 \Big( \hslash u_1\mc{S}(u_1 \hslash Q_1D_1) \frac{z_1}{z_1 - z_2} \Big) +\bar{U}_2 \Big(  \hslash u_2 \mc{S}(u_2 \hslash Q_2 D_2) \frac{z_1}{z_2 - z_1} \Big) \bigg) + \textup{const}\,.
	\end{equation}
	For $n = 1$ and $g > 0$ we have:
	\begin{equation}\label{ShapeOfHg1}
	\begin{split}
		H_{g,1} 
		&
		=
		[\hslash^{2g}]  \bigg( \hslash \bar{U}_1 1 + \sum_{j=1}^\infty D^{j-1}_1 L^{j+1}_{0,1} D_1y(z_1) +\int_{0}^{z_1}\frac{\hat{y}(z) - y(z)}{z} dz 
		\\
		&
		+ \int_{0}^{z_1} \frac{Q(z)}{z} \Big( \frac{1}{\mc{S} (\hslash \del_y)} \hat{\psi}(y) - \psi (y)\Big) \Big|_{y = y(z)}  Dy(z)dz \bigg) + \textup{const}\,.
	\end{split}
	\end{equation}
	In each case the extra constant can be determined from the condition that $H_{g,n}$ vanishes at zero. These constants are not important for the argument below and can be ignored.
\end{proposition}

We will use this in a similar way to \cite[Section 4]{BDKS20}, so let us quote part of the first page of that section:\par

\begin{quote}
We begin with a few general observations related to the structure of the formulas (102)-(107) [the ones in \cref{ShapeOfHgn}] for $H_{g,n}$, $2g - 2 +n \geq 0$, and relevant for the both families of parameters. These formulas give manifestly rational functions, whose principal parts at the points $p_1, .\dotsc , p_N$ are odd with respect to the deck transformations. So, we have to show that these functions have no other poles in each variable $z_1, \dotsc , z_n$.\par
Consider a particular $H_{g,n}(z_1, \dotsc , z_n)$. From the shape of the formula it is clear that its possible poles in the variable $z_1$ in addition to $p_1, \dotsc, p_N$ are either at the diagonals $z_1 -z_i = 0$, $i \neq 1$(but it is known from \cite{BDKS22} that these functions have no poles at the diagonals $z_i -z_j = 0$), or at $\infty$, or at the special points related to the specific form of the operator $\bar{U}_1$ for Family I and Family II. A bit more special case is the case of $H_{g,1}$, where we have to analyse some extra terms as well. \par
Note that it is in fact sufficient to analyse the pole structure just for $H_{g,1}$, $g \geq 1$, since this case subsumes the corresponding analysis of the pole structure for $H_{g,n}$, $n \geq 2$. Indeed, the factors of the form $w_{k,l}$ and $\hslash u_k\mc{S}(u_k \hslash Q_k D_k) \frac{z_i}{z_k- z_i}$ do not contribute any poles to the resulting expressions, as all diagonal poles get cancelled and these factors are regular at infinity as well. Therefore, the possible extra poles can only occur at the special points of $\bar{U}_i$, which enters the formula for $H_{g,1}$ in exactly the same way as formulas for $H_{g,n}$ for other values of $n$. The argument for the $n = 1$ case includes analysis of the singularities of $\bar{U}_1$, and once we show that it has no poles outside $p_1, \dots , p_N$, it immediately implies the same statement for any $n \geq 2$ as well.
\end{quote}

For Atlantes Hurwitz numbers, the same strategy applies, with the one difference that the middle term on the first line and the term on the second line of \cref{ShapeOfHg1} contribute poles at $ \infty$.

Denote $ d_i \coloneqq \deg P_i$, $ e_j \coloneqq \deg R_j$. As we only need to consider $ n = 1$, let us specialise some formulae, omitting subscripts $ n = i = 1$ where convenient. 

\begin{align}
		\bar{U} f
		&
		\coloneqq \sum_{s=0}^\infty \sum_{j=1}^\infty D^{j-1} \bigg( \frac{L_s^j}{Q} [u^s] \frac{e^{u (\mc{S}(u \hslash Q D) -1) \frac{ R_1 (z) }{ R_2 (z) } } }{u \hslash \mc{S}(u \hslash )} f \bigg)\,, \label{U1f}
		\\
		L_s^j
		&
		\coloneqq \Big( [v^j] ( \del_y + v \psi'(y))^s e^{v \big( \frac{ \mc{S}(v \hslash \del_y)}{\mc{S} (\hslash \del y)}-1 \big)  \big( P_1(y) + \log \frac{P_2 (y)}{P_3 (y)} \big) } \Big) \Big|_{y = \frac{ R_1(z) }{ R_2 (z)} }\,,
		\\
		H_{g,1} 
		&
		= \tau^1_g + \tau^2_g + \tau^3_g + \textup{const} \,,
		\\
		\tau^1_g
		&
		=
		[\hslash^{2g}] \hslash \bar{U} 1 \,,
		\\
		\tau^2_g
		&
		=
		[\hslash^{2g}] \sum_{j=1}^\infty D^{j-1} L^{j+1}_0 D \frac{ R_1 (z) }{ R_2 (z) } \,,
		\\
		\tau^3_g
		&
		=
		\Big( [u^{2g}] \frac{1}{ \mc{S} (u) } \Big) \bigg( \del_y^{2g-1}  \Big( P_1 (y) + \log \Big( \frac{ P_2 (y) }{ P_3 (y) } \Big) \Big) \bigg) \bigg|_{y = \frac {R_1 (z) }{ R_2 (z) } } \,.
\end{align}

We also introduce the following notation: for some polynomials $ \tilde{P}_j(z)$,

\begin{align}
	\psi (y( z)) 
	&
	= \frac{ \tilde{P}_1 (z) }{ \big( R_2 ( z) \big)^{d_1} } + \log \bigg( \frac{ \tilde{P}_2 (z) }{ \tilde{P}_3 (z) } \bigg) \,,
	\\
	\check{Q}(z)
	&
	\coloneqq R_2^{d_1 + 1} \tilde{P}_2 \tilde{P}_3 + z \Big( d_1 R_2' \tilde{P}_1 \tilde{P}_2 \tilde{P}_3 - R_2 \tilde{P}_1' \tilde{P}_2 \tilde{P}_3 + R_2^{d_1 + 1} \tilde{P}_2 \tilde{P}_3' - R_2^{d_1 + 1} \tilde{P}_2' \tilde{P}_3 \Big)
	\\
	&
	= c \prod_{j = 1}^N (z - p_j )
	\intertext{such that}
	Q(z)
	&
	= \frac{\check{Q}(z)}{R_2^{d_1 + 1} \tilde{P}_2 \tilde{P}_3 }\,.
\end{align}

Now consider the pole contributions of the $ \tau^j_g$ separately, starting with $ \tau^1_g$.

For a given power of $ \hslash$, the sum over $s$ in \cref{U1f} has an upper bound as each $ u_1$ must come with at least $ \hslash^{2/3}$. For a similar reason, the sum over $ j$ is finite. As we also have that
\begin{equation}
	\left( \frac{ \mc{S}(v \hslash \del_y)}{\mc{S} (\hslash \del y)}-1 \right)  \left( P_1(y) + \log \frac{P_2 (y)}{P_3 (y)} \right) 
\end{equation}
is a series in $ \hslash $ with coefficients rational functions in $y$, $ \tau^1_g$ is a rational function in $ z_1$. Its set of possible poles consists of the $ p_j$, the zeros of $ \tilde{P}_2$, $ \tilde{P}_3$, and $R_2$, and $ z = \infty$. We will show that the $ \tau^j_g$ have no poles at the zeros of $ \tilde{P}_2$, $ \tilde{P}_3$, and $R_2$, and that $ \tau^1_g$ and $\tau^2_g $ also have no zeroes at $ z = \infty$.

\begin{lemma} \label{BDKSLemma4.3}
	$\tau^1_g$ has no poles at the zeroes of $R_2 (z)$.
\end{lemma}

\begin{proof}
	This is the analogue of \cite[Lemma~4.3]{BDKS20}. The only difference is that
	\begin{equation}
		e^{v( \mc{S} ( v \hslash \del_y) - 1)P_1(y)} 
	\end{equation}
	should be replaced by
	\begin{equation}
		e^{v\big( \frac{\mc{S} ( v \hslash \del_y)}{\mc{S}(\hslash \del_y)} - 1\big)P_1(y)} \,.
	\end{equation}
	Let us show how this works.
	
	Let $B$ be a zero of $R_2 (z)$. For $z \to B $, we have $ y (z) = R_1 (z) / R_2 (z) \to \infty$, and if $ B$  is a simple zero of $ R_2 $, then it is a simple pole of $y (z)$. Let us count the order of the pole of
	\begin{equation}\label{BDKS139}
		\sum_{s=0}^\infty \sum_{j=1}^\infty D^{j-1} \bigg( \frac{1}{Q} \Big( [v^j] ( \del_y + v \psi'(y))^s e^{v \big( \frac{ \mc{S}(v \hslash \del_y)}{\mc{S} (\hslash \del y)}-1 \big)  \big( P_1(y) + \log \frac{P_2 (y)}{P_3 (y)} \big) } \Big) \Big|_{y = \frac{ R_1(z) }{ R_2 (z)} } [u^s] \frac{e^{u (\mc{S}(u \hslash Q D) -1) \frac{ R_1 (z) }{ R_2 (z) } } }{u \hslash \mc{S}(u \hslash )} \bigg)\,,
	\end{equation}
	at $z_1 = B$. To this end, two immediate observations are in order:
	\begin{itemize}
		\item Firstly, note that $ e^{v \big( \frac{ \mc{S}(v \hslash \del_y)}{\mc{S} (\hslash \del y)}-1 \big)  \log \frac{P_2 (y)}{P_3 (y)} } $ does not contribute to the pole at infinity in $y$, and, therefore, to the pole in $ z $ at $ z = B $, and can be safely ignored in this computation.
		\item Secondly, note that $Q^{-1} $ has zero of order $ d_1 + 1 $ at $ z = B $ and each application of $ D = Q^{-1} z \del_z $ decreases the degree of the pole in $ z $ at $ B $ by $ d_1$. The total effect of the factor $ Q^{-1} $ and $D^{j-1} $ is the decrease of the order of pole by $d_1 j + 1$.
	\end{itemize}
	Therefore, the order of the pole of \cref{BDKS139} is equal to the order of pole at $ z = B $ of
	\begin{equation}\label{BDKS140}
		(z - B) \sum_{s=0}^\infty \bigg( ( \del_y + v \psi'(y))^s e^{v \big( \frac{ \mc{S}(v \hslash \del_y)}{\mc{S} (\hslash \del y)}-1 \big) P_1(y) } \Big|_{y = \frac{ R_1(z) }{ R_2 (z)} } [u^s] \frac{e^{u (\mc{S}(u \hslash Q D) -1) \frac{ R_1 (z) }{ R_2 (z) } } }{u \hslash \mc{S}(u_1 \hslash )} \bigg) \bigg|'_{v = (z - B)^{d_1} } \,,
	\end{equation}
	where by $ \big|'$, we mean that we only select the terms with $\deg_v \geq 1 $ before the substitution $ v = (z - B )^{d_1}$. Note also that
	\begin{itemize}
		\item Since $ y (z) $ has a simple pole at $ z = B $, each $\del_y $ decreases the order of pole in the resulting expression by $1$.
		\item Multiplication by $ \psi' (y) $ increases the order of pole by $ d_1 - 1$.
	\end{itemize}
	Taking into account these two observations and that each $ v $ factor decreases the order of pole by $ d_1 $, we see that each application of the operator $ \del_y + v \psi' (y) $ decreases the order of pole in the resulting expression by $1$. Therefore, the order of the pole of \cref{BDKS140} is equal to the order of pole at $ z = B $ of
	\begin{equation}\label{BDKS141}
		(z - B) \bigg( e^{v \big( \frac{ \mc{S}(v \hslash \del_y)}{\mc{S} (\hslash \del y)}-1 \big) P_1(y) } \Big|_{y = \frac{ R_1(z) }{ R_2 (z)} } \frac{e^{u (\mc{S}(u \hslash Q D) -1) \frac{ R_1 (z) }{ R_2 (z) } } }{u \hslash \mc{S}(u \hslash )} \bigg) \bigg|''_{ \substack{ v = (z - B)^{d_1} \\ u = z - B } } \,,
	\end{equation}
	where by $ \big|'' $ we mean that we only select the terms with $ \deg v \geq 1 $ and regular in $ u $ before the substitutions $ v = ( z - B)^{d_1}$, $ u = z - B$.\par
	Here, the first exponent is different from \cite[Lemma~4.3]{BDKS20}, and this slightly changes the argument.\par
	In this first exponent, specifically in $\frac{ \mc{S} (v \hslash \del_y ) }{ \mc{S} (\hslash \del_y ) } - 1$, each $ \hslash \del_y $ does not increase the order of the pole at $ z = B $, and actually decreases it (\cite{BDKS20} only argues that $ v \hslash \del_y$ does not increase it; we do not have or need the $ v$); since $ v P_1 (y) $ has no pole at $ z = B $, this means that the whole first exponential is regular. In the second exponent, in $ \mc{S} ( u \hslash z \del_z ) - 1 $, each $ u \hslash z \del_z $  preserves the order of the pole at $ z = B $; since $ u R_1 ( z ) / R_2 (z) $ has no pole at $ z = B$, this means that the whole second exponential is regular. Finally, $ (z - B) / ( u \hslash \mc{S} ( u \hslash ) ) $ is also regular at $ z = B $ in this expression.\par
	Thus, \cref{BDKS141} is regular at $ z = B $, and therefore \cref{BDKS139} is regular at $ z = B $ as well.
\end{proof}

\begin{lemma} \label{BDKSLemma4.4}
	$\tau^1_g$ is regular at the zeros of $ \tilde{P}_2$ that are not zeros of $ R_2$.
\end{lemma}

\begin{proof}
	Analogous to \cite[Lemma~4.4]{BDKS20}. The same kind of modification of the exponent should be applied to the left-hand side of \cite[(143)]{BDKS20}, but the right-hand side and the rest of the proof hold verbatim.
\end{proof}

\begin{lemma} \label{BDKSLemma4.5}
	$\tau^1_g$ is regular at the zeros of $ \tilde{P}_3 $ that are not zeros or $R_2 $.
\end{lemma}

\begin{proof}
	Analogous to \cite[Lemma~4.5]{BDKS20} in the same way that \cref{BDKSLemma4.4} is analogous to \cite[Lemma~4.4]{BDKS20}.
\end{proof}

\begin{lemma} \label{BDKSLemma4.6}
	$\tau^1_g$ is regular at $ z = \infty $.
\end{lemma}

\begin{proof}
	Analogous to \cite[Lemma~4.6]{BDKS20}. If $ e_1 \leq e_2$, by the same degree counting, all parts are regular. If $ e_1 > e_2$, we should again do the same substitution as at the start of the proof of \cref{BDKSLemma4.3}. As here we also have that $ \hslash \del_y$ decreases the pole order, and not just $ v \hslash \del_y$, the penultimate paragraph of the proof of \cite[Lemma~4.6]{BDKS20} goes through. The rest holds without any changes.
\end{proof}

\begin{lemma} \label{BDKSLemma4.7}
	$\tau^2_g + \tau^3_g$ is regular at the zeros of $ R_2 (z)$.
\end{lemma}

\begin{proof}
	Analogous to \cite[Lemma~4.7]{BDKS20}: $\tau^3_g$ is regular, and for $ \tau^2_g$ the same change as in the start of the proof of \cite[Lemma~4.6]{BDKS20} must be adopted. Again, this is sufficient for pole counting, as $ \hslash \del_y$ is applied at least once and hence cancels the simple pole of $ R_1 (z) / R_2 (z) $.
\end{proof}

\begin{lemma} \label{BDKSLemma4.89}
	$ \tau^2_g + \tau^3_g $ is regular at zeros of $ \tilde{P}_2 $ or $ \tilde{P}_3$ that are not poles of $ R_2$.
\end{lemma}

\begin{proof}
	Analogous to \cite[Lemmata~4.8\&4.9]{BDKS20}. Again, the same modification as before works.
\end{proof}

\begin{proof}[Proof of \cref{NoPolesForn>1}]
	By \cref{BDKSLemma4.3,BDKSLemma4.4,BDKSLemma4.5,BDKSLemma4.6}, the only possible poles of $ \tau^1_g$ are at the $p_j$. As the poles of $ \frac{1}{Q}$ at the $ p_j$ are odd with respect to the deck transformations and iterative application of $D_1$ preserves this, this suffices to prove that $ \tau^1_g \in \Theta$.\par
	Similarly, by \cref{BDKSLemma4.7,BDKSLemma4.89}, the only possible poles of $ \tau^2_g$ and $ \tau^3_g$ are at the $ p_j$ and $ z = \infty $. Poles at the $ p_j$ can only be introduced by and application of $ D$; the first such appearance will give a simple pole, while repeated applications of $D$ will preserve the oddness of the principal parts with respect to deck transformations. This proves the claim for the $H_{g,1}$.\par
	For $n > 1$, the only terms are from $ \bar{U}_i$ applications, and by the extended quote above, they are similar to the analogous term in $ H_{g,1}$, which is $ \tau^1$. As this lies in $\Theta$, so do the $ H_{g,n}$ for $ n > 1$.\par
	So the only possible poles of $H_{g,1}$ are at the $ p_j $, or, for $ n = 1$, at the point $ z_1 = \infty $, coming from $ \tau_g^{2}$ and $\tau_g^{3}$, i.e. the middle term of the first line, and the second line of \cref{ShapeOfHg1}.
\end{proof}

\begin{proof}[Proof of \cref{PoleHg1}]
	The two terms in the statement of the corollary correspond to $ \tau^2_g$ and $ \tau^3_g $, respectively.\par
	The pole order counting for the first term is analogous to \cite[Lemma~4.10]{BDKS20}, but now only for $ \tau^2_g$. As for \cref{BDKSLemma4.6}, do the case distinction whether or not $ e_1 > e_2$. If not, i.e. if $ e_1 \leq e_2$, all terms are clearly regular. So let us consider the case $e_1 > e_2$ and write out $ \tau^2_g$:
	\begin{equation}
		\tau^2_g
		=
		[\hslash^{2g}] \sum_{j=1}^\infty D^{j-1} \Big( [v^{j+1}] e^{v \big( \frac{ \mc{S}(v \hslash \del_y)}{\mc{S} (\hslash \del y)}-1 \big)  \big( P_1(y) + \log \frac{P_2 (y)}{P_3 (y)} \big) } \Big) \Big|_{y = \frac{ R_1(z) }{ R_2 (z)} } D \frac{ R_1 (z) }{ R_2 (z) } \,.
	\end{equation}
	Derivatives of logarithms, and their exponentials, can never have poles at $ \infty$, so for calculating pole order we may as well consider
	\begin{equation}
		[\hslash^{2g}] \sum_{j=1}^\infty D^{j-1} \Big( [v^{j+1}] e^{v \big( \frac{ \mc{S}(v \hslash \del_y)}{\mc{S} (\hslash \del y)}-1 \big) P_1(y) } \Big) \Big|_{y = \frac{ R_1(z) }{ R_2 (z)} } D \frac{ R_1 (z) }{ R_2 (z) } \,.
	\end{equation}
	By definition, $ y(z) =\frac{R_1(z)}{R_2(z)}$ has pole order $ e_1 - e_2$ at $ \infty$, and the pole order of $ P_1(y(z)) $ is $ d_1 (e_1-e_2)$. Every $ \del_y$ lowers the pole order by $ e_1 - e_2$ while every $ D $ lowers it by $ d_1 (e_1-e_2)$. So for the pole order we may also consider
	\begin{equation}
		[\hslash^{2g}] \sum_{j=1}^\infty z^{(j-1)d_1(e_2-e_1)} \Big( [v^{j+1}] e^{v \big( \frac{ \mc{S}(v \hslash \del_y)}{\mc{S} (\hslash \del y)}-1 \big) P_1(y) } \Big) \Big|_{y = z^{e_1 -e_2} } z^{(1-d_1)(e_1 - e_2)} \,.
	\end{equation}
	For $ g \geq 2$, we see that we can replace each $v$ by $ z^{d_1(e_2-e_1)}$, and divide the result by its square. Then the product $ vP_1(y) $ in the exponent has pole order $ 0$, and the non-trivial terms of $ \mc{S}(v\hslash \del_y$ always give sub-leading pole order, so we may ignore them.\par
	Similarly, we may replace each $ \del_y$ by $ z^{e_2 - e_1}$, and we see that it is always coupled with $ \hslash$, so we get this $ 2g$ times. So the pole order is
	\begin{equation}
		2g(e_2-e_1) -2d_1(e_2-e_1) +(1-d_1)(e_1 - e_2) = (1+d_1 - 2g)(e_1 - e_2)\,.
	\end{equation}
	We can only achieve this pole order by taking all powers of $\hslash $ from the $ \mc{S}(\hslash \del_y)$ in the denominator. Any time we would take an $ \hslash $ from the $ \mc{S}(v \hslash \del_y)$ in the numerator, the extra $v$ pairs with an extra $ D$ to lower the pole order by $ d_1 (e_1 -e_2)$, making it negative. So only the first term in the expansion $ \mc{S}( v\hslash \del_y) = 1 + \mc{O}(v^2)$ contributes. This replacement gives the formula
	\begin{equation}
		\begin{split}
			\tau^2_g 
			&\sim 
			[\hslash^{2g}] \sum_{j=1}^\infty D^{j-1} \Big( [v^{j+1}] e^{v \big( \frac{ 1}{\mc{S} (\hslash \del y)} -1\big) \psi (y)} \Big) \Big|_{y = y(z)} Dy(z)
			\\
			&\sim
			[\hslash^{2g}] \sum_{k=0}^{g-2} \frac{1}{(k+2)!}D^k \Big( \big( \frac{ 1}{\mc{S} (\hslash \del y)} -1\big) \psi (y) \Big)^{k+2} \Big|_{y = y(z)} Dy(z)\,,
		\end{split}
	\end{equation}
	where $ \sim $ means`equal polar part' and we restricted the sum because each factor of the $ (k+2)$-nd power contributes at least $\hslash^2$.\par
	For the $ \tau^3_g$ term, we get
	\begin{equation}
		[\hslash^{2g}] \int_{0}^{z_1} \frac{Q(z)}{z} \Big( \frac{1}{\mc{S}(\hslash \del_y)} -1 \Big) \psi (y) \Big|_{y = y(z)} Dy(z) dz = \frac{(2^{1-2g}-1)B_{2g}}{(2g)!} \del_y^{2g-1} \psi (y) \Big|_{y = y(z_1)}\,,
	\end{equation}
	and adding both terms give the corollary.
\end{proof}

\bibliographystyle{elsarticle-num}
\bibliography{Refs}{}

\begin{thebibliography}{10}
\expandafter\ifx\csname url\endcsname\relax
  \def\url#1{\texttt{#1}}\fi
\expandafter\ifx\csname urlprefix\endcsname\relax\def\urlprefix{URL }\fi
\expandafter\ifx\csname href\endcsname\relax
  \def\href#1#2{#2} \def\path#1{#1}\fi

\bibitem{BE17}
V.~Bouchard, B.~Eynard, {Reconstructing WKB from topological recursion}, J.
  \'{E}c. polytech. Math. 4 (2017) 845--908.
\newblock \href {http://arxiv.org/abs/1606.04498} {\path{arXiv:1606.04498}},
  \href {https://doi.org/10.5802/jep.58} {\path{doi:10.5802/jep.58}}.

\bibitem{EO07}
B.~Eynard, N.~Orantin, {Invariants of algebraic curves and topological
  expansion}, Commun. Number Theory Phys. 1~(2) (2007) 347--452.
\newblock \href {http://arxiv.org/abs/math-ph/0702045}
  {\path{arXiv:math-ph/0702045}}, \href
  {https://doi.org/10.4310/CNTP.2007.v1.n2.a4}
  {\path{doi:10.4310/CNTP.2007.v1.n2.a4}}.

\bibitem{CEO06}
L.~Chekhov, B.~Eynard, N.~Orantin, Free energy topological expansion for the
  2-matrix model, J. High Energy Phys. 2006~(12) (2006) 1--31.
\newblock \href {http://arxiv.org/abs/math-ph/0603003}
  {\path{arXiv:math-ph/0603003}}, \href
  {https://doi.org/10.1088/1126-6708/2006/12/053}
  {\path{doi:10.1088/1126-6708/2006/12/053}}.

\bibitem{Eyn11}
B.~{Eynard}, {Intersection numbers of spectral curves} (2011) 55\href
  {http://arxiv.org/abs/1104.0176} {\path{arXiv:1104.0176}}, \href
  {https://doi.org/10.48550/arXiv.1104.0176}
  {\path{doi:10.48550/arXiv.1104.0176}}.

\bibitem{DOSS14}
P.~{Dunin-Barkowski}, N.~Orantin, S.~Shadrin, L.~Spitz, Identification of the
  {G}ivental formula with the spectral curve topological recursion procedure,
  Comm. Math. Phys. 328~(2) (2014) 669--700.
\newblock \href {http://arxiv.org/abs/1211.4021} {\path{arXiv:1211.4021}},
  \href {https://doi.org/10.1007/s00220-014-1887-2}
  {\path{doi:10.1007/s00220-014-1887-2}}.

\bibitem{EO07a}
B.~Eynard, N.~Orantin, Weil--{P}etersson volume of moduli spaces,
  {M}irzakhani's recursion and matrix models (2007) 1--9\href
  {http://arxiv.org/abs/0705.3600} {\path{arXiv:0705.3600}}, \href
  {https://doi.org/10.48550/arXiv.0705.3600}
  {\path{doi:10.48550/arXiv.0705.3600}}.

\bibitem{ABCDGLW23}
J.~E. Andersen, G.~Borot, S.~Charbonnier, V.~Delecroix, A.~Giacchetto,
  D.~Lewa{\'{n}}ski, C.~Wheeler, {Topological recursion for Masur-Veech
  volumes}, J. Lond. Math. Soc. 107~(1) (2023) 254–332.
\newblock \href {http://arxiv.org/abs/1905.10352} {\path{arXiv:1905.10352}},
  \href {https://doi.org/10.1112/jlms.12686} {\path{doi:10.1112/jlms.12686}}.

\bibitem{BM08}
V.~Bouchard, M.~Mari{\~{n}}o, Hurwitz numbers, matrix models and enumerative
  geometry, in: From {H}odge theory to integrability and {TQFT} tt*-geometry,
  Vol.~78 of Proc. Sympos. Pure Math., Amer. Math. Soc., Providence, RI, 2008,
  pp. 263--283.
\newblock \href {http://arxiv.org/abs/0709.1458} {\path{arXiv:0709.1458}},
  \href {https://doi.org/10.1090/pspum/078/2483754}
  {\path{doi:10.1090/pspum/078/2483754}}.

\bibitem{EMS11}
B.~{Eynard}, M.~{Mulase}, B.~{Safnuk}, {The Laplace transform of the
  cut-and-join equation and the Bouchard-Mari{\~n}o conjecture on Hurwitz
  numbers.}, {Publ. Res. Inst. Math. Sci.} 47~(2) (2011) 629--670.
\newblock \href {http://arxiv.org/abs/0907.5224} {\path{arXiv:0907.5224}},
  \href {https://doi.org/10.2977/PRIMS/47} {\path{doi:10.2977/PRIMS/47}}.

\bibitem{BHLM14}
V.~Bouchard, D.~Hernandez~Serrano, X.~Liu, M.~Mulase, {Mirror symmetry for
  orbifold Hurwitz numbers}, J. Diff. Geom. 98~(3) (2014) 375--423.
\newblock \href {http://arxiv.org/abs/1301.4871} {\path{arXiv:1301.4871}}.

\bibitem{DKPS23}
P.~Dunin-Barkowski, R.~Kramer, A.~Popolitov, S.~Shadrin, Loop equations and a
  proof of {Z}vonkine's {$qr$}-{ELSV} formula, Ann. Sci. \'{E}c. Norm.
  Sup\'{e}r. (4) 56~(4) (2023) 1199--1229.
\newblock \href {http://arxiv.org/abs/1905.04524} {\path{arXiv:1905.04524}},
  \href {https://doi.org/10.24033/asens.2553} {\path{doi:10.24033/asens.2553}}.

\bibitem{BKMP08}
V.~Bouchard, A.~Klemm, M.~Mari{\~{n}}o, S.~Pasquetti, Remodeling the {B}-model,
  Communications in Mathematical Physics 287~(1) (2008) 117--178.
\newblock \href {https://doi.org/10.1007/s00220-008-0620-4}
  {\path{doi:10.1007/s00220-008-0620-4}}.

\bibitem{EO15}
B.~Eynard, N.~Orantin, {Computation of open Gromov--Witten invariants for toric
  Calabi--Yau 3-folds by topological recursion, a proof of the BKMP
  conjecture}, {Commun. Math. Phys.} 337~(2) (2015) 483--567.
\newblock \href {http://arxiv.org/abs/1205.1103} {\path{arXiv:1205.1103}},
  \href {https://doi.org/10.1007/s00220-015-2361-5}
  {\path{doi:10.1007/s00220-015-2361-5}}.

\bibitem{NS14}
P.~Norbury, N.~Scott, {Gromov-Witten invariants of
  \texorpdfstring{$\mathbb{P}^1$}{P1} and Eynard-Orantin invariants}, Geometry
  \& Topology 18~(4) (2014) 1865--1910.
\newblock \href {https://doi.org/10.2140/gt.2014.18.1865}
  {\path{doi:10.2140/gt.2014.18.1865}}.

\bibitem{FLZ17}
B.~Fang, C.-C. Liu, Z.~Zong, The {Eynard--Orantin} recursion and equivariant
  mirror symmetry for the projective line, Geometry \& Topology 21~(4) (2017)
  2049--2092.
\newblock \href {https://doi.org/10.2140/gt.2017.21.2049}
  {\path{doi:10.2140/gt.2017.21.2049}}.

\bibitem{FLZ20}
B.~Fang, C.-C.~M. Liu, Z.~Zong, {On the Remodeling Conjecture for Toric
  Calabi-Yau 3-Orbifolds}, J. Amer. Math. Soc. 33~(1) (2020) 135--222.
\newblock \href {http://arxiv.org/abs/1604.07123} {\path{arXiv:1604.07123}},
  \href {https://doi.org/10.1090/jams/934} {\path{doi:10.1090/jams/934}}.

\bibitem{GKLS22}
A.~Giacchetto, R.~Kramer, D.~Lewa{\'{n}}ski, A.~Sauvaget, {The Spin
  Gromov--Witten/Hurwitz correspondence for
  \texorpdfstring{$\mathbb{P}^1$}{P1}} (2022).
\newblock \href {http://arxiv.org/abs/2208.03259} {\path{arXiv:2208.03259}},
  \href {https://doi.org/10.48550/arXiv.2208.03259}
  {\path{doi:10.48550/arXiv.2208.03259}}.

\bibitem{BCDG19}
G.~Borot, S.~Charbonnier, N.~Do, E.~Garcia-Failde, Relating ordinary and fully
  simple maps via monotone {H}urwitz numbers, Electron. J. Combin. 26~(3)
  (2019) Paper No. 3.43, 24.
\newblock \href {http://arxiv.org/abs/1904.02267} {\path{arXiv:1904.02267}},
  \href {https://doi.org/10.37236/8634} {\path{doi:10.37236/8634}}.

\bibitem{BCGLS21}
G.~{Borot}, S.~{Charbonnier}, E.~{Garcia-Failde}, F.~{Leid}, S.~{Shadrin},
  Functional relations for higher-order free cumulants (2021).
\newblock \href {http://arxiv.org/abs/2112.12184} {\path{arXiv:2112.12184}},
  \href {https://doi.org/10.48550/arXiv.2112.12184}
  {\path{doi:10.48550/arXiv.2112.12184}}.

\bibitem{BBCC24}
G.~Borot, V.~Bouchard, N.~K. Chidambaram, T.~Creutzig, Whittaker vectors for
  {$\mathcal{W}$}-algebras from topological recursion, Selecta Math. (N.S.)
  30~(2) (2024) Paper No. 33, 91.
\newblock \href {http://arxiv.org/abs/2104.04516} {\path{arXiv:2104.04516}},
  \href {https://doi.org/10.1007/s00029-024-00921-x}
  {\path{doi:10.1007/s00029-024-00921-x}}.

\bibitem{BEM17}
R.~Belliard, B.~Eynard, O.~Marchal, Integrable differential systems of
  topological type and reconstruction by the topological recursion, Ann. Henri
  Poincar\'{e} 18~(10) (2017) 3193--3248.
\newblock \href {http://arxiv.org/abs/1610.00496} {\path{arXiv:1610.00496}},
  \href {https://doi.org/10.1007/s00023-017-0595-9}
  {\path{doi:10.1007/s00023-017-0595-9}}.

\bibitem{EGMO21}
B.~Eynard, E.~Garcia-Failde, O.~Marchal, N.~Orantin, Quantization of classical
  spectral curves via topological recursion, Comm. Math. Phys. 405~(5) (2024)
  Paper No. 116, 118.
\newblock \href {http://arxiv.org/abs/2106.04339} {\path{arXiv:2106.04339}},
  \href {https://doi.org/10.1007/s00220-024-04997-6}
  {\path{doi:10.1007/s00220-024-04997-6}}.

\bibitem{BDKS20}
B.~Bychkov, P.~Dunin-Barkowski, M.~Kazarian, S.~Shadrin, Topological recursion
  for {K}adomtsev--{P}etviashvili tau functions of hypergeometric type, J.
  Lond. Math. Soc. (2) 109~(6) (2024) Paper No. e12946, 57.
\newblock \href {http://arxiv.org/abs/2012.14723v4}
  {\path{arXiv:2012.14723v4}}, \href {https://doi.org/10.1112/jlms.12946}
  {\path{doi:10.1112/jlms.12946}}.

\bibitem{BHLMR13}
V.~Bouchard, J.~Hutchinson, P.~Loliencar, M.~Meiers, M.~Rupert, A generalized
  topological recursion for arbitrary ramification, Annales Henri
  Poincar{\'{e}} 15~(1) (2013) 143--169.
\newblock \href {https://doi.org/10.1007/s00023-013-0233-0}
  {\path{doi:10.1007/s00023-013-0233-0}}.

\bibitem{BE13}
V.~Bouchard, B.~Eynard, {Think globally, compute locally}, Journal of High
  Energy Physics 2013~(2) (2013) 1--34.
\newblock \href {http://arxiv.org/abs/1211.2302} {\path{arXiv:1211.2302}},
  \href {https://doi.org/10.1007/JHEP02(2013)143}
  {\path{doi:10.1007/JHEP02(2013)143}}.

\bibitem{CN19}
L.~Chekhov, P.~Norbury, Topological recursion with hard edges, Internat. J.
  Math. 30~(3) (2019) 1950014, 29.
\newblock \href {http://arxiv.org/abs/1702.08631} {\path{arXiv:1702.08631}},
  \href {https://doi.org/10.1142/S0129167X19500149}
  {\path{doi:10.1142/S0129167X19500149}}.

\bibitem{DN18a}
N.~Do, P.~Norbury, Topological recursion for irregular spectral curves, J.
  Lond. Math. Soc. (2) 97~(3) (2018) 398--426.
\newblock \href {http://arxiv.org/abs/1412.8334} {\path{arXiv:1412.8334}},
  \href {https://doi.org/10.1112/jlms.12112} {\path{doi:10.1112/jlms.12112}}.

\bibitem{BBCCN18}
G.~Borot, V.~Bouchard, N.~K. Chidambaram, T.~Creutzig, D.~Noshchenko, Higher
  {A}iry structures, \texorpdfstring{$\mathcal{W}$}{W}-algebras and topological
  recursion, Mem. Amer. Math. Soc. 296~(1476) (2024) v+108.
\newblock \href {http://arxiv.org/abs/1812.08738} {\path{arXiv:1812.08738}},
  \href {https://doi.org/10.1090/memo/1476} {\path{doi:10.1090/memo/1476}}.

\bibitem{BKS24}
G.~Borot, R.~Kramer, Y.~Sch\"{u}ler, Higher {A}iry structures and topological
  recursion for singular spectral curves, Ann. Inst. Henri Poincar\'{e} D
  11~(1) (2024) 1--146.
\newblock \href {http://arxiv.org/abs/2010.03512v3}
  {\path{arXiv:2010.03512v3}}, \href {https://doi.org/10.4171/aihpd/168}
  {\path{doi:10.4171/aihpd/168}}.

\bibitem{BE09}
M.~Berg{\`e}re, B.~Eynard, Determinantal formulae and loop equations (2009).
\newblock \href {http://arxiv.org/abs/0901.3273} {\path{arXiv:0901.3273}},
  \href {https://doi.org/10.48550/arXiv.0901.3273}
  {\path{doi:10.48550/arXiv.0901.3273}}.

\bibitem{GS11}
S.~Gukov, P.~Sulkowski, {A}-polynomial, {B}-model, and quantization, Journal of
  High Energy Physics 02~(070) (2011) 1--57.
\newblock \href {http://arxiv.org/abs/1108.0002} {\path{arXiv:1108.0002}},
  \href {https://doi.org/10.1007/JHEP02(2012)070}
  {\path{doi:10.1007/JHEP02(2012)070}}.

\bibitem{N15}
P.~Norbury, Quantum curves and topological recursion, in: String-{M}ath 2014,
  Vol.~93 of Proc. Sympos. Pure Math., Amer. Math. Soc., Providence, RI, 2016,
  pp. 41--65.
\newblock \href {http://arxiv.org/abs/1502.04394} {\path{arXiv:1502.04394}}.

\bibitem{EG23}
B.~Eynard, E.~Garcia-Failde, From topological recursion to wave functions and
  {PDE}s quantizing hyperelliptic curves, Forum Math. Sigma 11 (2023) Paper No.
  e99, 42.
\newblock \href {http://arxiv.org/abs/1911.07795} {\path{arXiv:1911.07795}},
  \href {https://doi.org/10.1017/fms.2023.96} {\path{doi:10.1017/fms.2023.96}}.

\bibitem{MO22}
O.~Marchal, N.~Orantin, Quantization of hyper-elliptic curves from
  isomonodromic systems and topological recursion, J. Geom. Phys. 171 (2022)
  Paper No. 104407, 44.
\newblock \href {http://arxiv.org/abs/1911.07739} {\path{arXiv:1911.07739}},
  \href {https://doi.org/10.1016/j.geomphys.2021.104407}
  {\path{doi:10.1016/j.geomphys.2021.104407}}.

\bibitem{DNOPS15}
P.~{Dunin-Barkowski}, P.~{Norbury}, N.~{Orantin}, A.~{Popolitov}, S.~{Shadrin},
  Dubrovin's superpotential as a global spectral curve, J. Inst. Math. Jussieu
  (2017) 1--49\href {http://arxiv.org/abs/1509.06954}
  {\path{arXiv:1509.06954}}, \href {https://doi.org/10/c3f8}
  {\path{doi:10/c3f8}}.

\bibitem{BEMS10}
G.~Borot, B.~Eynard, M.~Mulase, B.~Safnuk, {A matrix model for simple Hurwitz
  numbers, and topological recursion}, Journal of Geometry and Physics 61~(2)
  (2010) 522--540.
\newblock \href {http://arxiv.org/abs/0906.1206} {\path{arXiv:0906.1206}},
  \href {https://doi.org/10.1016/j.geomphys.2010.10.017}
  {\path{doi:10.1016/j.geomphys.2010.10.017}}.

\bibitem{BDKS22}
B.~Bychkov, P.~Dunin-Barkowski, M.~Kazarian, S.~Shadrin, Explicit closed
  algebraic formulas for {O}rlov--{S}cherbin {$n$}-point functions, J. \'{E}c.
  polytech. Math. 9 (2022) 1121--1158.
\newblock \href {http://arxiv.org/abs/2008.13123} {\path{arXiv:2008.13123}},
  \href {https://doi.org/10.5802/jep.202} {\path{doi:10.5802/jep.202}}.

\bibitem{MSS13}
M.~Mulase, S.~Shadrin, L.~Spitz, The spectral curve and the {S}chr{\"o}dinger
  equation of double {H}urwitz numbers and higher spin structures, Commun. Num.
  Theor Phys. 07 (2013) 125--143.
\newblock \href {http://arxiv.org/abs/1301.5580} {\path{arXiv:1301.5580}},
  \href {https://doi.org/10.4310/CNTP.2013.v7.n1.a4}
  {\path{doi:10.4310/CNTP.2013.v7.n1.a4}}.

\bibitem{ALS16}
A.~Alexandrov, D.~Lewanski, S.~Shadrin, {Ramifications of Hurwitz theory, KP
  integrability and quantum curves}, J. High Energy Phys. 2016~(5) (2016)
  1--30.
\newblock \href {http://arxiv.org/abs/1512.07026} {\path{arXiv:1512.07026}},
  \href {https://doi.org/10.1007/JHEP05(2016)124}
  {\path{doi:10.1007/JHEP05(2016)124}}.

\bibitem{ABDKS24}
A.~Alexandrov, B.~Bychkov, P.~Dunin-Barkowski, M.~Kazarian, S.~Shadrin,
  {Symplectic duality via log topological recursion} (2024).
\newblock \href {http://arxiv.org/abs/2405.10720} {\path{arXiv:2405.10720}},
  \href {https://doi.org/10.48550/arXiv.2405.10720}
  {\path{doi:10.48550/arXiv.2405.10720}}.

\bibitem{ABDKS22}
A.~Alexandrov, B.~Bychkov, P.~Dunin-Barkowski, M.~Kazarian, S.~Shadrin, A
  universal formula for the \texorpdfstring{$x-y$}{x-y} swap in topological
  recursion (2022).
\newblock \href {http://arxiv.org/abs/2212.00320} {\path{arXiv:2212.00320}}.

\bibitem{ABDKS23}
A.~Alexandrov, B.~Bychkov, P.~Dunin-Barkowski, M.~Kazarian, S.~Shadrin, {Log
  topological recursion through the prism of $x-y$ swap} (2023).
\newblock \href {http://arxiv.org/abs/2312.16950} {\path{arXiv:2312.16950}},
  \href {https://doi.org/10.48550/arXiv.2212.00320}
  {\path{doi:10.48550/arXiv.2212.00320}}.

\bibitem{BP15b}
K.~Biswas, R.~P{\'e}rez-Marco, Log-{R}iemann surfaces (2015).
\newblock \href {http://arxiv.org/abs/1512.03776} {\path{arXiv:1512.03776}},
  \href {https://doi.org/10.48550/arXiv.1512.03776}
  {\path{doi:10.48550/arXiv.1512.03776}}.

\bibitem{PM19}
R.~P{\'e}rez-Marco, E{\~n}e product in the transalgebraic class (2019).
\newblock \href {http://arxiv.org/abs/1912.08557} {\path{arXiv:1912.08557}},
  \href {https://doi.org/10.48550/arXiv.1912.08557}
  {\path{doi:10.48550/arXiv.1912.08557}}.

\bibitem{BP15}
K.~Biswas, R.~P{\'e}rez-Marco, Log-{R}iemann surfaces, {C}aratheodory
  convergence and {E}uler's formula, in: Geometry, groups and dynamics, Vol.
  639 of Contemp. Math., Amer. Math. Soc., Providence, RI, 2015, pp. 197--203.
\newblock \href {http://arxiv.org/abs/1011.0535} {\path{arXiv:1011.0535}},
  \href {https://doi.org/10.1090/conm/639/12826}
  {\path{doi:10.1090/conm/639/12826}}.

\bibitem{BP15a}
K.~Biswas, R.~P{\'e}rez-Marco, Uniformization of simply connected finite type
  log-{R}iemann surfaces, in: Geometry, groups and dynamics, Vol. 639 of
  Contemp. Math., Amer. Math. Soc., Providence, RI, 2015, pp. 205--216.
\newblock \href {http://arxiv.org/abs/1011.0812} {\path{arXiv:1011.0812}},
  \href {https://doi.org/10.1090/conm/639/12827}
  {\path{doi:10.1090/conm/639/12827}}.

\bibitem{BP13}
K.~Biswas, R.~P{\'e}rez-Marco, Uniformization of higher genus finite type
  log-{R}iemann surfaces (2013).
\newblock \href {http://arxiv.org/abs/1305.2339} {\path{arXiv:1305.2339}},
  \href {https://doi.org/10.48550/arXiv.1305.2339}
  {\path{doi:10.48550/arXiv.1305.2339}}.

\bibitem{CGHJK}
R.~M. Corless, G.~H. Gonnet, D.~E.~G. Hare, D.~J. Jeffrey, D.~E. Knuth, On the
  {Lambert} {W} function, Advances in Computational Mathematics 5 (1996)
  329--359.
\newblock \href {https://doi.org/10.1007/BF02124750}
  {\path{doi:10.1007/BF02124750}}.

\bibitem{BBCKS23}
G.~Borot, V.~Bouchard, N.~K. Chidambaram, R.~Kramer, S.~Shadrin, Taking limits
  in topological recursion (2023).
\newblock \href {http://arxiv.org/abs/2309.01654} {\path{arXiv:2309.01654}},
  \href {https://doi.org/10.48550/arXiv.2309.01654}
  {\path{doi:10.48550/arXiv.2309.01654}}.

\bibitem{sagemath}
{The Sage Developers}, \href{https://www.sagemath.org}{{S}ageMath, the {S}age
  {M}athematics {S}oftware {S}ystem ({V}ersion 9.3)} (2021).
\newline\urlprefix\url{https://www.sagemath.org}

\bibitem{E04}
B.~Eynard, Topological expansion for the 1-{H}ermitian matrix model correlation
  functions, Journal of High Energy Physics 11~(031) (11 2004).
\newblock \href {http://arxiv.org/abs/hep-th/0407261}
  {\path{arXiv:hep-th/0407261}}, \href
  {https://doi.org/10.1088/1126-6708/2004/11/031}
  {\path{doi:10.1088/1126-6708/2004/11/031}}.

\bibitem{M90}
M.~L. Mehta, Random Matrices, 2nd Edition, Academic Press, New York, 1990.
\newblock \href {https://doi.org/10.1016/C2009-0-22297-5}
  {\path{doi:10.1016/C2009-0-22297-5}}.

\bibitem{B1895}
H.~F. Baker, \href{https://books.google.ca/books?id=k17KzQEACAAJ}{Examples of
  the application of {N}ewton's polygon to the theory of singular points of
  algebraic functions}, Trans. Cambridge Philos. Soc. 15 (1895) 403--450.
\newline\urlprefix\url{https://books.google.ca/books?id=k17KzQEACAAJ}

\bibitem{SSZ15}
S.~Shadrin, L.~Spitz, D.~Zvonkine, Equivalence of {ELSV} and
  {B}ouchard-{M}ari{\~n}o conjectures for $r$-spin {H}urwitz numbers, Math.
  Ann, 361~(3-4) (2015) 611--645.
\newblock \href {http://arxiv.org/abs/1306.6226} {\path{arXiv:1306.6226}},
  \href {https://doi.org/10.1007/s00208-014-1082-y}
  {\path{doi:10.1007/s00208-014-1082-y}}.

\bibitem{KMMM95}
S.~M. Kharchev, A.~V. Marshakov, A.~D. Mironov, A.~Y. Morozov, Generalized
  {K}azakov-{M}igdal-{K}ontsevich model: group theory aspects, Internat. J.
  Modern Phys. A 10~(14) (1995) 2015--2051.
\newblock \href {https://doi.org/10.1142/S0217751X9500098X}
  {\path{doi:10.1142/S0217751X9500098X}}.

\bibitem{OS01a}
A.~Y. Orlov, D.~M. Scherbin, Multivariate hypergeometric functions as
  {$\tau$}-functions of {T}oda lattice and {K}adomtsev-{P}etviashvili equation,
  Vol. 152/153, 2001, pp. 51--65, advances in nonlinear mathematics and
  science.
\newblock \href {https://doi.org/10.1016/S0167-2789(01)00158-0}
  {\path{doi:10.1016/S0167-2789(01)00158-0}}.

\bibitem{OS01b}
A.~Y. Orlov, D.~M. Scherbin, Hypergeometric solutions of soliton equations,
  Teoret. Mat. Fiz. 128~(1) (2001) 84--108.
\newblock \href {https://doi.org/10.1023/A:1010402200567}
  {\path{doi:10.1023/A:1010402200567}}.

\bibitem{HO15}
J.~Harnad, A.~Y. Orlov, {Hypergeometric \texorpdfstring{$ \tau
  $}{tau}-functions, {H}urwitz numbers and enumeration of paths}, Comm. Math.
  Phys. 338~(1) (2015) 267--284.
\newblock \href {http://arxiv.org/abs/1407.7800} {\path{arXiv:1407.7800}},
  \href {https://doi.org/10.1007/s00220-015-2329-5}
  {\path{doi:10.1007/s00220-015-2329-5}}.

\bibitem{DLN16}
N.~{Do}, O.~{Leigh}, P.~{Norbury}, {Orbifold Hurwitz numbers and Eynard-Orantin
  invariants}, Math. Res. Lett. 23~(5) (2016) 1281--1327.
\newblock \href {http://arxiv.org/abs/1212.6850} {\path{arXiv:1212.6850}},
  \href {https://doi.org/10.4310/MRL.2016.v23.n5.a3}
  {\path{doi:10.4310/MRL.2016.v23.n5.a3}}.

\bibitem{KPS22}
R.~Kramer, A.~Popolitov, S.~Shadrin, Topological recursion for monotone
  orbifold {H}urwitz numbers: a proof of the {D}o-{K}arev conjecture, Ann. Sc.
  Norm. Super. Pisa Cl. Sci. (5) 23~(2) (2022) 809--827.
\newblock \href {http://arxiv.org/abs/1909.02302} {\path{arXiv:1909.02302}},
  \href {https://doi.org/10.2422/2036-2145.201909\_010}
  {\path{doi:10.2422/2036-2145.201909\_010}}.

\bibitem{ACEH20}
A.~Alexandrov, G.~Chapuy, B.~Eynard, J.~Harnad, Weighted {H}urwitz numbers and
  topological recursion, Comm. Math. Phys. 375~(1) (2020) 237--305.
\newblock \href {http://arxiv.org/abs/math-ph/1806.09738}
  {\path{arXiv:math-ph/1806.09738}}, \href
  {https://doi.org/10.1007/s00220-020-03717-0}
  {\path{doi:10.1007/s00220-020-03717-0}}.

\bibitem{BS17}
G.~Borot, S.~Shadrin, Blobbed topological recursion: properties and
  applications, Math. Proc. Camb. Phil. Soc. 162 (2017) 39--87.
\newblock \href {http://arxiv.org/abs/1502.00981} {\path{arXiv:1502.00981}},
  \href {https://doi.org/10.1017/S0305004116000323}
  {\path{doi:10.1017/S0305004116000323}}.

\bibitem{Juc74}
A.-A.~A. Jucys, {Symmetric polynomials and the center of the symmetric group
  ring}, Reports on Mathematical Physics 5~(1) (1974) 107--112.

\bibitem{FH59}
H.~Farahat, G.~Higman, The centres of symmetric group rings, Proc. Roy. Soc.
  London Ser. A 250 (1959) 212--221.
\newblock \href {https://doi.org/10.1098/rspa.1959.0060}
  {\path{doi:10.1098/rspa.1959.0060}}.

\bibitem{C16}
A.~Chotai, Extending the {WKB}-topological recursion connection, Master's
  thesis, University of Alberta (08 2016).
\newblock \href {https://doi.org/10.7939/R3MW28K7G}
  {\path{doi:10.7939/R3MW28K7G}}.

\bibitem{Chio08}
A.~Chiodo, Towards an enumerative geometry of the moduli space of twisted
  curves and \texorpdfstring{$r$}{r}th roots, Compos. Math. 144~(6) (2008)
  1461--1496.
\newblock \href {http://arxiv.org/abs/math/0607324}
  {\path{arXiv:math/0607324}}, \href {https://doi.org/10/dnt2wg}
  {\path{doi:10/dnt2wg}}.

\bibitem{OP06a}
A.~Okounkov, R.~Pandharipande, Gromov-{W}itten theory, {H}urwitz theory, and
  completed cycles, Ann. of Math. (2) 163~(2) (2006) 517--560.
\newblock \href {http://arxiv.org/abs/math/0204305}
  {\path{arXiv:math/0204305}}, \href
  {https://doi.org/10.4007/annals.2006.163.517}
  {\path{doi:10.4007/annals.2006.163.517}}.

\bibitem{OP06}
A.~Okounkov, R.~Pandharipande, {The equivariant Gromov-Witten theory of
  \texorpdfstring{$\mathbb{P}^1$}{P1}}, Annals of Mathematics. Second Series
  163~(2) (2006) 561--605.
\newblock \href {http://arxiv.org/abs/math/0207233}
  {\path{arXiv:math/0207233}}, \href
  {https://doi.org/10.4007/annals.2006.163.561}
  {\path{doi:10.4007/annals.2006.163.561}}.

\bibitem{KS17}
M.~Kontsevich, Y.~Soibelman, Airy structures and symplectic geometry of
  topological recursion, in: Topological recursion and its influence in
  analysis, geometry, and topology, Vol. 100 of Proc. Sympos. Pure Math., Amer.
  Math. Soc., Providence, RI, 2018, pp. 433--489.
\newblock \href {http://arxiv.org/abs/1701.09137} {\path{arXiv:1701.09137}},
  \href {https://doi.org/10.1090/pspum/100/01765}
  {\path{doi:10.1090/pspum/100/01765}}.

\bibitem{ABCO24}
J.~E. Andersen, G.~Borot, L.~O. Chekhov, N.~Orantin, The {ABCD} of topological
  recursion, Adv. Math. 439 (2024) Paper No. 109473, 105.
\newblock \href {http://arxiv.org/abs/1703.03307} {\path{arXiv:1703.03307}},
  \href {https://doi.org/10.1016/j.aim.2023.109473}
  {\path{doi:10.1016/j.aim.2023.109473}}.

\end{thebibliography}

\end{document}